\documentclass[prodmode,acmtecs]{acmsmall} % Aptara syntax
\usepackage{verbatim}
\usepackage{comment}
\usepackage{graphics}
\usepackage{graphicx}
\usepackage{subfig}
\usepackage{xr}
\usepackage{lipsum}
\usepackage{float}
\usepackage{subfloat}
\usepackage{mathptmx}      % use Times fonts if available on your TeX system
\usepackage{epstopdf}
\usepackage{multirow}
\usepackage{amsmath}
\usepackage{mwe}
\usepackage{setspace}

\newcommand{\specialcell}[2][c]{%
	\begin{tabular}[#1]{@{}c@{}}#2\end{tabular}}

\begin{document}
% Page heads
\markboth{S. \v{S}\'{c}epanovi\'{c}  et al.}{Semantic homophily in online communication: evidence from Twitter}
% Title portion
\title{Semantic homophily in online communication: evidence from Twitter}
\author{SANjA \v{S}\'{C}EPANOVI\'{C}
\affil{Aalto University}
IGOR MISHKOVSKI
\affil{University Ss. Cyril and Methodius}
BRUNO GON\c{C}ALVES
\affil{New York University}
NGUYEN TRUNG HIEU
\affil{University of Tampere}
PAN HUI
\affil{Hong Kong University of Science and Technology}
}

\begin{abstract}
People are observed to assortatively connect on a set of traits. This phenomenon, termed assortative mixing or sometimes homophily, can be quantified through assortativity coefficient in social networks. Uncovering the exact causes of strong assortative mixing found in social networks has been a research challenge. Among the main suggested causes from sociology are the tendency of similar individuals to connect (often itself referred as homophily) and the social influence among already connected individuals. Distinguishing between these tendencies and other plausible causes and quantifying their contribution to the amount of assortative mixing has been a difficult task, and proven not even possible from observational data. However, another task of similar importance to researchers and in practice can be tackled, as we present here: understanding the exact mechanisms of interplay between these tendencies and the underlying social network structure. Namely, in addition to the mentioned assortativity coefficient, there are several other static and temporal network properties and substructures that can be linked to the tendencies of homophily and social influence in the social network and we herein investigate those.

Concretely, we tackle a computer-mediated \textit{communication network} (based on Twitter mentions) and a particular type of assortative mixing that can be inferred from the semantic features of communication content that we term \textit{semantic homophily}. 
Our work, to the best of our knowledge, is the first to offer an in-depth analysis on semantic homophily in a communication network and the interplay between them. We quantify diverse levels of semantic homophily, identify the semantic aspects that are the drivers of observed homophily, show insights in its temporal evolution and finally, we present its intricate interplay with the communication network on Twitter. By analyzing these mechanisms we increase understanding on what are the semantic aspects that shape and how they shape the human computer-mediated communication. In addition, our analysis framework presented on this concrete case can be easily adapted, extended and applied on other type of social networks and for different types of homophily.
\end{abstract}


\begin{bottomstuff}
\textbf{Author's addresses}: S. \v{S}\'{c}epanovi\'{c}, Aalto University, Department of Computer Science, Espoo, 02150, Finland; I. Mishkovski, University Ss. Cyril and Methodius, Faculty of Computer Science, Skopje, 1000, Macedonia; H. Nguyen Trung, University of Tampere, Tampere, 33100, Finland; P. Hui, Hong Kong University of Science and Technology, Department of Computer Science, Clear Water Bay, Kowloon, Hong Kong; B. Gon\c{c}alves, New York University, Center for Data Science, New York, 10003, U.S.; 
\end{bottomstuff}
\maketitle

\section{Introduction}

\linespread{1.5}

	\textbf{Homophily} \cite{lazarsfeld1954friendship,McPherson2001} (sometimes referred as selection \cite{leenders1997longitudinal,crandall2008feedback}) represents a tendency of individuals who are similar on some traits to connect to each other (become friends, follow each other, communicate etc.) in a social network. \textbf{Social influence} (peer pressure) is in a way an inverse tendency for people to become similar on some traits or to adopt certain behavior from their social contacts. Both, homophily and social influence result in a higher correlation (assortative mixing) on certain traits between connected than between random users in a network. This \textit{assortative mixing} property (also in some studies referred as social correlation \cite{anagnostopoulos2008influence}) is repeatedly confirmed in social network analysis literature \cite{bollen2011happiness,de2010birds,anagnostopoulos2008influence,aral2012identifying,tang2013exploiting}. A question remains, to what extent is the observed assortative mixing a result of an underlying homophily that shapes the formation of the network or of the social influence taking place in an already formed network \cite{leenders1997longitudinal}. A third factor that could be the cause of social correlation is a common \textbf{external influence}. Moreover, a combination of these factors is often at play. For instance, an external factor might have non-homogeneous adoption in the network because friends could have a higher common latent propensity for it and adopt it to a larger extent than non-friends. 
	Distinguishing between these factors as the main  causes of assortative mixing has been a challenge, and proven not even possible from observational data \cite{shalizi2011homophily}. 
	
	Extensive research is conducted in sociology on homophily in social networks as abstractions of diverse groups in society (see the seminal review by McPherson et al.\ \cite{McPherson2001}). Classical paper \cite{lazarsfeld1954friendship} introduced two basic levels or dimensions of homophily:
	status and value homophily.
	\textbf{Status homophily} relates to any formal or perceived status among individuals. It includes some of the most important social dimensions, such as \textit{race, ethnicity, sex, age, education, occupation} and \textit{religion}.
	\textbf{Value homophily} relates to our internal states that might shape the future behavior; for example: \textit{abilities (intelligence), aspirations}, and \textit{attitudes (political orientation)}, regardless of the differences in status. 
	
	In addition to individuals connecting to similar individuals, another suggested mechanism driving homophily is the process of \textit{tie (link) dissolution} over time that happens more often among non-similar individuals. However, both mechanisms, of similarity and dissimilarity are not enough to explain a particular clustered (community) structure found in social networks. Sociologists have proposed that instead of only being driven by similarity, a tie is actually often formed around a specific \textbf{focus of homophily} \cite{feld1981focused}. McPherson et al.\ \cite{McPherson2001} offer a nice overview on possible different foci, and below we briefly discuss some of them. \textit{Geographical proximity} is considered one of the most important foci of homophily, simply put, because we are more likely to have contacts with the people who are geographically closer to us. The ties induced by proximity in space are often weak; however, they leave more potential for stronger ties formation. It is worth noting that the advent of new technologies over time did not remove this pattern of geographical homophily and recent empirical research on online social networks finds that people online still tend to connect more
	often to geographically close people (in Twitter network \cite{Kulshrestha2012,DeChoudhury2011}; in Microsoft IMS \cite{Leskovec2008}; 
	in Facebook social graph \cite{Ugander2011}; in mobile phone communication \cite{blondel2010regions}). The only study we found that reports no significant effects of geographical homophily tackles organization-individual relationship on Twitter \cite{sun2017link}. Another important focus that causes homophily are \textit{family ties}. Family ties are
	an interesting focus of formation that causes people who are similar on some aspects and as well who are dissimilar on certain other aspects to connect. For this reason, when it comes to family ties, we find the largest geographic, age, sex and educational heterophily;
	but at the same time, the largest race, religious and ethnic homophily.
	\textit{Organizational foci} turns to be the most important cause of ties that are not relatives nor family-bound. These foci include schoolmates, colleagues from work and voluntary organizations.
	A more implicit cause of homophily shows to be \textit{network position}. Research finding exist that holding a same position inside an organization will
	induce larger homophily between individuals than it would be the case if the ties were random \cite{lincoln1979work}. 
	Another, more internal, focus for homophily lies inside perceived similarity and shared knowledge, and it is termed \textit{cognitive processes}. It is particularly notable among teenagers who tend to
	connect to those who are perceived to be more similar on some of the internal traits. Looking back at the described homophily traits and foci, it is not easy to make a clear distinction between the consequences of homophily and the causes or origins of it. Whether cognitive processes focus among teenagers causes them to become friends with similar ones; or whether the friend teenagers influence each other and hence become similar on a value homophily level?
	
	\subsection{Terminology} 
	{\textbf{Communication network:}}
	In this study, the social network of interest is a \textit{computer-mediated communication} \cite{thurlow2004computer} \textit{network} from Twitter. It is formed of nodes representing Twitter users, and the directed links representing the tweets in which they mention (reply to) each other. The tweet content is also included. Hence, our network can be seen as a subtype of previously introduced \textit{interaction networks} on Facebook \cite{wilson2009user}. 
	Throughout the rest of this study we simply use the term $\mathtt{communication\ network}$ referring to this definition. While in general communication refers to exchanging of information, we recognize the potential of Twitter mentions to carry two different forms of communication. In the first form, the source is directly addressing the receiver, and in the second form, there is a sort of authority attribution where the source comments to the rest of the Twitter users about the receiver (this could be a critique as well).
	\textbf{Communication intensity (CI)} in our network denotes the weight on the links i.e., the number of mentions between a pair of users.

	{\textbf{Semantic homophily:}}  Importantly, in many related studies the term \textit{homophily} is used with the meaning of \textit{assortative mixing} as we introduced it here (one possible reason being described indistinguishability of presented phenomena). We also use the term \textbf{semantic homophily} when talking about assortative mixing on semantic aspects of communication. In the light of introduced definitions, a more precise term to use would be \textit{semantic assortative mixing}. However, we select to talk about semantic homophily in order to be consistent with the related studies and also since we do not focus on distinguishing between homophily and social influence. Hence, using an umbrella term semantic homophily to cover both tendencies is simpler. When at some point we talk about one of the tendencies in particular, we then point that out. In order to analyze semantic homophily, we tackle following \textit{semantic aspects of communication}:
	 \begin{itemize}
	 \item \textbf{semantic relatedness (SR)} between the tweet contents of two users. SR is a more general metric compared to semantic similarity \cite{harispe2015semantic} since in addition to similarity, it includes also any other relation between the terms, such as antonymes (opposite terms) \cite{lehrer1982antonymy} and meronymes (a term is a part of or member of the other) \cite{murphy2003semantic}. For instance, the term \textit{airplane} is similar to the term \textit{spacecraft}. The same term is related to \textit{car}, \textit{train} or \textit{wing}, but not similar to them. SR relation between tweets of a pair of users is quantified by a value ranging from 0 (not related at all) to 1 (maximally related);	 
	\item \textbf{sentiment} of user tweet content. The sentiment value ranges from -1 (negative) to 1 (positive);	 
	  \item the most important \textbf{entities} (people, companies, organizations, cities, geographic features etc.), \textbf{concepts} (abstract ideas in the text: \textit{for example, if an article mentions CERN and the Higgs boson, it will have Large Hadron Collider as a concept even if the term is not mentioned explicitly in the page} \cite{AlchemyAPI}) and \textbf{taxonomy} (a hierarchy that helps to classify the content into its most likely topic category) of user tweets content. 
	\end{itemize}
	
	\textbf{Communication propensity ($\hat{{cp}}$)} is defined as function of some property and represents the extent to which the observed communication and its intensity diverge from what would be expected in a uniformly random setting with respect to that property. We investigate communication propensity in our network with respect to SR threshold in the network (formula is given in Section \ref{sec:SR_comm}).
	
	{\textbf{Social capital:}} Among a variety of definitions from sociology \cite{portes2000social,bourdieu2011forms}, one that translates well to our case introduces social capital as the actual and potential resources that are linked to the ego's social network and relationships.
	Hence, in similarity to the previous study  on socio-semantic networks \cite{roth2010social}, we define social capital in our $\mathtt{communication\ network}$ as the total \textbf{number of contacts} (degree in the unweighted network) or the total \textbf{communication intensity} (degree in the weighted network). Moreover, we can divide the social capital, defined as such, in both cases to \textbf{popularity} (if we look at in-degree) and \textbf{communication activity} (if looking at out-degree). To sum up, thanks to our network being directed and weighted, we can introduce \textit{four types of social capital} in it: (i) popularity in terms of number of communication contacts and (ii) popularity in terms of communication intensity and (iii) activity in terms of number of contacts and (iv) activity in terms of communication intensity.
	  
{\textbf{Semantic capital}} denotes the amount of diversity of user tweet content with respect to the introduced semantic attributes, similarly as in \cite{roth2010social}.

\textbf{Relative status} of two users can be defined for both social and semantic capital and represents the difference of their respective status ranks. Finally, for a single user we define \textbf{status inconsistency} \cite{lenski1954status,rogers1970homophily} as a relative difference between his/her ranking among all users on social and semantic capital. Status inconsistent individuals tend to be highly ranked on some aspects and lowly ranked on others. This is suggested to be an attribute of individuals who are drivers of social change \cite{lenski1954status}. Status inconsistency can be defined on a communication \textit{link}, as well, as a measure of inconsistency of both participating users (we give a formal definition in Section \ref{sec:soc_sem_interplay}).

\subsection{Contributions} 
In this study, we offer a deeper understanding on the mechanisms of semantic homophily and how they are shaping the structure and properties of the underlying $\mathtt{communication\ network}$.

 While homophily has been identified in a diverse set of social networks, most of the studies investigated friendship, followers or citation type of ties. Interaction ties are more suitable for inferring meaningful social relationships \cite{wilson2009user}. Our analysis is on the \textbf{communication ties} formed from Twitter mentions (replies), that are a subtype of interaction ties. The ties in our network are not only formed once (such as friendship and followership), but they require an active engagement over time. The nature of the mention network is fundamentally different from follower/friendship network in Twitter \cite{bliss2012twitter}. For instance, the reciprocity of the followers network is found  to be around 22\% \cite{kwak2010twitter} which is lower compared to the other social networks. The reciprocity of our mention network is 64\%, considerably higher. When a user A follows a user B it simply states some type of potential interest in what B has to say. Depending on the different time zones and the number of other users that A is already following s/he might not even get to see any of B's tweets. In the case of our $\mathtt{communication\ network}$ we can clearly point to interactions and information diffusion between users (when the user A mentions the user B), instead of simply speculating about it when using the friendship/follower network. While retweet network allows for similar information diffusion analysis, its nature is also shown to be importantly different from mention network \cite{conover2011political}. Finally, observance of communication interruption in time allows us to define a \textit{tie dissolution (link decommission)}. As discussed in \cite{bliss2012twitter}, considering link decommission resolves issues of analyzing social network with stale links without current functional role. 

The focus of our work is on \textbf{semantic homophily}. While several other studies have tackled some aspects of semantic homophily, as we discuss in the related work, 
to the best of our knowledge this is the first study aiming towards a comprehensive picture on the role of semantic homophily in communication. We offer an in-depth and detailed investigation of semantic homophily: from quantification and qualitative assessment, through temporal evolution to its interplay with community structure of $\mathtt{communication\ network}$. 

\begin{figure}
	\centering
	\includegraphics[width=0.99\linewidth]{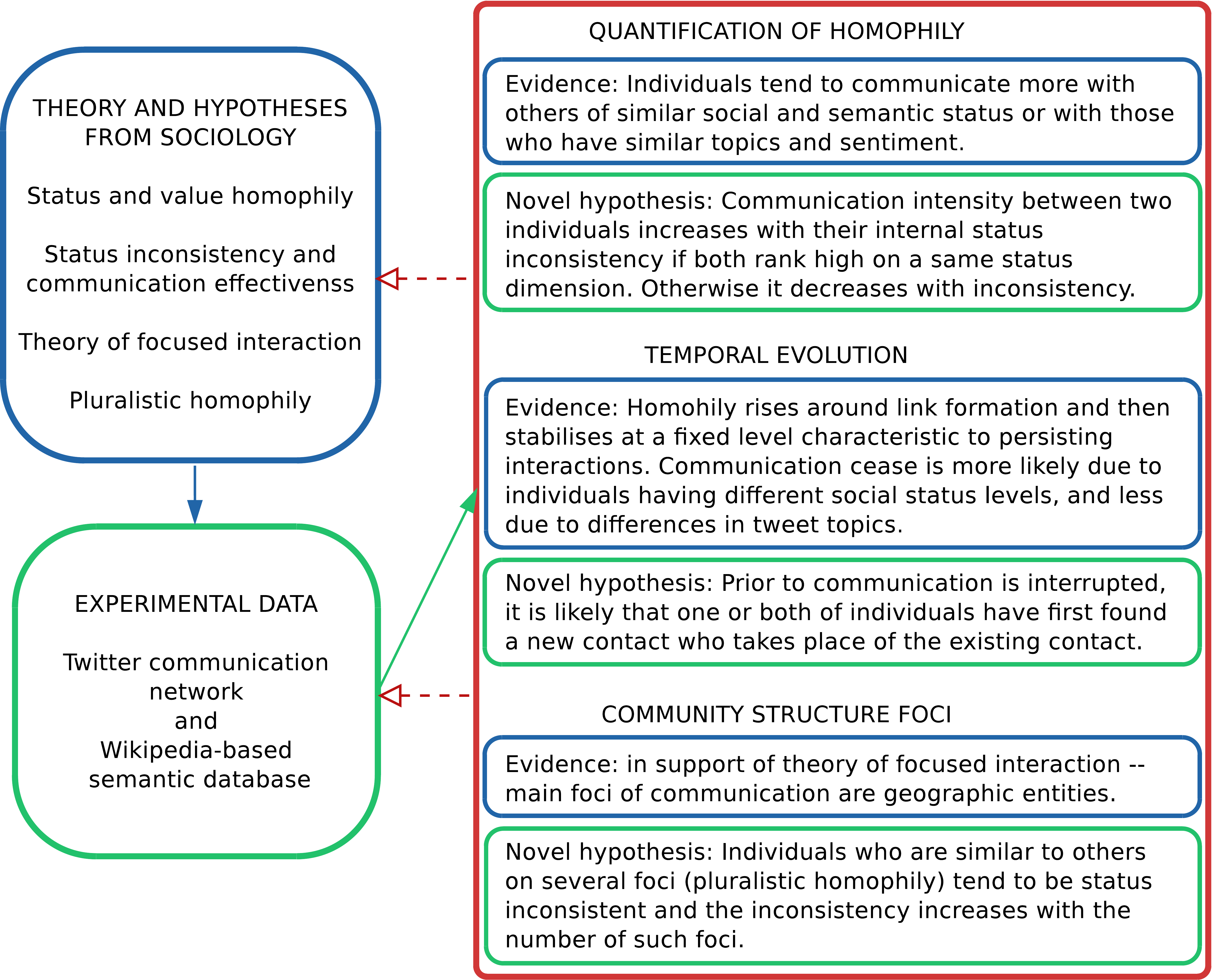}
	\caption{\textbf{General framework and main contributions of our study.} In blue frames we denote the evidence found in Twitter experimental data for the existing theories from sociology. During data analysis, we also find evidence pointing to some novel hypotheses, presented in green fames. However, such evidence should be evaluated and confirmed in several other datasets before any general conclusions about semantic homophily in communication can be reached.}
	\label{fig:contributions}
\end{figure}

Fig.\ \ref{fig:contributions} presents the general framework for our study and lists several main contributions. For a full list of our contributions, we refer the reader to Table \ref{t:contr_summ} in Discussion \ref{sec:discussion}. As depicted in Fig.\ \ref{fig:contributions}, we operate on experimental datasets (from Twitter and Wikipedia), while at the same time building on existing sociological findings and theories. By testing for existence of status and value homophily, we confirm that these general theories from sociology hold in a communication social network. In addition, we identify the aspects of homophily that are specific for communication, compared to other types of social networks. A natural method to asses homophily in communication is through semantic aspects of it. 

At first we \textbf{quantify} diverse aspects of semantic homophily in the network. We start by uncovering the subtle relationship between SR among users and the intensity of their communication. Next we introduce measures of social and semantic status of users and show that communication network exhibits assortativity on those metrics. This confirms sociological theories on status level homophily. We also show that such status correlation increases with strength of ties in communication. In addition, analysis of the interplay between two types of capital reveals large status heterogeneity among users. Accordingly, we find that status inconsistency of one or both communicating parties correlates with intensity of communication. 

Next we focus on \textbf{temporal evolution} of semantic homophily. We detect temporal increase in average semantic relatedness among users and investigate new links formation as a possible cause. However, we also find a number of links that get decommissioned in time. After comparing relative statuses of users who stop communication, we present evidence that decommission is more due to status than to value heterophily. 
Finally, the analysis on the 
 \textbf{community structure} of the $\mathtt{communication\ network}$ (structural communities) reveals the semantic foci around which such communities are formed (functional communities). In this way, we find evidence for Feld's theory of focused organization of social ties \cite{feld1981focused}  and also identify some of such foci around which communication ties are formed. In the end, we delve into the mechanisms of pluralistic homophily (assortative mixing as a result of several foci), and describe specificity of users who have such a position in $\mathtt{communication\ network}$.		 
	
The rest of the paper is organized as follows. Section \ref{sec:rw} presents related research literature. In Section \ref{sec:data} we describe two Web datasets (from Twitter and Wikipedia) used for analysis, as well as the framework of analysis consisting of a communication (Section \ref{sec:comm_layer}) and semantic (Section \ref{sec:semantic_layer}) layer. Quantification of different forms of homophily in our network is presented in Section \ref{sec:quantifying}: social status homophily in \ref{sec:da_mention_network} and semantic status and value homophily in \ref{sec:sem_cap}. Insights on the relationship between these forms of capital, and relative status and status inconsistency are given in \ref{sec:soc_sem_interplay}. The relationship between semantic relatedness and communication are reported in Subsection \ref{sec:SR_comm}. 
Temporal aspects of semantic homophily, from link formation and dissolution to persisting interactions, are discussed in Section \ref{sec:edges_temporal_SR}. Community structure and focused organization of social ties are the topic in Section \ref{sec:comm_str}. Pluralistic homophily is also characterized in this section. The article concludes with a discussion and final remarks on future research directions in Section \ref{sec:discussion}.

\section{Related work}
\label{sec:rw}
% epistemic socio-semantic networks
Knowledge networks representing scientific collaboration and blogger citations are studied in \cite{roth2010social}. This study is similar to ours in that the joint dynamics and co-evolution of the social and socio-semantic structures is analyzed in these knowledge networks. Our work is different since we focus on another type of a network (communication). Hence, we respond in part to the call by Roth and Cointet \cite{roth2010social} to analyze some of the epistemic patterns, which they found in the scientist and blogger communities, in other type of communities. Moreover, while they only investigate social link formation, we are also able to investigate \textit{link decommission (disconnection)}, thanks to the type of the network we analyze. Therefore, our work offers an additional understanding on the temporal interplay between semantic and social structures. Another important difference is that we offer considerably deeper semantic aspects analysis. Compared to a hand-picked set of categories used in \cite{roth2010social}, our Wikipedia-based database in combination with Alchemy API provide us with richer insights on entities, categories, taxonomy and also sentiment of communication.

% Twitter homophily
A recent study on Twitter analyzes homophily on the status (defined as the difference in the follower counts) and the value (tweet contents, common followees, location, age etc.) levels \cite{sun2017link}. There are several important differences to our work: the focus of their study is on reciprocal followers network (instead of mention network in our case), homophily is analyzed on the organization-individual relationship (whereas we focus on individual-individual relationship) and there is no focus on community analysis or temporal aspects of homophily as in our study.

% temporal homophily
However, there are several previous studies in online settings that have analyzed the \textit{temporal interplay between homophily and social ties}. Crandal et al.\ \cite{crandall2008feedback} find that the homophily between two Wikipedia admin users sharply rises some time before the tie formation and after that continues to slowly grow. This is interpreted so that, at first, homophily plays a role in the tie formation, but after that, the tie plays a role in the continuous increase of homophily. Another similar study on Flickr \cite{zeng2013social}, finds more subtle insights: the users who have similar popularity (defined as the average number of favorites for their photos) are more likely to diverge in similarity after the tie formation; while the similarity continues to grow for the users who have a larger popularity difference. This is explained by the tendency of users to stay unique and diverse in their uploaded content from equally popular users. Besides focusing on a different type of social ties -- communication, our work extends these previous studies with the insights on interplay of homophily and \textit{tie (link) decommission} that they have not investigated. In addition, we also uncover the relationship between introduced social and semantic forms of capital and homophily around the time of link formation and decommission.

% community, geography, language
Significant homophilous foci on Facebook \cite{barnett2015predicting} are found to be geographic proximity, language, civilization, and migration.
The analysis performed on 3 online datasets: Last.fm, Flickr and aNobii \cite{aiello2012friendship}, presents how homophily information can be used for link prediction. The authors present best accuracy in the case of aNobii (92\%) when combining multiple features: in-degree, activity, number of distinct tags, assortativity of users in terms of topics etc. A conclusion is that the distinct language groups present in the aNobii dataset, which are quite homogenous and non-mixing, support the prediction accuracy. 
Halberstam et al.\ \cite{halberstam2014homophily} analyze communication on Twitter (comprising both retweets and mentions of political candidates) in similarity to us, however, with a different aim -- to understand information diffusion. They find a greater degree of homophily exhibited and also more connections per node in larger communities.

% homophily and or influence
Below we mention several other studies that have tackled homophily in online settings, but with a different focus from us. A number of studies are conducted toward \textit{distinguishing between influence and homophily} \cite{aral2009distinguishing,la2010randomization,anagnostopoulos2008influence} reporting different levels and proportions of the two traits in online social networks. For example, De Choudhury et al.\ \cite{de2010birds} quantified the impact of various types of homophily on influence on Twitter. Users were given homophilous traits based on attributes such as: location, information roles they take (generators, mediators and receptors), content creation (meformer, informer) and activity behavior (number of tweets per period of time). However, it is later shown that in empirical settings these tendencies are indistinguishable due to confounding effects \cite{shalizi2011homophily}. A couple of more recent papers tackled this research challenge in controlled experiments. The experiment on Facebook found that the probability for a user to share a link increases with the number of friends who shared the same link even without the user being exposed to their link shares \cite{bakshy2012role}. Hence this controlled experiment confirmed homophily or some unobserved common external influence taking place in the network.
 
%%%%%%%%%%%%%%%%%%%%%%%%%%%%%%%%%%%%%%%%%%%%%%%%%%%%%%%%%%%%%%%%%%%%%%%%%%%%%%%%%%%%%%%%%%%
\section{Datasets and framework for analysis}
\label{sec:data}
%%%%%%%%%%%%%%%%%%%%%%%%%%%%%%%%%%%%%%%%%%%%%%%%%%%%%%%%%%%%%%%%%%%%%%%%%%%%%%%%%%%%%%%%%%%%%%%%%%%%%
 \subsection{Communication layer: Twitter mention network}
\label{sec:comm_layer}
%%%%%%%%%%%%%%%%%%%%%%%%%%%%%%%%%%%%%%%%%%%%%%%%%%%%%%%%%%%%%%%%%%%%%%%%%%%%%%%%%%%%%%%%%%%%%%%%%%%%%
 \begin{table}
	\centering
	\caption{Twitter dataset filtering steps statistics}\label{t:twitter}
	\begin{tabular}[t]{c|l|l} \hline
		\textbf{\textit{Dataset}}&\textbf{\textit{mentions}}&\textbf{\textit{users}}\\ \hline
		original download &12 441 636&  547 368\\ 
		English language &2 527 990& 284 100\\ 
		users $>20$ tweets & 1 344 692 & 29 616\\ 
		\textbf{internal replies} & \textbf{744 821} & \textbf{26 717} \\
		\hline\end{tabular}	
\end{table}
 Our initial dataset contains $12,441,636$ mentions (tweets including @username) among $547,368$ users over the course of $6$ months (May-Nov 2011). \textit{All internal mentions} are included, meaning, each time when a user from our dataset mentions a user from outside, we did not keep such tweets, but all the mentions among the users in the dataset are present. 
 
 In order to have a well suited dataset for the intended analysis, we perform several cleaning and filtering steps described below. The initial dataset includes tweets in several languages, so we  filter it to select only English tweets and from the users who mostly tweet in English. We use NLTK Python library \cite{bird2009natural} in this step. After the language filtering, the dataset is reduced to $20\%$ of its original size in terms of tweets, while the number of users halved. For semantic analysis, individual tweets are often too small and noisy, so the next step involves filtering the remaining users based on their total number of tweets. Upon research and pre-test with the semantic knowledge database that we built (described in the following subsection), a threshold of minimum $20$ tweets is selected. After this step, the dataset contains $29,616$ users. Finally, again keeping only the internal replies withing this group of users, we end up with $26,717$ users in our final dataset for analysis (see Table \ref{t:twitter}).
 
 From the final filtered dataset we build our analysis target, the $\mathtt{communication\ network}$, $G=(V,E,W)$. The nodes $u_i, u_j \in V$ represent Twitter users; they are connected with a directed edge $e_{ij} = (u_i, u_j) \in E$ if a user $u_i$ mentions $u_j$, and the edge is assigned the weight $w_{ij} = (u_i, u_j) \in W$ equal to the communication intensity (total number of such mentions). Properties of the $\mathtt{communication\ network}$ are given in Table \ref{t:ment_stats}. Finally, at some points we will look at undirected and/or unweighted versions of the presented network. When we do so, it will be pointed out, otherwise, whenever we discuss $\mathtt{communication\ network}$ it refers to the weighted and directed network described here.
%%%%%%%%%%%%%%%%%%%%%%%%%%%%%%%%%%%%%%%%%%%%%%%%%%%%%%%%%%%%%%%%%%%%%%%%%%%%%%%%%%%%%%%%%%%%%%%%%%%%%
%%%%%%%%%%%%%%%%%%%%%%%%%%%%%%%%%%%%%%%%%%%%%%%%%%%%%%%%%%%%%%%%%%%%%%%%%%%%%%%%%%%%%%%%%%%%%%%%%%%%%
\begin{table}	
	\centering
	\caption{$\mathtt{Communication\ network}$ statistics}
	\begin{minipage}{0.30\textwidth}
		
		\begin{tabular}[t]{c|l} \hline
			\textbf{\textit{Network parameter}}&\textbf{\textit{value}}\\ \hline
			Nodes & 26 717 \\ 
			Edges & 99 910 \\ 
			Avg weighted deg.& 55.75 \\ 
			Avg clustering coeff.& 0.051  \\  
			\hline\end{tabular}
		
	\end{minipage}\quad \quad \quad \quad \quad \quad
	\begin{minipage}{0.30\textwidth}
		
		\begin{tabular}[t]{c|l} \hline
			\textbf{\textit{Network parameter}}&\textbf{\textit{value}}\\ \hline
			Max out-degree  & 1358 \\ 
			Max in-degree  & 3228 \\  
			Diameter & 29 \\ 
			Density & 0.00014 \\ 
			\hline\end{tabular}
	\end{minipage}	
	\label{t:ment_stats}
\end{table}
%%%%%%%%%%%%%%%%%%%%%%%%%%%%%%%%%%%%%%%%%%%%%%%%%%%%%%%%%%%%%%%%%%%%%%%%%%%%%%%%%%%%%%%%%%%%%%%%%%%%%
%%%%%%%%%%%%%%%%%%%%%%%%%%%%%%%%%%%%%%%%%%%%%%%%%%%%%%%%%%%%%%%%%%%%%%%%%%%%%%%%%%%%%%%%%%%%%%%%%%%%%
%%%%%%%%%%%%%%%%%%%%%%%%%%%%%%%%%%%%%%%%%%%%%%%%%%%%%%%%%%%%%%%%%%%%%%%%%%%%%%%%%%%%%%%%%%%%%%%%%%%%%
\subsection{Semantic layers: Semantic enrichment of communication network} 
\label{sec:semantic_layer}
%%%%%%%%%%%%%%%%%%%%%%%%%%%%%%%%%%%%%%%%%%%%%%%%%%%%%%%%%%%%%%%%%%%%%%%%%%%%%%%%%%%%%%%%%%%%%%%%%%%%%
On top of the communication layer, we extract another, \textit{semantic layer} from the Twitter data. Concretely, we apply two semantic analysis procedures that  enrich our $\mathtt{communication\ network}$ in terms of \textit{node} and \textit{edge attributes}.
The first procedure is based on 
 \textbf{Wikipedia semantic relatedness database} that we build from a whole English Wikipedia dump according to the Explicit Semantic Relatedness (ESA) algorithm \cite{gabrilovich2009wikipedia,gabrilovich2007computing}. The second procedure employs an existing, \textbf{natural language processing API, AlchemyAPI} \cite{AlchemyAPI} from IBM. Wikipedia SR database provides enrichment for both, edges (SR between tweets of two users) and nodes (extracted Wikipedia concepts relevant to the user tweets -- see following paragraph for details). AlchemyAPI provides an additional set of node attributes: concepts, entities, taxonomy and sentiment of the user tweets. We describe both procedures and the enrichment they provide in more detail in the following.

%%%%%%%%%%%%%%%%%%%%%%%%%%%%%%%%%%%%%%%%%%%%%%%%%%%%%%%%%%%%%%%%%%%%%%%%%%%%%%%%%%%%%%%%%%%%%%%%%%%%%
\subsubsection{Wikipedia Semantic Relatedness database}
\label{sec:SR_db}
%%%%%%%%%%%%%%%%%%%%%%%%%%%%%%%%%%%%%%%%%%%%%%%%%%%%%%%%%%%%%%%%%%%%%%%%%%%%%%%%%%%%%%%%%%%%%%%%%%%%%
The semantic layer includes a network of users featuring semantic relatedness (SR) between their tweets collections as edge weights, we refer to it as the $\mathtt{SR\ network}$. The $\mathtt{SR\ network}$ is based on SR knowledge database built using a Wikipedia XML dump from April 2015 (for details see Methods \ref{sec:methods}). In addition to SR scores, from the Wikipedia SR database, for each user we can also obtain their corresponding Wikipedia \textbf{concept vectors $CV$s}. $CV$s are formed of relevant Wikipedia concepts (articles) describing semantically user tweet contents. 

In a somewhat computationally demanding task, we calculate the SR scores between \textit{all the user pairs} (not just those who communicate and are connected in $\mathtt{communication\ network}$), resulting in a full $\mathtt{SR\ network}$. Distribution of SR values of the full $\mathtt{SR\ network}$ is shown in Fig. \ref{fig:sr_distributions} (right). During the analysis, we also apply different thresholds ($SR_{th}$) on the edge weights and obtain several SR sub-networks, which we denote $\mathtt{SR_{th}\ network}$s.

\subsubsection{AlchemyAPI}
\label{sec:method_taxonomy}
%%%%%%%%%%%%%%%%%%%%%%%%%%%%%%%%%%%%%%%%%%%%%%%%%%%%%%%%%%%%%%%%%%%%%%%%%%%%%%%%%%%%%%%%%%%%%%%%%%%%%
AlchemyAPI \cite{AlchemyAPI} performs natural language processing (NLP) and machine learning (ML) analysis. We send individual user tweets collections for analysis and AlchemyAPI returns semantic meta-data from the content. Not all are relevant for our study but we utilize following: sentiment score, taxonomy, concepts, entities and keywords. Hence, based on the output, we assign a set of attributes to users: the overall sentiment of his/her tweets (a real number between -1 for fully negative and 1 for fully positive), the taxonomy hierarchy representing topics, concepts, entities and keywords found relevant in the tweets. For each of the elements in the output, AlchemyAPI also returns corresponding relevance score, that we utilize to filter for most relevant semantic attributes. 

Based on the evaluations in the literature, we believe that AlchemyAPI is a suitable
choice to support our work. In \cite{meehan2013context} it was shown that the sentiment analysis obtained from AlchemyAPI achieved accuracy of 86\% on a corpus of 5,370 tweets employed by an intelligent recommendation system for tourism. The AlchemyAPI's performance on a number of datasets and in different contexts was also shown in \cite{rizzo2011nerd} and \cite{saif2012semantic}, where AlchemyAPI outperfomed  Zemanta\footnote{{http://blog.zemanta.com/}}, OpenCalais\footnote{http://www.opencalais.com/}, Extractiv\footnote{http://extractiv.com/} and DBpedia Spotlight\footnote{https://github.com/dbpedia-spotlight/dbpedia-spotlight} in extracting and categorizing named entities. However, besides the evaluations stated above, and the benchmark analysis done in \cite{ribeiro2016sentibench}, we might consider using sentence-level methods, as VADER \cite{hutto2014vader}, SentiStrength \cite{thelwall2013heart} or Umigon \cite{levallois2013umigon} on our Twitter dataset as our future work.

%%%%%%%%%%%%%%%%%%%%%%%%%%%%%%%%%%%%%%%%%%%%%%%%%%%%%%%%%%%%%%%%%%%%%%%%%%%%%%%%%%%%%%%%%%%%%%%%%%%%%
\section{Quantifying semantic homophily}
\label{sec:quantifying}
%%%%%%%%%%%%%%%%%%%%%%%%%%%%%%%%%%%%%%%%%%%%%%%%%%%%%%%%%%%%%%%%%%%%%%%%%%%%%%%%%%%%%%%%%%%%%%%%%%%%%

%%%%%%%%%%%%%%%%%%%%%%%%%%%%%%%%%%%%%%%%%%%%%%%%%%%%%%%%%%%%%%%%%%%%%%%%%%%%%%%%%%%%%%%%%%%%%%%%%%%%%
\subsection{Semantic relatedness and communication}
\label{sec:SR_comm}
\begin{figure}[htp]
	\subfloat{\includegraphics[width=0.5\linewidth]{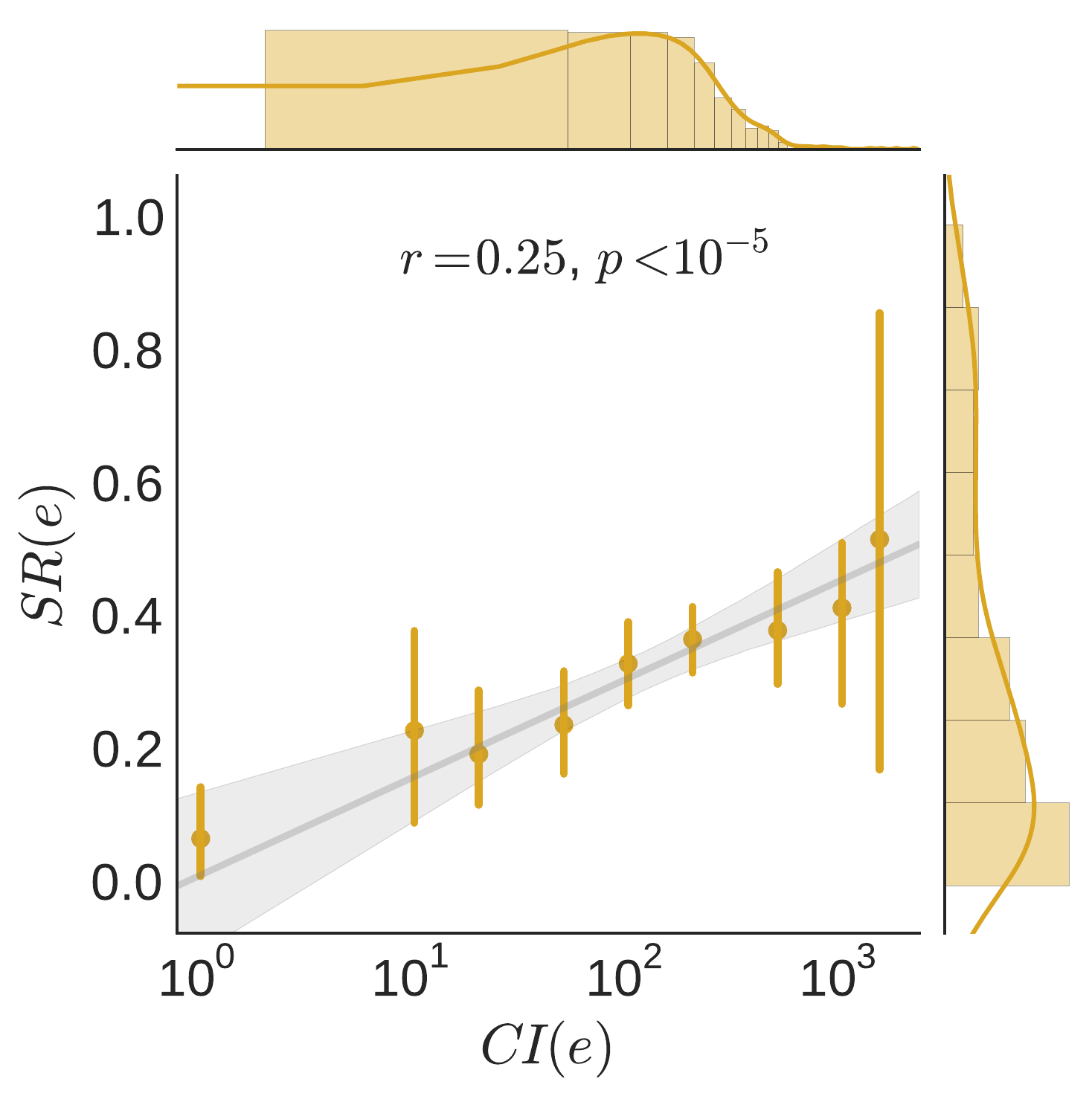}}
	\subfloat{\includegraphics[width=0.47\linewidth]{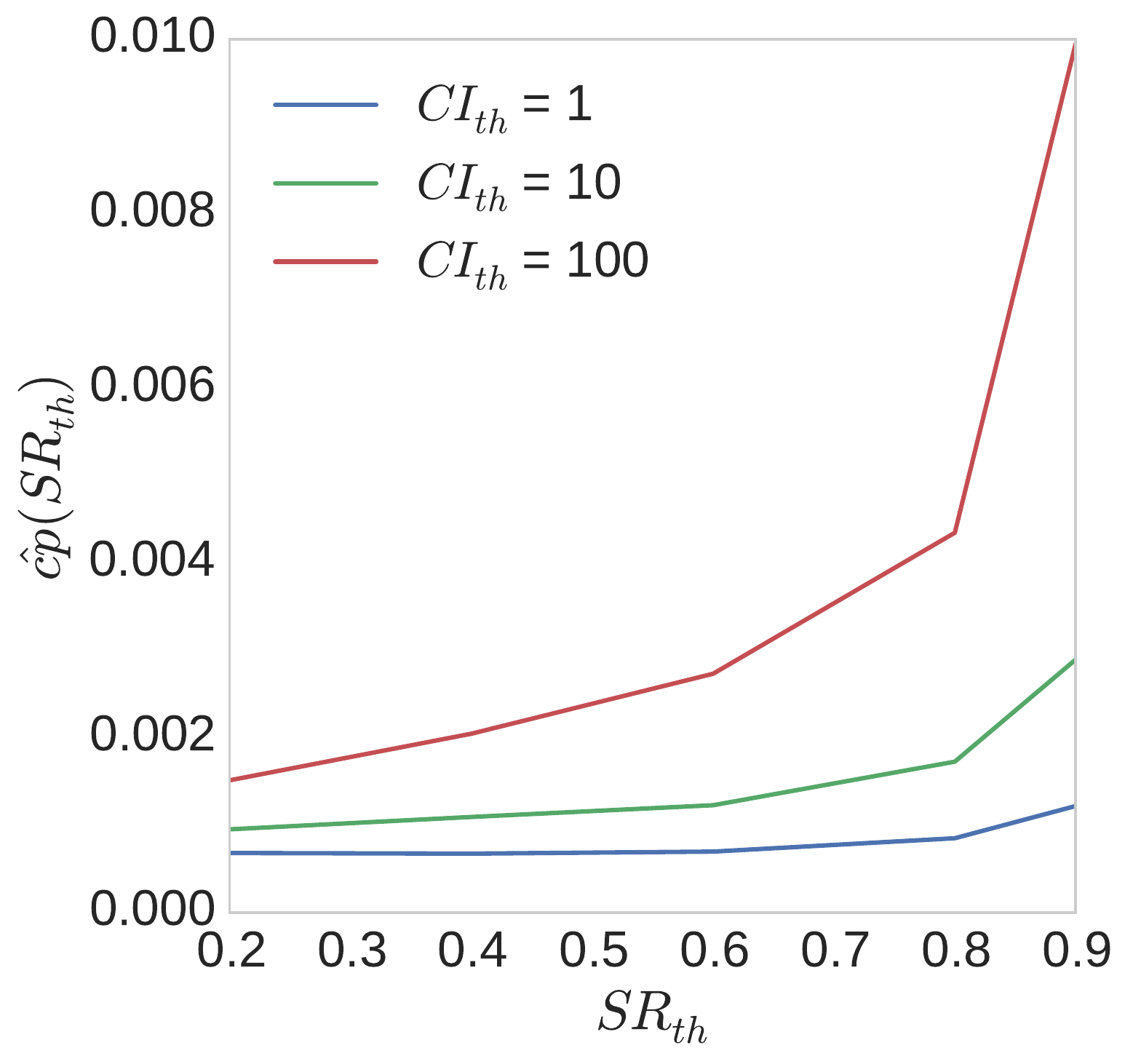}}
	\caption{\textbf{Interplay between SR and communication}: (a) correlation of link SR value ($SR(e)$) and its communication intensity ($CI(e)$); we apply logarithmic binning to account for long-tailed distribution of $CI(e)$; average value and standard deviation are shown for each bin; (b) communication propensity with respect to SR ($\hat{{cp}}(SR_{th})$) for different minimum communication intensity ($CI_{th}$) of links \label{fig:comm_prop}}
\end{figure}

%%%%%%%%%%%%%%%%%%%%%%%%%%%%%%%%%%%%%%%%%%%%%%%%%%%%%%%%%%%%%%%%%%%%%%%%%%%%%%%%%%%%%%%%%%%%%%%%%%%%%
%%%%%%%%%%%%%%%%%%%%%%%%%%%%%%%%%%%%%%%%%%%%%%%%%%%%%%%%%%%%%%
We start by investigating interplay between $SR$ for a pair of users and their $CI$ by asking: \textit{whether higher communication intensity is linked to a higher semantic relatedness}? Fig.\ \ref{fig:comm_prop} (left) displays the correlation when we apply logarithmic binning to account for long-tailed distribution of $CI(e)$. However, we find that user pairs exist who communicate quite intensively but have low relatedness of their tweet contents and also on the opposite -- some users with relatedness close to $1$ seldom communicate. Our result is comparable those in the study that evaluated similar relationship in retweet and follower Twitter graphs \cite{mitzlaff2014social}. Next we turn to another way of assessing the interplay between the two communication aspects.   

We calculate \textbf{communication propensity} ($\hat{{cp}}$) with respect to SR threshold ($SR_{th}$) as the extent to which observed communication and its intensity diverge from what would be expected in an uniformly random setting. To illustrate, $\mathtt{SR_{0.2}\ network}$ has $\sim40M$ links, or $\sim9\%$ out of all the possible $\sim438M$ links in full $\mathtt{SR\ network}$. Hence, in a uniformly random setting, we would expect a similar percent of communication links in $\mathtt{SR_{0.2}\ network}$. However, we find this percent to be $3$ times higher. Precisely, we apply the dyadic propensity formula defined in \cite{roth2005generalized} to calculate $\hat{{cp}}$:
$$
\hat{{cp}}(SR_{th}) = L_{comm}(SR_{th}) / L_{tot}(SR_{th}),
$$
where $L_{comm}(SR_{th})$ is the number of links in $\mathtt{communication\ network}$ with $SR$ value higher than the threshold and $L_{tot}(SR_{th})$ is the number of total possible such links. We also evaluate in the same way existence of links with a minimum communication intensity threshold ($CI_{th}$). Fig.\ \ref{fig:comm_prop} presents the results. Communication propensity increases with the increase in both $SR$ and $CI$ thresholds. The increase reveals presence of semantic homophily in the network with respect to SR. After both presented analyses, we conclude that the correlation between $SR$ and $CI$ is not simple and linear, but it is strongly captured by the subtle aspects of $\mathtt{communication\ network}$.

\begin{figure}
	\centering
	\subfloat{\includegraphics[width=0.5\textwidth]{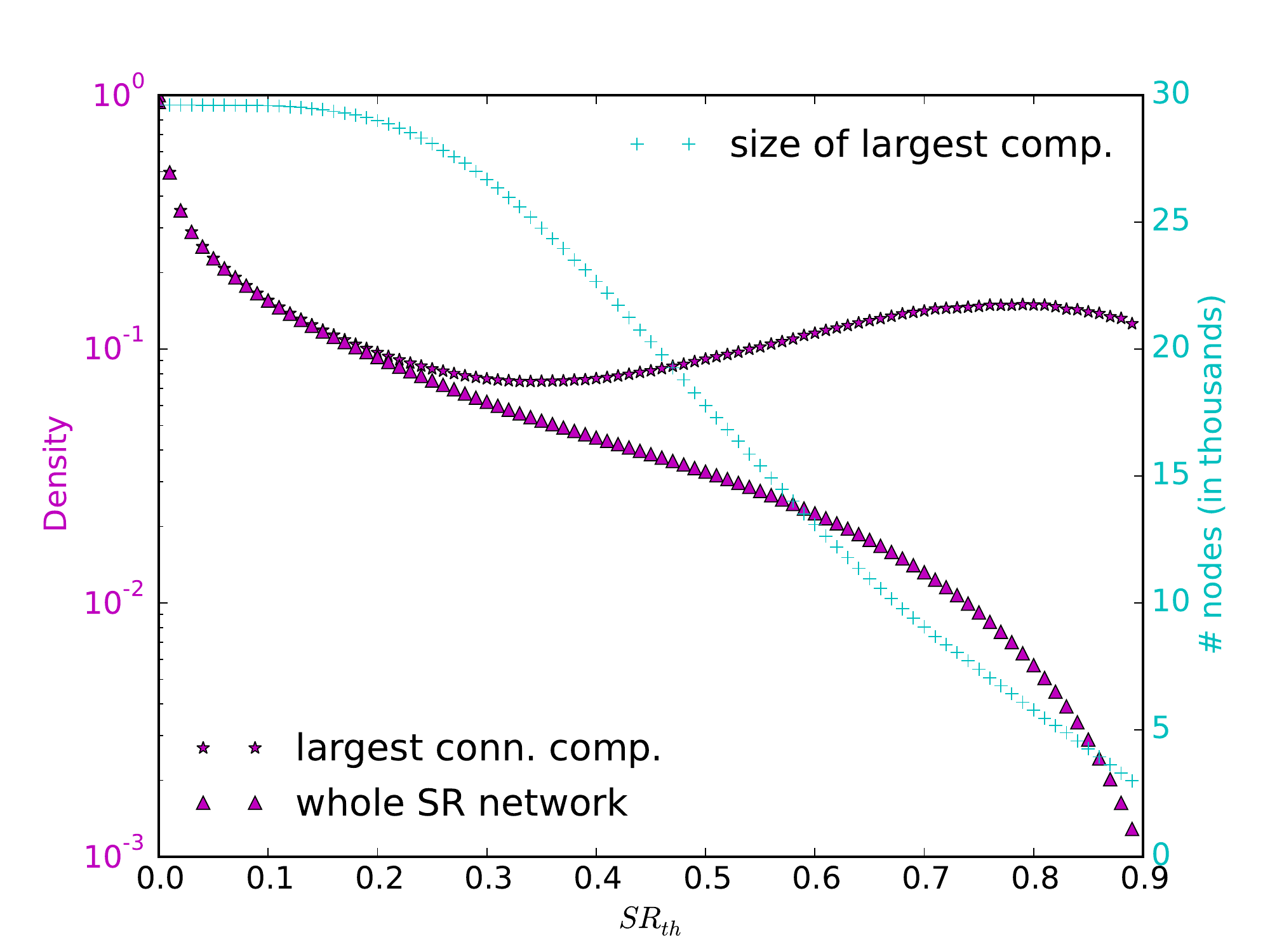}}
	\subfloat{\includegraphics[width=0.5\textwidth]{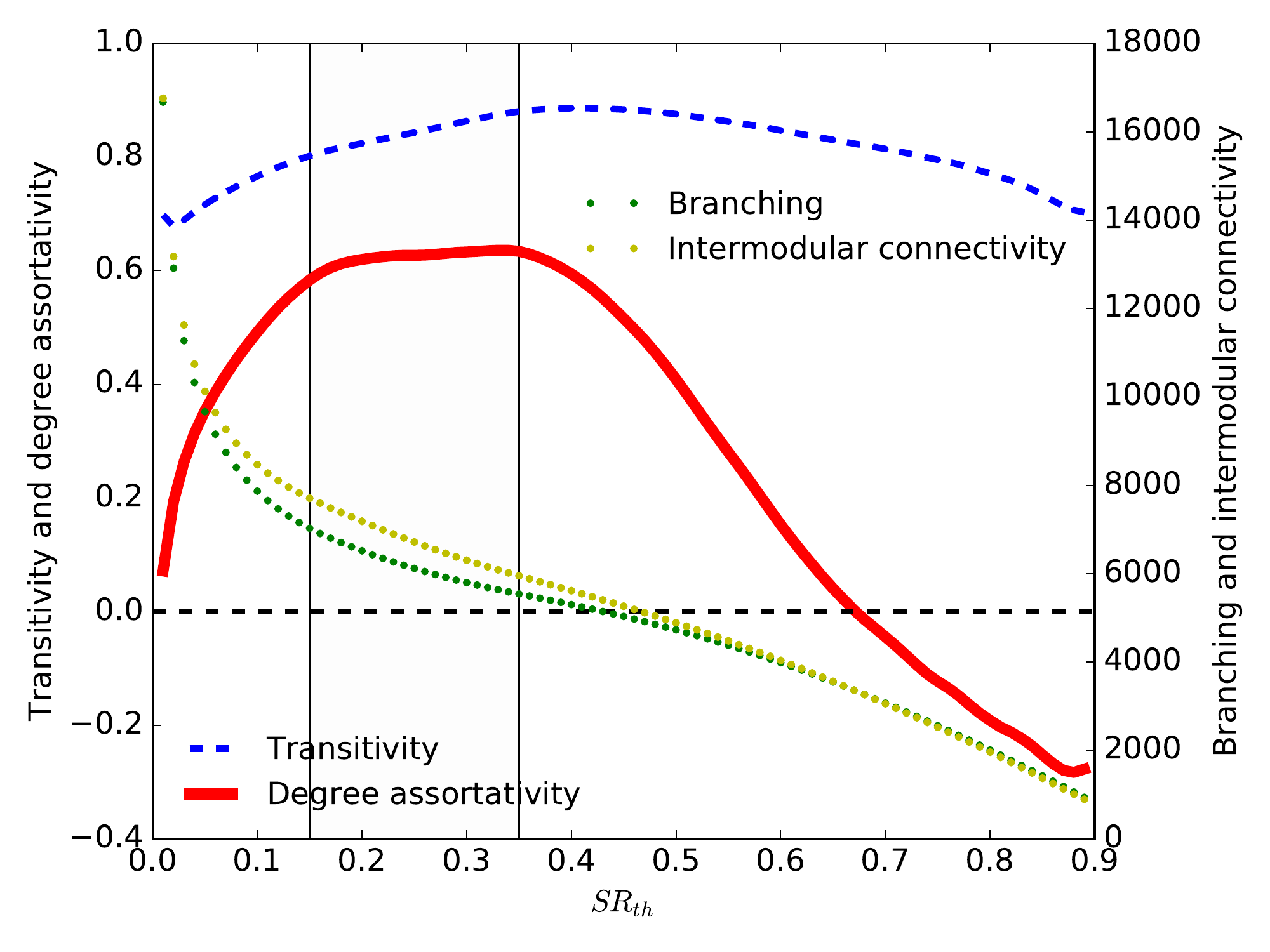}}
	\caption{\textbf{Properties of $\mathtt{SR\ network}$ in function of $SR_{th}$}: (left) size of the largest connected component, its density and overall network density; (right) branching factor, intermodular connectivity and transitivity as three ingredients for network degree assortativity \protect\cite{estrada2011combinatorial}\label{fig:sr_properties}}
\end{figure}

This section we conclude with several results on the \textit{properties of the full $\mathtt{SR\ network}$}. It is important to point out that for this analysis we take the network built from the data for the whole $6$ months period. Such results inform us about semantic relatedness metrics of a random group of people (not necessarily ever communicating). Fig.\ \ref{fig:sr_properties} (left) reveals that when thresholding the $\mathtt{SR\ network}$ near SR value $0.25$, the largest connected component still has around $85\%$ of the nodes and its density stabilizes, even it starts to grow, whereas the overall density in the network is significantly reduced. 

In Fig.\ \ref{fig:sr_properties} (right) we plot the \textit{degree assortativity} \cite{PhysRevLett.89.208701} in $\mathtt{SR\ network}$ as a function of SR threshold. We detect an interesting changing pattern from positive to negative degree assortativity. In order to make sure that this pattern is specific to real-world SR metric, we randomize the SR values on $\mathtt{SR\ network}$ in several ways and find no pattern in such cases. Hence, we conclude that a structurally important change in human $\mathtt{SR\ network}$ takes place when we consider different SR threshold. 

Fig.\ \ref{fig:sr_properties} (right) also shows the values for branching factor, intermodular connectivity and network transitivity (clustering coefficient), as it has been proven that they together define degree assortativity value \cite{estrada2011combinatorial}. In the interval ($0.15, 0.35$) $\mathtt{SR\ network}$ obeys highest assortativity and transitivity. In this way we find lower and upper bounds for the threshold that can be used to remove the noise generated when building the SR knowledge database. For these values we also obtain the best community matching between $\mathtt{SR\ network}$ and $\mathtt{communication\ network}$, as described in Section \ref{sec:comm_str}. From an application point of view, these findings might be important to consider while designing other semantic relatedness and similarity metrics, in particular when choosing a suitable threshold to distinguish significantly related and not related users.

%%%%%%%%%%%%
%%%%%%%%%%%%%%%%%%%%%%%%%%%%%%%%%%%%%%%%%%%%%%%%%%%%%%%%%%%%%%%%%%%%%%%%%%%%%%%%%%%%%%%%%%%%%%%%%%%%%
\subsection{Forms of social capital and degree assortativity}
\label{sec:da_mention_network}
As we introduced earlier, a basic measure of assortative mixing in a network is the assortativity coefficient \cite{newman2003mixing} or simply assortativity. This coefficient is calculated as Pearson correlation between the value of a property on a node and the average value of that property on its neighbors. Hence the assortativity value ranges from 1 in a perfectly assortative network to -1 in a perfectly dissasortative network. Any discrete or scalar attribute of nodes can be used to calculate this coefficient. 

We start by calculating assortativity based on node degree, an inherent node attribute of any network. Positive degree assortativity \cite{PhysRevLett.89.208701} is suggested to be fundamental to social networks and to distinguish  them from other types of networks \cite{newman2003social}. 

\paragraph{Undirected network variants}
We start by looking at an undirected variant of our $\mathtt{communication\ network}$. Such an abstraction provides us with social capital in terms of number of contacts (unweighted) and total communication intensity (weighted network case). When we look at mutual edges, then we tackle \textit{strong communication ties}, and when including all edges, then we also consider \textit{weak communication ties} \cite{granovetter1973strength}.
The values of degree assortativity coefficient ($r$) in different variants of the $\mathtt{communication\ network}$ are presented in Table \ref{t:mention_assortativity_coeffs}. Using jackknife method as in \cite{newman2003mixing} we calculate and present also the standard deviation for each measurement  to verify statistical significance of the results. Below we discuss and interpret the cases when our networks exhibits assortativity.
\begin{itemize}
	\item Undirected unweighted network including all edges is \textbf{slightly disassortative} with $r = -0.015$. 
	
	\item Undirected unweighted network with only mutual edges is on the other hand \textbf{highly assortative} with $r = 0.414$ (similar result reported in \cite{bliss2012twitter}). 
	This result shows that \textit{the more strong contacts you have, the more strong contacts they themselves tend to have.}
	
	\item Undirected weighted network including all edges is \textbf{slightly disassortative} with $r = -0.014$.

\item Undirected weighted network with only mutual edges is again \textbf{highly assortative} with $r = 0.474$. 
This result shows that \textit{the stronger communication intensity you have, the stronger communication intensity your contacts tend to have.}
\end{itemize}

\paragraph{Directed network variants}
In directed networks, four types of degree assortativity can be calculated, as introduced in \cite{piraveenan2012assortative}. These four types of assortativity coefficients show if the degree of a source node is correlated with the degree of the target nodes, hence tackling relational analysis between source and receiver in communication \cite{rogers1970homophily}. As shown in Table \ref{t:mention_assortativity_coeffs} the \textbf{in-in} is the only negative of the four coefficients in our network. This is in agreement with the findings for assortativity in directed followers Twitter network \cite{myers2014information}, except for \textbf{out-in} coefficient which is also found negative in the followers graph and it is slightly positive in our case. The authors (ibid.) argue that Twitter exhibits negative assortativity coefficients, unlike other social networks, because of its role as an information network, too. Below we interpret the results in our network.
\begin{itemize}

\item Looking at \textbf{in-in} coefficient, there was \textbf{no assortativity} with $r = -0.001$ in the unweighted network. This value increases to $r = -0.015$ in the weighted network case and becomes statistically significant. It is still low so we do not interpret it. 

\item Low positive \textbf{in-out} degree assortativity tells that: \textit{the more popular you are the more active those who you contact tend to be (both in terms of number of contacts and in terms of communication intensity)}.

\item Positive \textbf{out-in} degree assortativity
is low ($0.038$) so we do not interpret it.

\item The highest coefficient is for \textbf{out-out} degree assortativity, informing us that \textit{the higher the number of users whom you contact, the higher the number of users they also tend to contact (or the more intensively you are communicating, the more intensively those who you contact also tend to be communicating)}.
\end{itemize}

\begin{table}[htp]
	\centering
		\caption{{\textbf{Degree} assortativity $r$ coefficients in the $\mathtt{communication\ network}$. Standard deviation $s$ calculated using jackknife method \protect\cite{newman2003mixing} is also presented }}
	 \label{t:mention_assortativity_coeffs}
	\begin{tabular}{l|l|c|c||l|c|c} \hline
		\multicolumn{2}{l|}{\textbf{undirected networks}}&\textbf{r}&\textbf{s}&{\textbf{directed networks}}&\textbf{r}&\textbf{s}\\ \hline
		\multirow{4}{*}{\textbf{unweighted}}&\multirow{2}{*}{\textbf{mutual edges}}&\multirow{2}{*}{0.414}&\multirow{2}{*}{0.010}&\textbf{in-in}&{\textit{-0.001}}&\textit{{0.002}}\\ \cline{5-7}
		& & & &\textbf{in-out}&{0.110}&{0.013}\\ \cline{5-7}
		&\multirow{2}{*}{\textbf{all edges}}&\multirow{2}{*}{-0.015}&\multirow{2}{*}{0.001}&\textbf{out-in}&{0.038}&{0.003}\\ \cline{5-7}
		& & & & \textbf{out-out}&{0.389}&{0.014}\\ \hline \hline
		
		\multirow{4}{*}{\textbf{weighted}}&\multirow{2}{*}{\textbf{mutual edges}}&\multirow{2}{*}{0.474}&\multirow{2}{*}{0.017} & \textbf{in-in}&{-0.015}&{0.002}\\ \cline{5-7}
		& & & & \textbf{in-out}&{0.207}&{0.020}\\ \cline{5-7}
	    & \multirow{2}{*}{\textbf{all edges}}&\multirow{2}{*}{-0.014}&\multirow{2}{*}{0.001}&\textbf{out-in}&{0.014}&{0.004}\\ \cline{5-7}
	    & & & & \textbf{out-out}&{0.338}&{0.026}\\ \hline
		\end{tabular}
\end{table}

\begin{figure}
	\centering
	\includegraphics[width=0.55\textwidth]{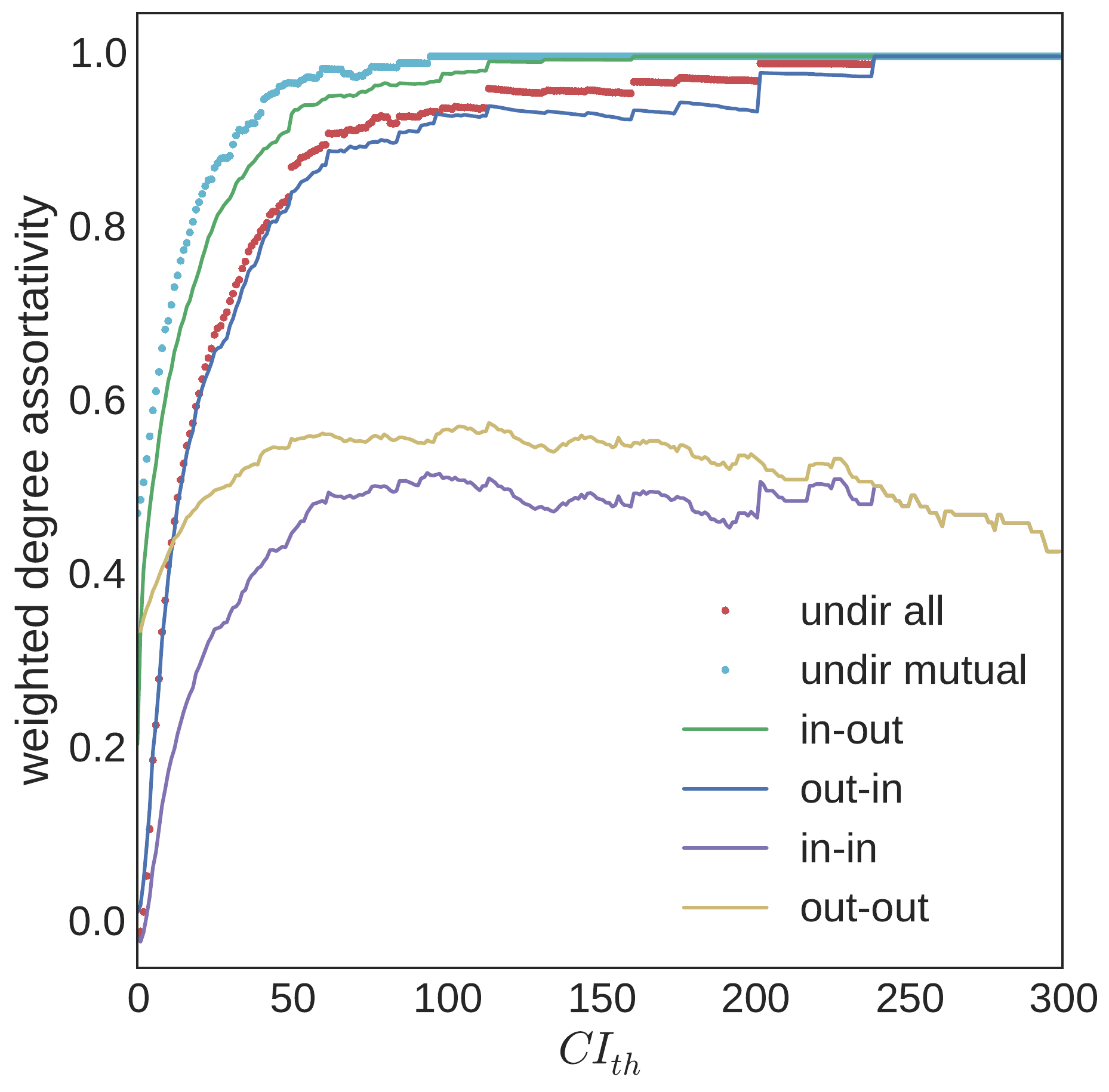}
	\caption{\textbf{Degree assortativity} as a function of communication intensity $CI$ in the ensemble of thresholded $\mathtt{communication\ networks}$. \textit{undir all} is degree assortativity in undirected network including all edges; \textit{undir mutual} in undirected network with only reciprocal; \textit{in(out)-in(out)} are the four types of coefficients in directed networks showing the correlation between \textit{in(out)}-degrees of source and receiver nodes \protect\cite{piraveenan2012assortative} }
	\label{fig:ment_assortativity}
\end{figure}

\paragraph{Assortativity as a function of communication intensity}
 We can create an ensemble of weighted $\mathtt{communication\ networks}$ by thresholding the original network on different minimum edge weights. Then we calculate the above presented coefficients in each thresholded network. Since weight on the edges represents intensity of communication, the result is degree assortativity as a function of the communication intensity, as shown in Fig.\ \ref{fig:ment_assortativity}. 

First insight is that already with a small threshold, the two assortativity coefficients that are in the original network found slightly negative (in undirected network with all edges and in directed network \textbf{in-in} coefficient) become positive. With the threshold larger than $20$ mentions, the networks are highly assortative on all the coefficients. This property exhibits one of the differences between often analyzed social networks based on unweighted, once formed links (such as friendship and followership) and the \textbf{weighted communication} network that we focus on. Bliss et al.\ \cite{bliss2012twitter} demonstrated temporal stability of degree assortativity in mutual mention network, while herein we exhibit its variability with respect to the minimum communication intensity. Coming back to the above mentioned negative assortativity results in the Twitter followers network \cite{myers2014information}, \textit{we argue that at higher communication intensity (requiring more time and effort than other interactions, such as following) the Twitter mention network serves more of a social than information role}. That is exhibited by the strong degree assortativity coefficients.

Moreover, looking at the higher communication intensity thresholds, we notice two more interesting patterns. Two directed assortativity coefficients (\textbf{in-in} and \textbf{out-out}) start to slowly decrease, while the four other coefficients asymptotically reach the maximum value $1$. In our concrete network case, the threshold of $239$ mentions is when the four coefficients all become equal by reaching the value $1$ and also the coefficients \textbf{in-in} and \textbf{out-out} become equal (at value $0.505$). While not shown in Fig.\ \ref{fig:ment_assortativity}, we calculated and those two coefficients continue to drop, while the others stay at the maximum value as we increase the threshold further.

To conclude, presented positive degree assortativity properties reveal presence of \textit{social status homophily} (users with higher status tend to assortatively connect) on different forms of social capital in the $\mathtt{communication\ network}$. We also find slight amounts of \textit{social status heterophily} in relation to weak ties and popularity, but this heterophily quickly gives place to strong homophily when there is higher communication intensity in the network.

%%%%%%%%%%%%%%%%%%%%%%%%%%%%%%%%%%%%%%%%%%%%%%%%%%%%%%%%%%%%%%%%%%%%%%%%%%%%%%%%%%%%%%%%%%%%%%%%%%%%%
\subsection{Forms of semantic capital and attribute assortativity }\label{sec:sem_cap}
%%%%%%%%%%%%%%%%%%%%%%%%%%%%%%%%%%%%%%%%%%%%%%%%%%%%%%%%%%%%%%%%%%%%%%%%%%%%%%%%%%%%%%%%%%%%%%%%%%%%%
% formula for homphily halberstam2014homophily
In this section, we investigate levels of assortative mixing  on semantic aspects in the $\mathtt{communication\ network}$. Besides degree, social networks are shown to exhibit assortativity on diverse nodes attributes \cite{bollen2011happiness,aiello2012friendship,eom2014generalized}. In line with such previous findings, we ask on which \textbf{semantic attributes} our Twitter $\mathtt{communication\ network}$ exhibits assortativity and to what extent. While social capital aspects presented in previous section reveal status homophily, some of the semantic capital aspects in this section exhibit {value} and some {status homophily}. Precisely, we look at assortativity on sentiment score and topics presence in the tweets, revealing \textit{semantic value homophily}. We also look at semantic capital, or the diversity with regard to the number of relevant entities, concepts and taxonomy levels found in the tweets and this analysis reveals \textit{semantic status homophily}. 
   \begin{figure*}
	\subfloat{\includegraphics[width=0.33\textwidth]{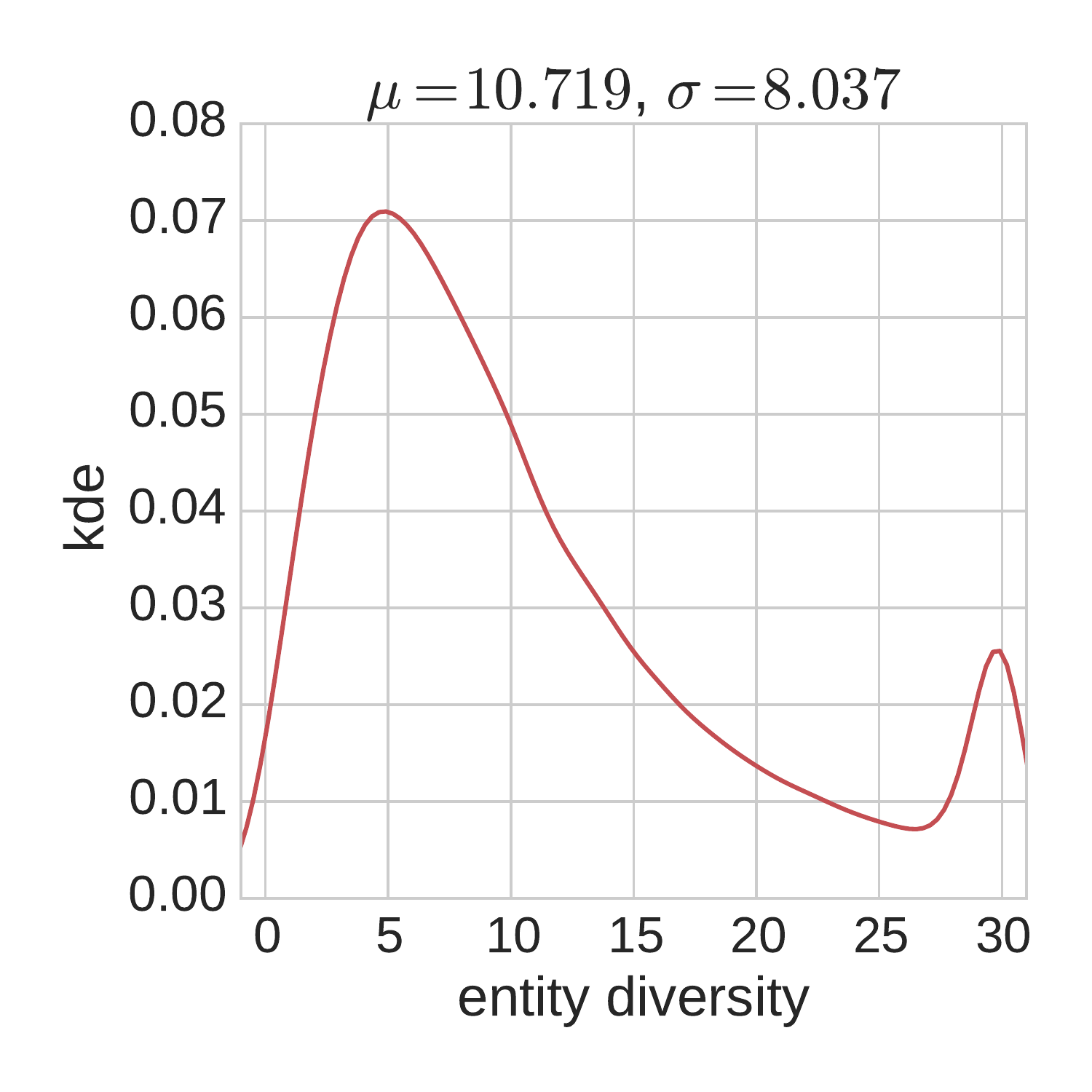}}
	%\hfill
	\subfloat{\includegraphics[width=0.33\textwidth]{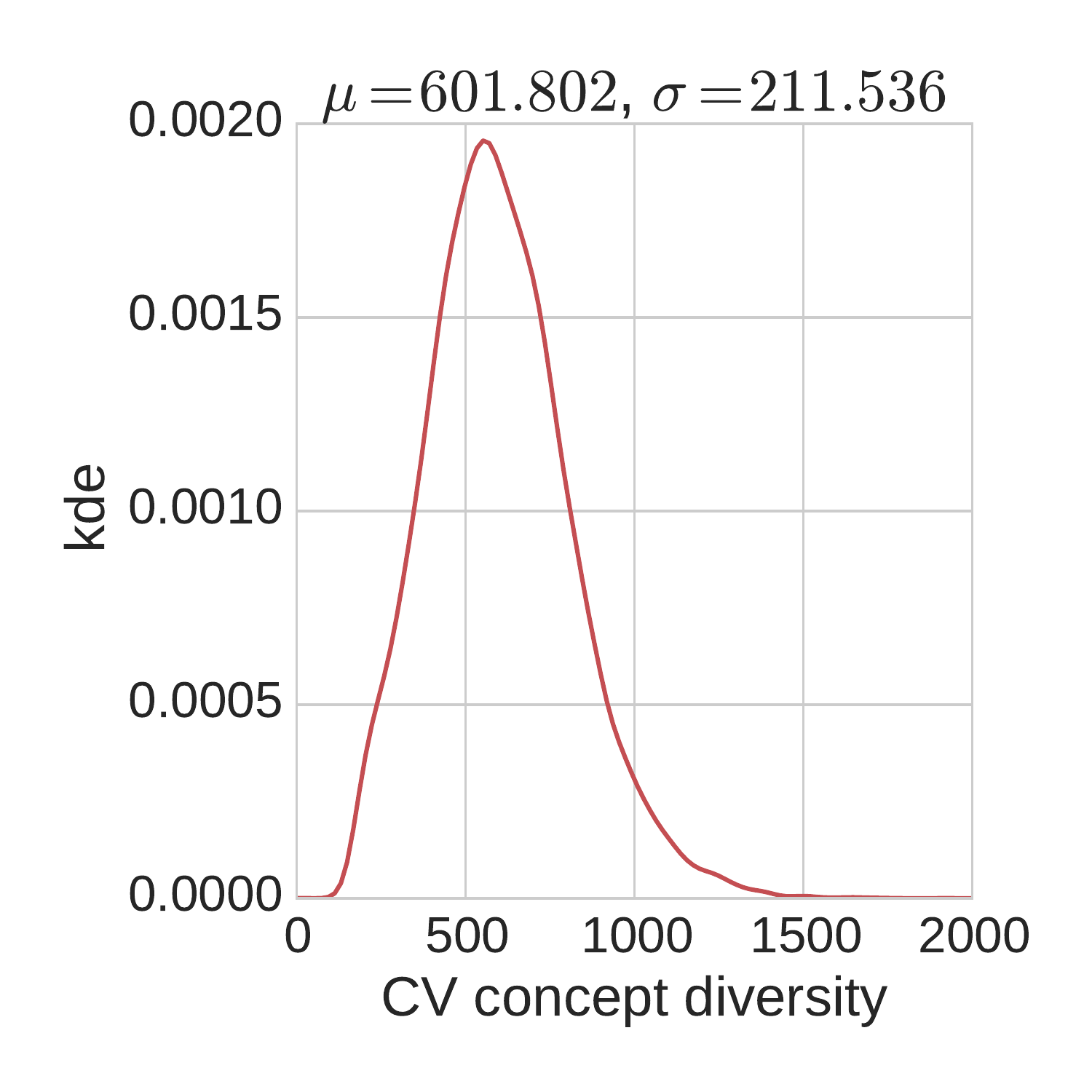}}
	\subfloat{\includegraphics[width=0.33\textwidth]{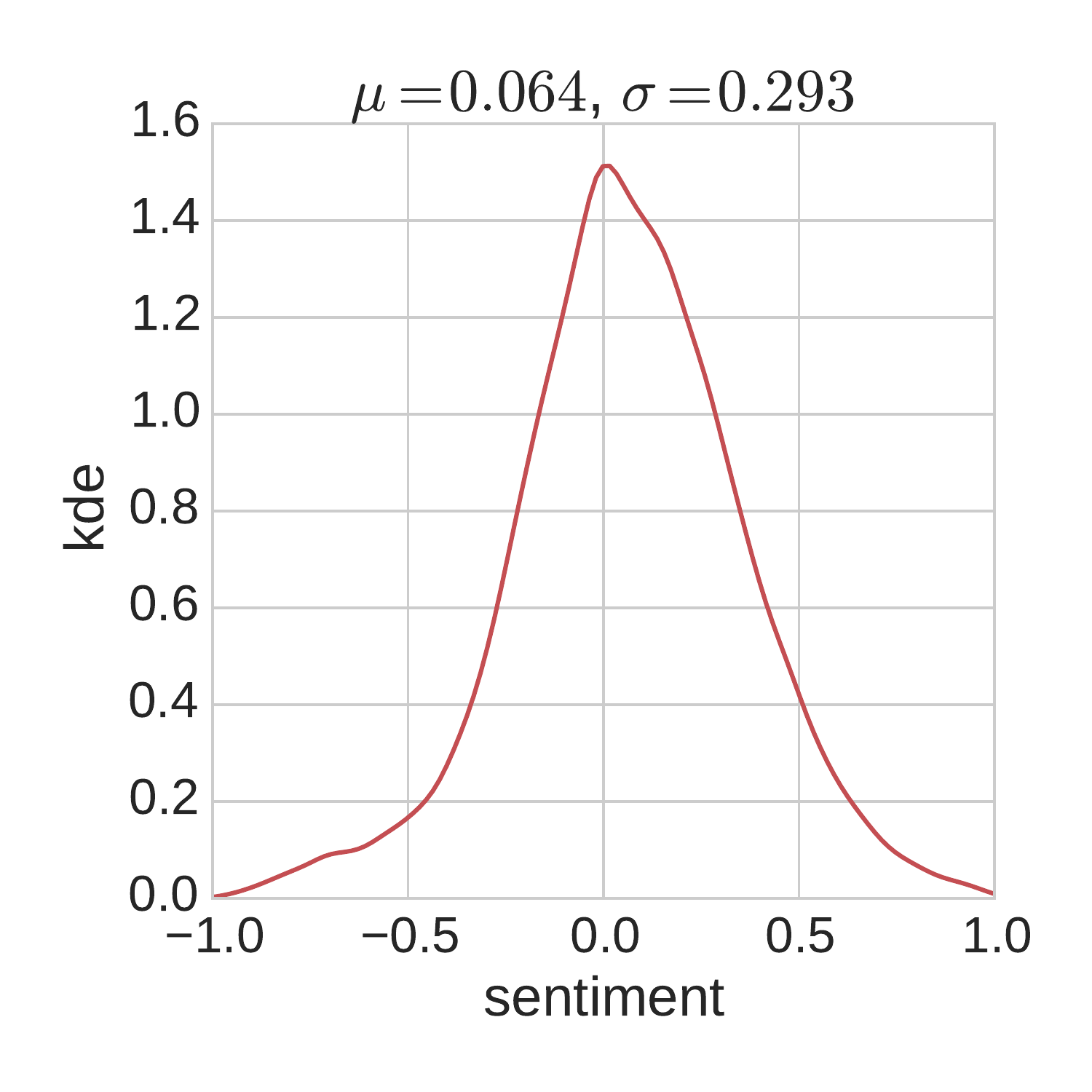}}
\caption{\label{fig:soc_cap_distr} \textbf{Semantic capital distributions}} 
\end{figure*}
Prior to looking at assortativity, it is useful to familiarize ourselves with the distributions of semantic capital and sentiment values for the whole user base. In Fig.\ \ref{fig:soc_cap_distr} we show kernel density estimates of their distributions displaying heterogeneity of entities and CVs diversity (see Section \ref{sec:SR_db} and Methods for description of CVs), and sentiment values among users. While most of the users tend to have around $5$ entities relevant to their tweet contents, we also find an important percent of users with nearly $30$ such entities. Similarly for concepts, a majority of users has $500-700$ concepts in their CVs, but we find also users with with $1500-2000$ concepts. As for sentiment, a majority of users tend to have neutral tweets sentiment, however, we also find users on both sides of the spectrum (negative and positive sentiment scores). Hence, we conclude that there is large \textbf{semantic capital heterogeneity} among our users (see \cite{roth2010social} for similar result in different types of networks).
 
\begin{table} [htp]
	\centering
	\caption{\textbf{Status and value homophily}: attributes assortativity $r$ in the unweighted $\mathtt{communication\ network}$. Standard deviation $s$ calculated using jackknife method is also presented }\label{t:mention_attr_assortativity}
	\begin{tabular}{c|llll|lllll} \hline
		\textit{level}&\multicolumn{4}{|c|}{\textit{status homophily}}&\multicolumn{4}{|c}{\textit{value homophily}}\\ \hline
		\multirow{2}{*}{\textbf{attr}}&\textbf{Wiki CVs}&\textbf{taxonomy}&\textbf{entity}&\textbf{concept} &\textbf{sentiment}&\textbf{topic}&\textbf{topic}&\textbf{topic}\\ 
		&\textbf{diversity}&\textbf{diversity}&\textbf{diversity}&\textbf{diversity}
		&\textbf{score}&\textbf{music}&\textbf{movies}&\textbf{sex}\\ \hline
		\multicolumn{9}{c}{{directed network, all edges}} \\ \hline 
		\textit{\textbf{r}}&$0.144$&$0.157$&$0.292$&$0.173$&$0.315$&$0.151$&$0.136$&$0.136$\\
		\textit{\textbf{s}}&$0.003$&$0.003$&$0.003$&$0.003$&$0.003$&$0.003$&$0.004$&$0.004$\\ \hline
		\multicolumn{9}{c}{{undirected network, mutual edges}} \\ \hline 
		\textit{\textbf{r}}&$0.269$&$0.282$&$0.398$&$0.289$&$0.452$&$0.269$&$0.244$&$0.253$\\
		\textit{\textbf{s}}&$0.006$&$0.005$&$0.005$&$0.005$&$0.005$&$0.006$&$0.005$&$0.006$\\
		\hline\end{tabular}
\end{table}

The results presented in Table \ref{t:mention_attr_assortativity} suggest the presence of both, \textbf{value} (topics of tweeting, sentiment) and \textbf{status} (semantic capital) homophily in the unweighted versions of the $\mathtt{communication\ network}$. We focus on the unweighted versions, since we first of all ask, whether there is a tendency among the users to have contact with other users who are similar to them on some semantic attributes (without looking at intensity of communication). This means that the answers to this question in the networks including only mutual edges will inform us about such correlation among strong contacts, while looking at networks with all edges included will inform us also about weak contacts. Once again, as with the degree assortativity, we find that mutual (reciprocal) $\mathtt{communication\ network}$ is importantly different compared to the network including also one-sided communication edges. Notably, it exhibits higher levels of assortativity on all the analyzed attributes.
 
As the observed correlation levels could be induced by existing degree assortativity, we also test the presence of assortativity after node attribute randomization. The assortativity value in such case is importantly lower, $0.07$ and so we conclude that indeed the $\mathtt{communication\ network}$ exhibits low to moderate levels of \textit{semantic status and value homophily}. Moreover, among analyzed semantic attributes, status homophily is the largest with respect to entity diversity and value homophily with respect to sentiment.

%%%%%%%%%%%
\subsection{Interplay between social and semantic capital} \label{sec:soc_sem_interplay}
%%%%%%%%%%%
After establishing the presence of status and value homophily in the $\mathtt{communication\ network}$ on different forms of social and semantic capital, we ask next about the relationship between these forms of capital. Whether the users who are richer in terms of social capital (and hence more network central) are also richer in terms of semantic capital (their tweets are semantically richer, or exhibit more diversity on semantic aspects)? With this analysis, we respond to the call by authors in \cite{roth2010social} to look for similar types of patterns as they have investigated in the bloggers and scientists networks. Indeed, we also find a wide range of possible combinations of joint values of social and semantic capital, as they have reported. In the end we conclude that the observed patterns in the Twitter $\mathtt{communication\ network}$ resemble more of the bloggers than the scientists network presented in \cite{roth2010social}. 

   \begin{figure*}
	\centering
	\subfloat{\includegraphics[width=0.33\textwidth]{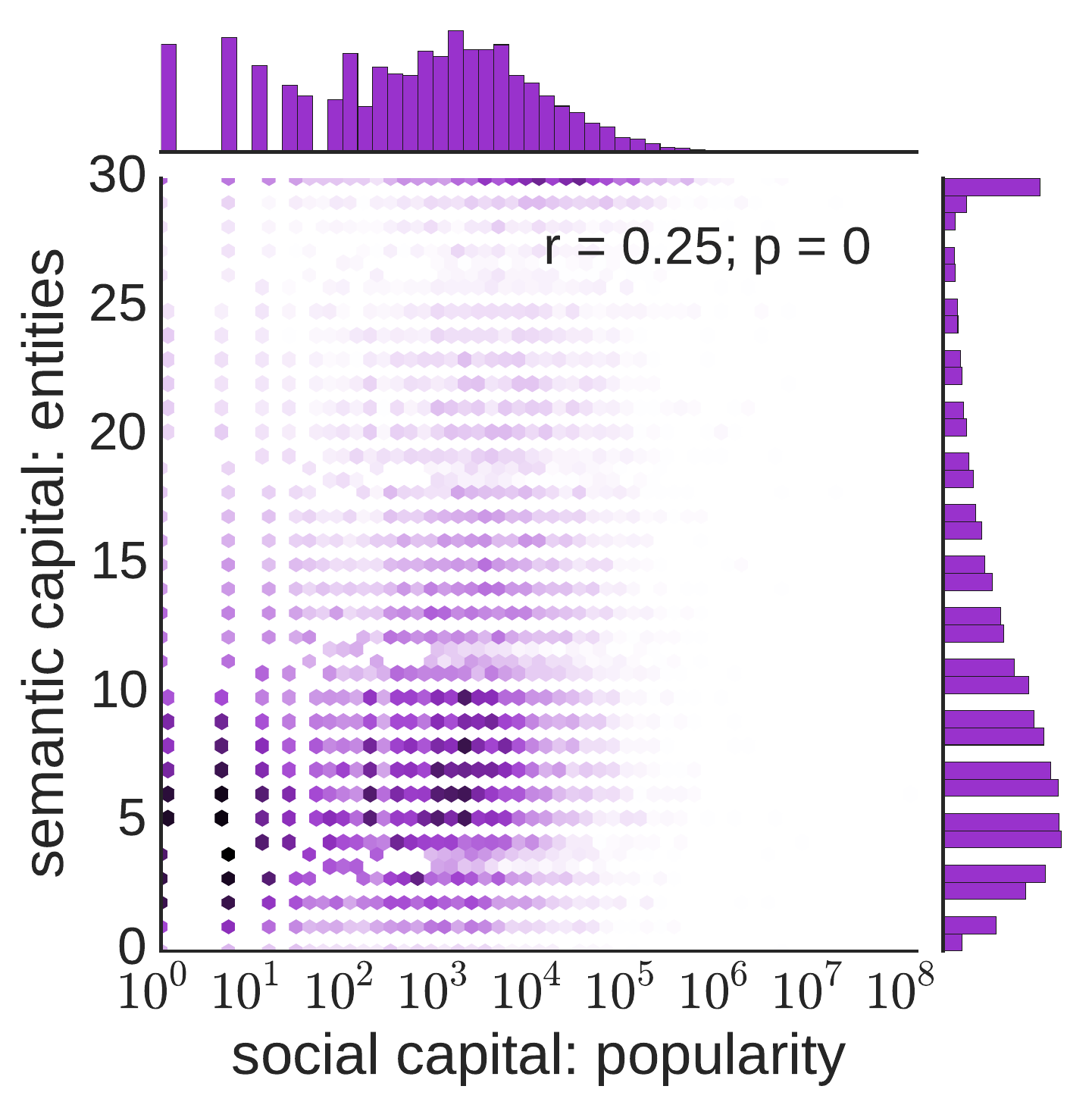}}
	\subfloat{\includegraphics[width=0.33\textwidth]{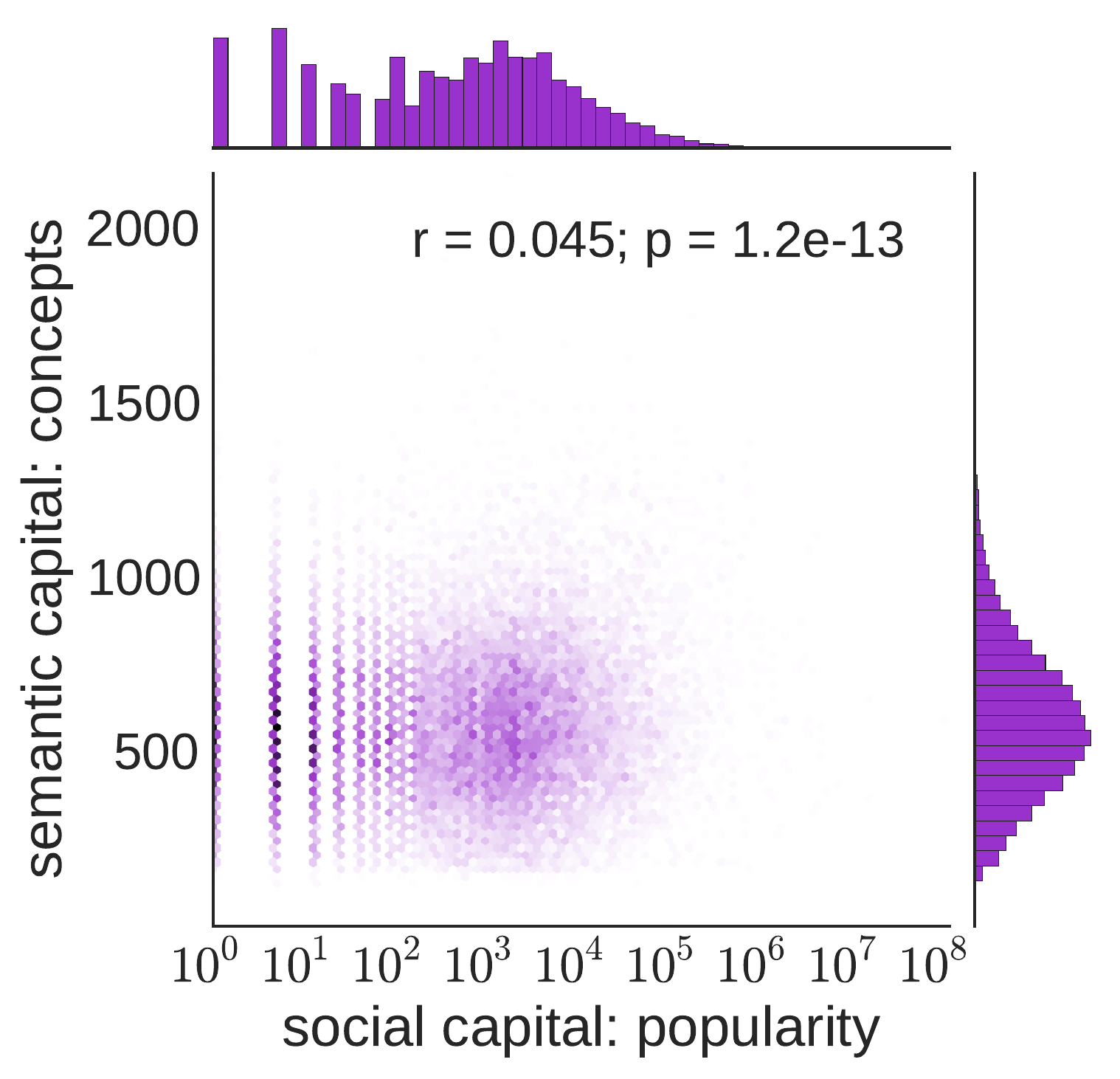}}
	\subfloat{\includegraphics[width=0.33\textwidth]{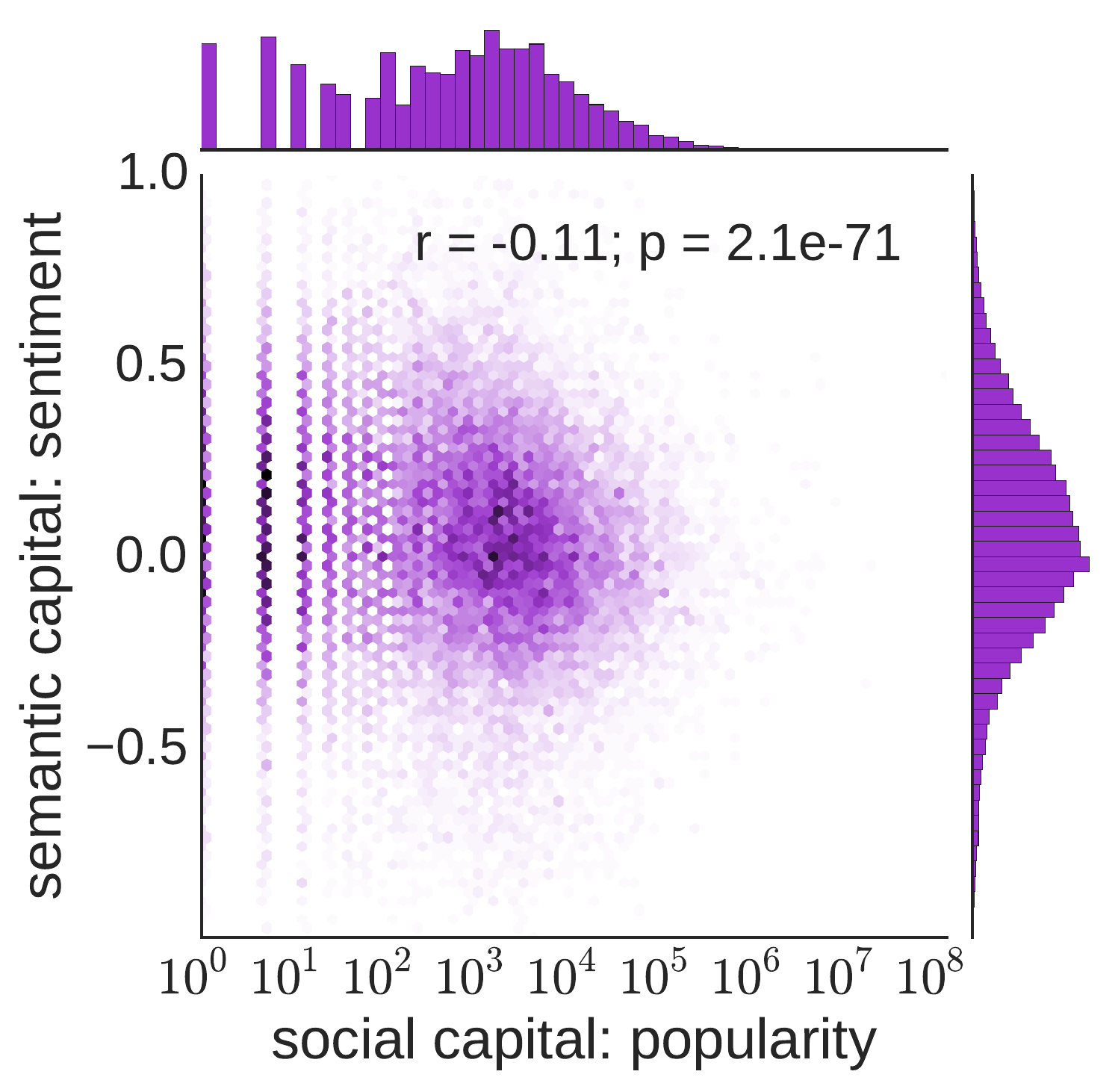}}
	\hfill
	\subfloat{\includegraphics[width=0.33\textwidth]{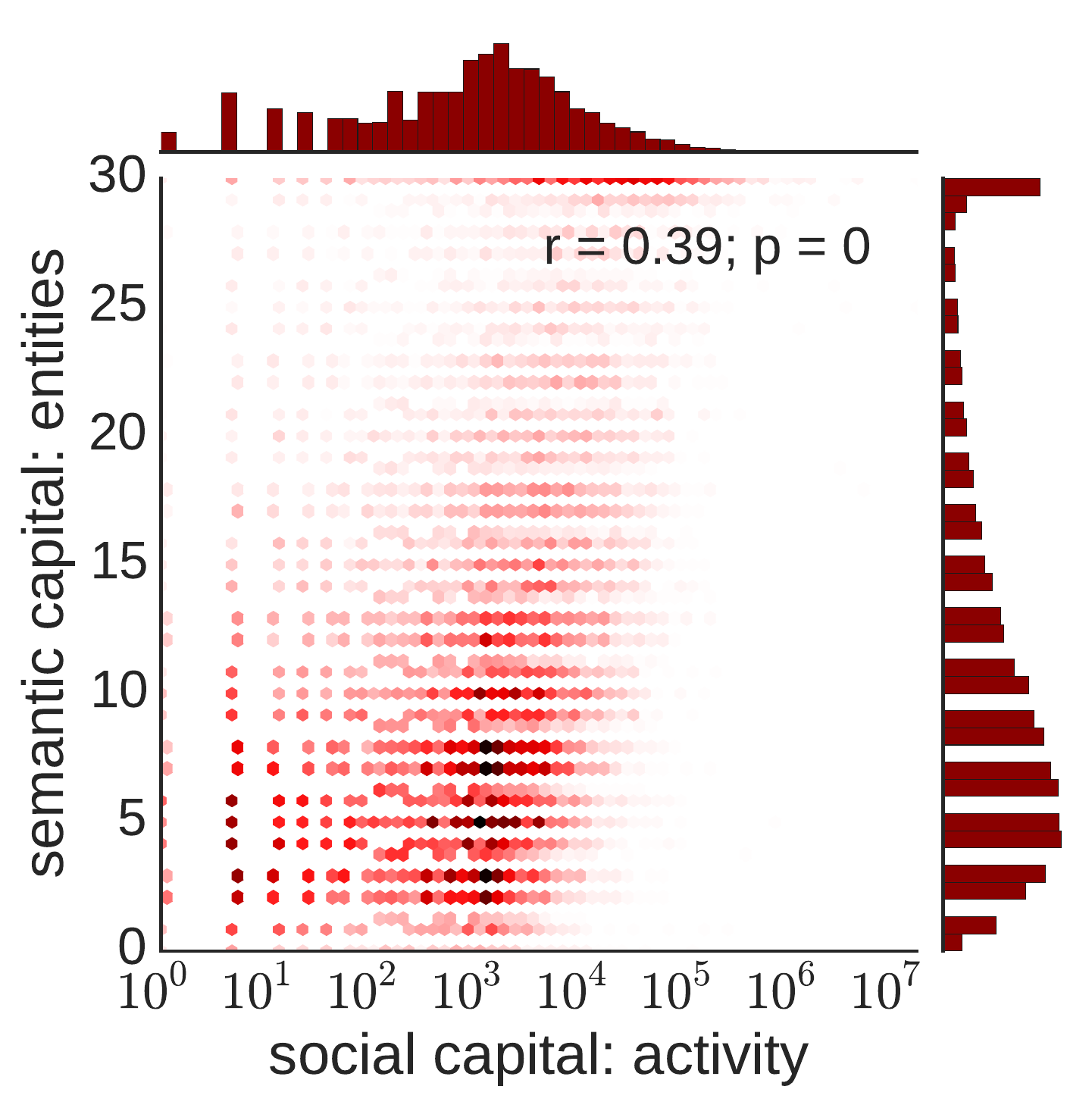}}
	\subfloat{\includegraphics[width=0.33\textwidth]{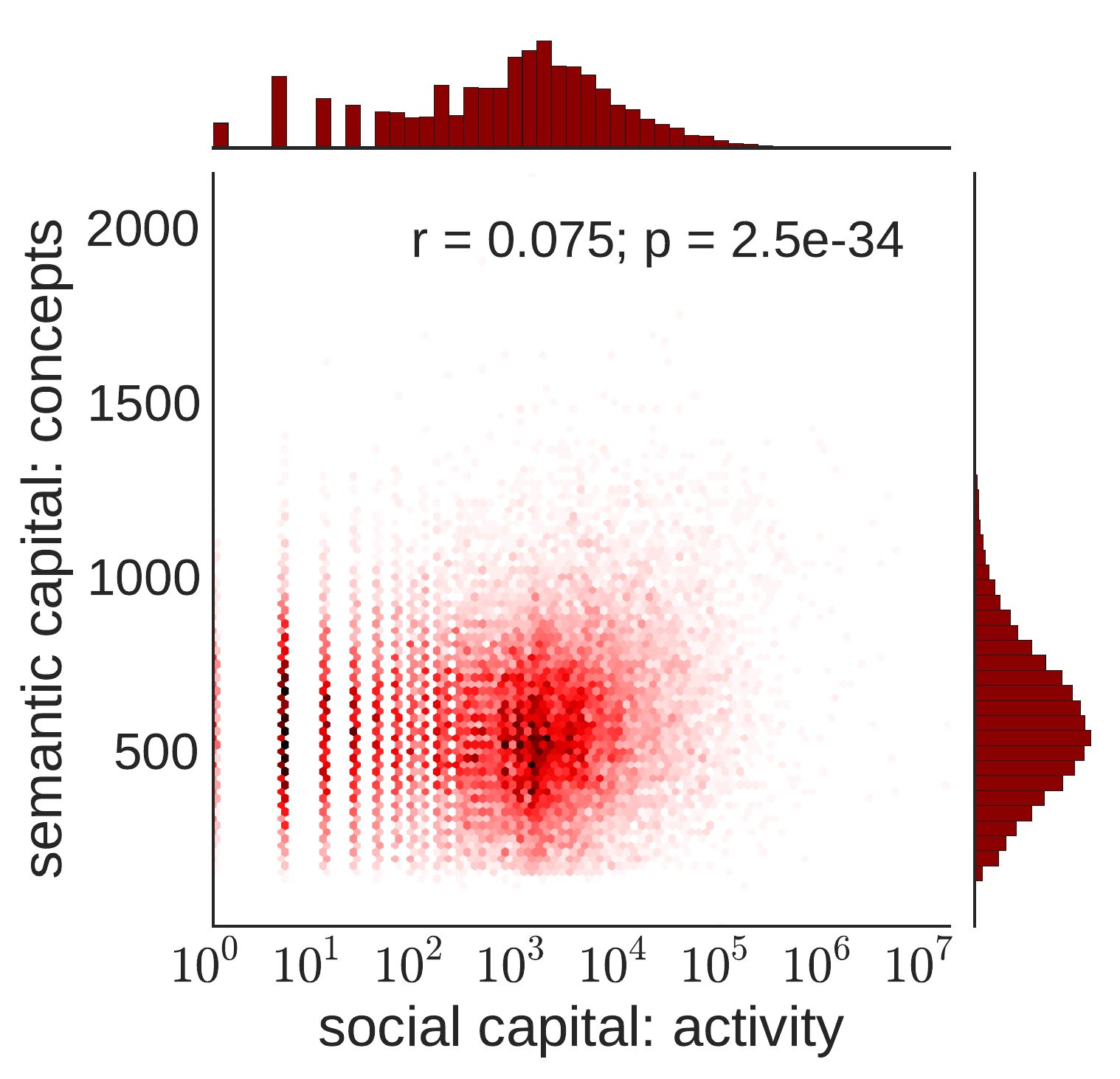}}
	\subfloat{\includegraphics[width=0.33\textwidth]{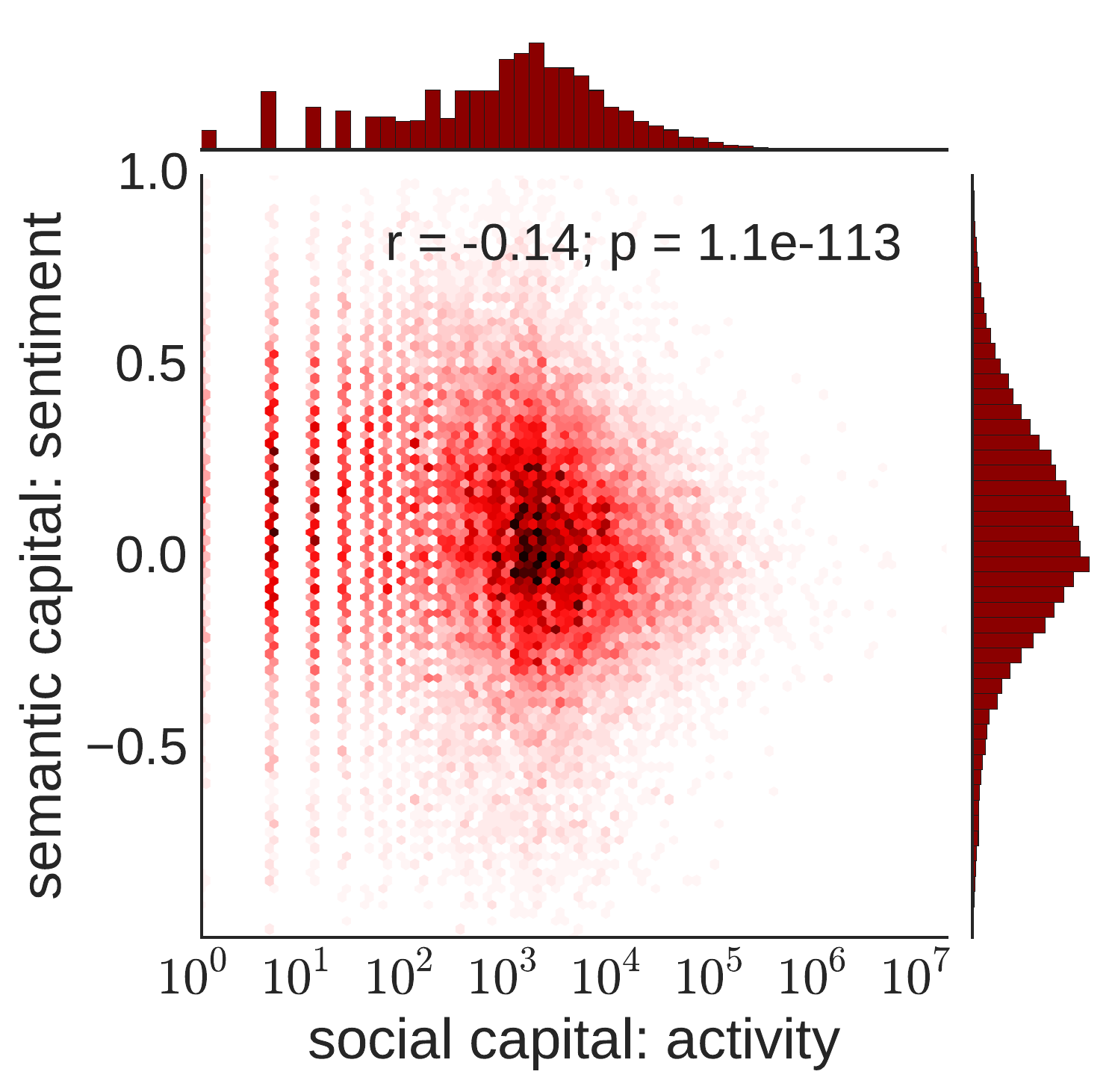}}
	\caption{\label{fig:soc_vs_sem} \textbf{Joint distributions of social and semantic capital}: the darkness of the hexagon corresponds to the frequency of users with the combination of social and semantic capital values} 
\end{figure*}
 Precisely, testing for different forms of social capital against different forms of semantic capital reveals no significant or low to medium correlations between the two. For the purpose of visualization, in Fig.\ \ref{fig:soc_vs_sem}, we show joint distributions for entity, concepts diversity and sentiment score on one side and communication intensity (of popularity and of activity) on the other.
 
 When it comes to \textbf{popularity} (weighted indegree), we observe a wide spectrum of semantic diversity in terms of entities for both, the users with low and high popularity. Most popular users tend to be slightly more likely to have high semantic diversity. On the other hand, most popular users are likely to have quite neutral sentiment in their tweets. However, users which are more positive or negative in their sentiment are likely to have modest to low popularity.
 
 When it comes to intensity of \textbf{communication activity} (weighted outdegree), we observe similar patterns that are a bit more pronounced for the socially richest users. Basically, most actively communicating users are likely to have higher semantic diversity in terms of entities (however, we still find a number of users with diverse tweet contents that are not actively communicating).  Semantic (entity) diversity has the highest correlations with {communication activity} (weighted outdegree; $r=0.397$) presented in Fig.\ \ref{fig:soc_vs_sem} and with weighted mutual degree ($r=0.396$). These values are similar to the value found in the bloggers network and lower compared to the scientists network in \cite{roth2010social}. 
 
 Sentiment has \textit{negative correlations} with both popularity and activity, also presented in Fig.\ \ref{fig:soc_vs_sem}. This means that with popularity and being active users tend to have a slightly more negative tweets sentiment. Finally, when it comes to diversity in terms of number of concepts present in their CVs, we do not find any differences between popular and active users. The richest users in terms of both types of social capital tend to have an average semantic capital (between $500$ and $1000$). Hence, we conclude that different forms of semantic capital have different patterns of interplay with social capitals.

Thanks to our network being directed and weighted, we are able to observe one additional pattern: while being particularly low for popularity (indegree), all the correlations increase for user activity (outdegree) and with communication intensity (weighted degrees). For instance, the correlation between entity diversity and (unweighted) indegree is only $0.051$. In this way, we exhibit that \textit{communication activity, intensity and stronger contacts} are more conductive of \textit{higher semantic capital}, compared to popularity and weaker contacts.

\begin{figure*}
	\centering
	\subfloat{\includegraphics[width=0.5\textwidth]{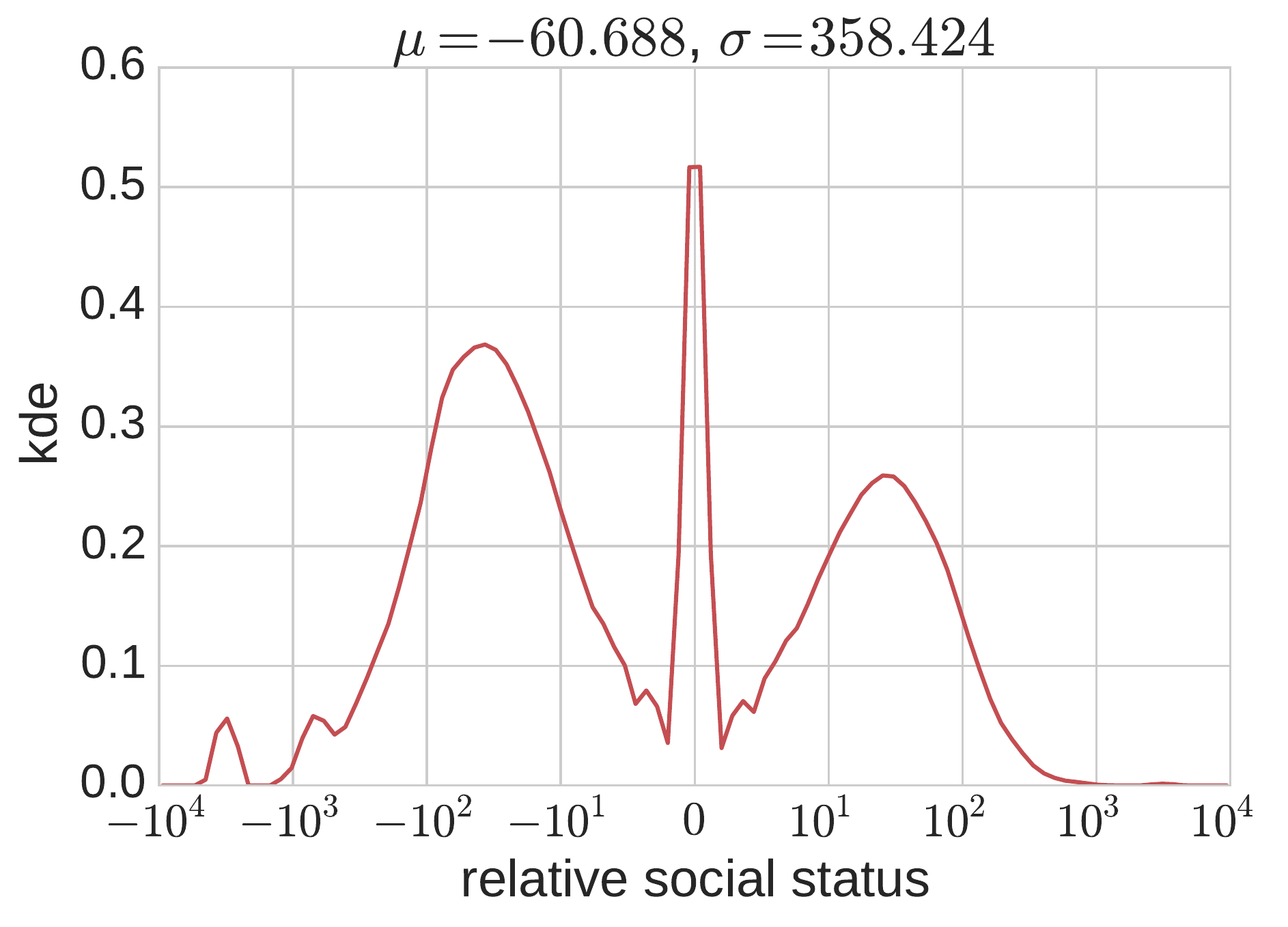}}
	\subfloat{\includegraphics[width=0.5\textwidth]{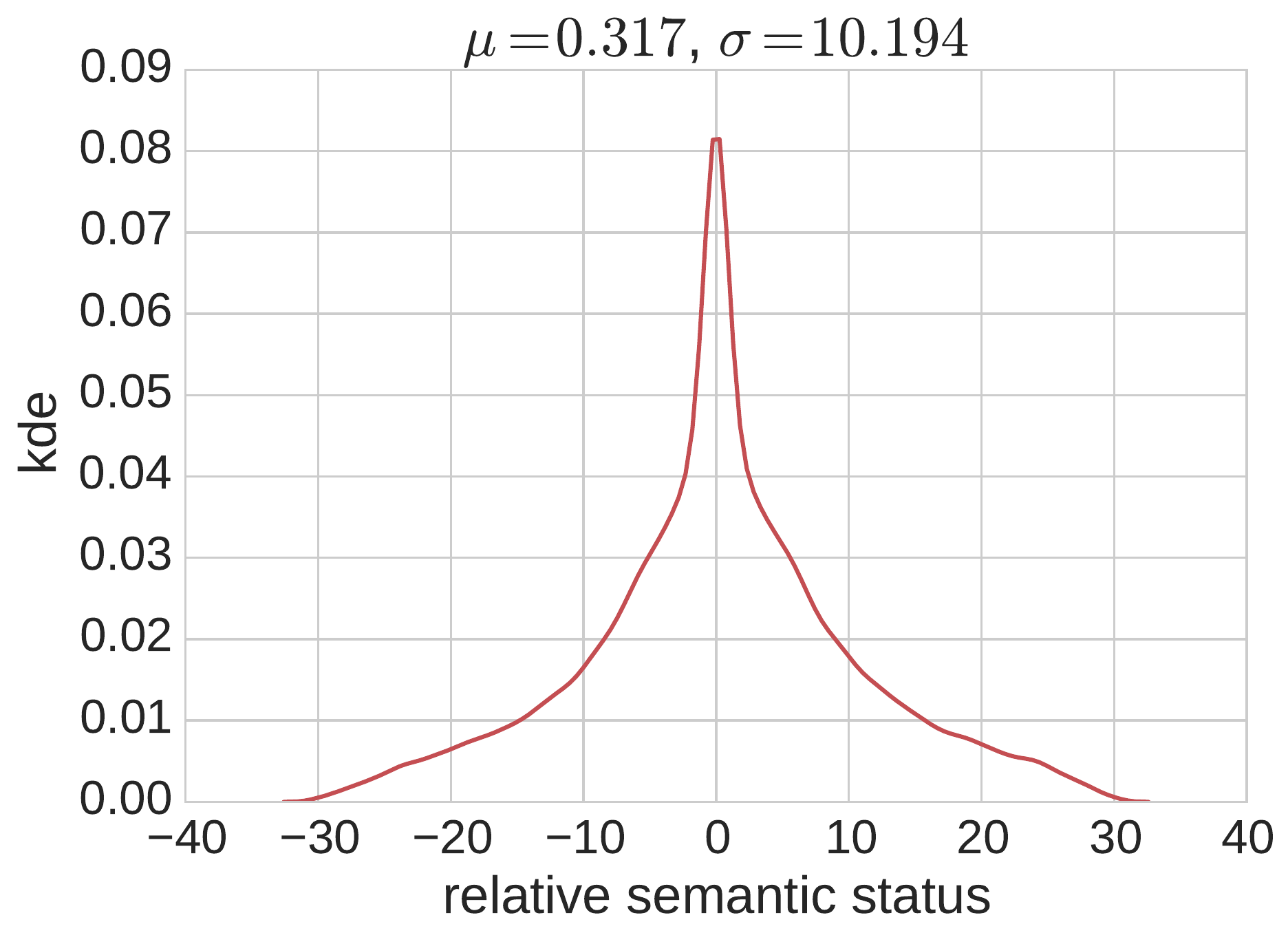}}
	%\subfloat{\includegraphics[width=0.33\textwidth]{popularity_vs_semantic_status-eps-converted-to.pdf}}
	%\hfill
	\caption{\label{fig:status_diff_comm} \textbf{Status differences in communication}: Kernel density estimates for distributions of (left) popularity difference and (right) semantic capital difference.} 
\end{figure*}

\paragraph{Relative status of source and receiver} An additional way to investigate the interplay between social and semantic capitals is in terms of \textbf{relative status} of source and receiver in communication. By relative status we mean the difference in status on a particular form of capital. Such definition is similar to the achieved status presented in \cite{sun2017link}. In Fig.\ \ref{fig:status_diff_comm} we show distributions of relative social status (popularity difference) and relative semantic status (entity diversity difference) between source and receiver. In particular, the distribution for relative social status exhibits a dominant peak at zero (users with similar status are most likely to communicate), but plotting it on a log scale reveals two additional interesting peaks at intervals ($-100,-10$) and ($10,100$). There is a higher likelihood for users with differences in social status belonging to these ranges to be talking to each other. The left peak is higher, and this together with the negative mean value for relative social status informs us that source users tend to be a bit less popular. There is also a small number of users mentioning considerably more popular users than themselves (leftmost part of the distribution). This happens to a smaller extent in the other direction, from more popular source users. When it comes to semantic capital, most of communication happens between those who have close to equal semantic capital.

For the joint distribution of social and semantic relative statuses we find (analyzed, not shown in a graph) a wide range of combinations. There is a small positive correlation between the two. As for the small number of users who initiate communication towards a considerably more popular users discussed above, we find that they tend to be semantically richer compared to the receiving users. We speculate that this \textit{semantic superiority might be a needed approach for such users to compensate for their lower popularity}.

\paragraph{Status inconsistency of source and receiver} Finally, we can tackle a sociological proposition that source and/or receiver \textbf{status inconsistency} can increase effectiveness of their communication \cite{rogers1970homophily}. \textit{Status inconsistency (internal heterophily of an individual) is defined in sociology as the relative lack of similarity in an individual's ranking on various indicators of social status}  \cite{lenski1954status}. Hence we introduce status inconsistency for Twitter users as a relative difference in their social and semantic capital ranks. We apply a similar formula to calculate \textbf{status inconsistency ($st_{inc}$)} as in  \cite{lenski1954status}:
\[
st_{inc} = 
\begin{cases}
- ( 1 - {r_{soc}} / {r_{sem}}), & \text{if } r_{soc} \leq r_{sem} \\      
  ( 1 - {r_{sem}} / {r_{soc}}),              & \text{otherwise;}
\end{cases}
\]
where $r_{soc}$ and $r_{sem}$ are users ranks in terms of social and semantic capital, respectively, among all users. This definition allows firstly to asses the amount of user status inconsistency (how close is $abs(st_{inc})$ to $1$), and second, it also encodes whether he/she has higher social ($st_{inc}$ is positive) or semantic ($st_{inc}$ is negative) status.

While we can not measure effectiveness of communication directly using our dataset, we allow \textit{communication intensity} to be a proxy for it. Our hypothesis in this regard is: the higher the communication intensity between a source and receiver, the higher potential for an effective communication. Now, for all the directed links ($e_{i,j}$) in our $\mathtt{communication\ network}$ we define \textbf{link inconsistency} using above introduced status inconsistency of the source ($st_{inc}(u_i)$) and the receiver ($st_{inc}(u_j)$) as their product:
\[
 st_{inc}(e_{i,j}) = st_{inc}(u_i) \cdot st_{inc}(u_j).
\]
This simple formula produces a higher absolute value for the links with higher total pair's inconsistency. The sign in this case indicates whether the source and receiver are ranked higher on the same sorts of capital ($st_{inc}(e_{i,j})$ positive) or different forms of capital ($st_{inc}(e_{i,j})$ negative).

\begin{figure*}
   	\centering
   	\subfloat{\includegraphics[width=0.5\textwidth]{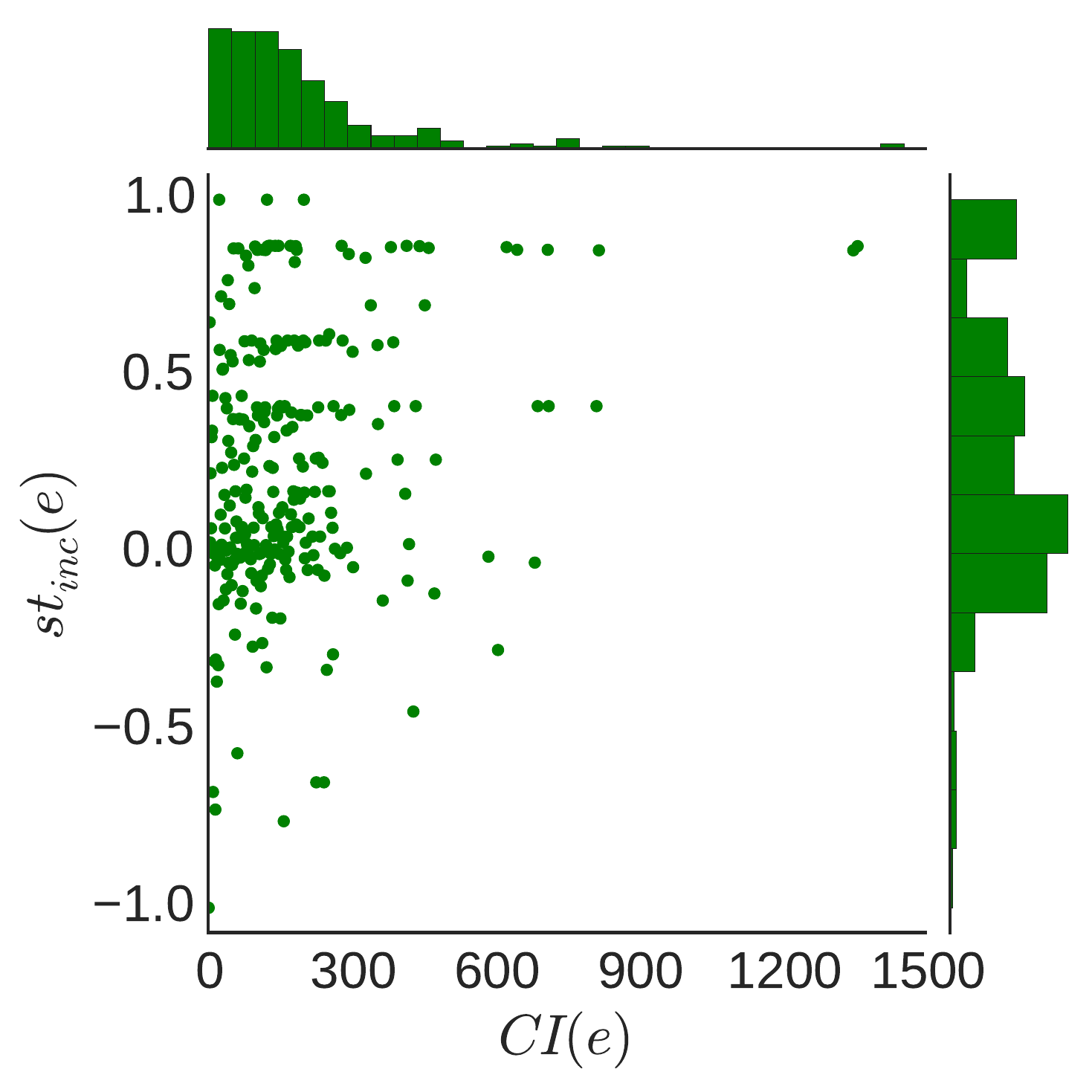}}
   	\subfloat{\includegraphics[width=0.5\textwidth]{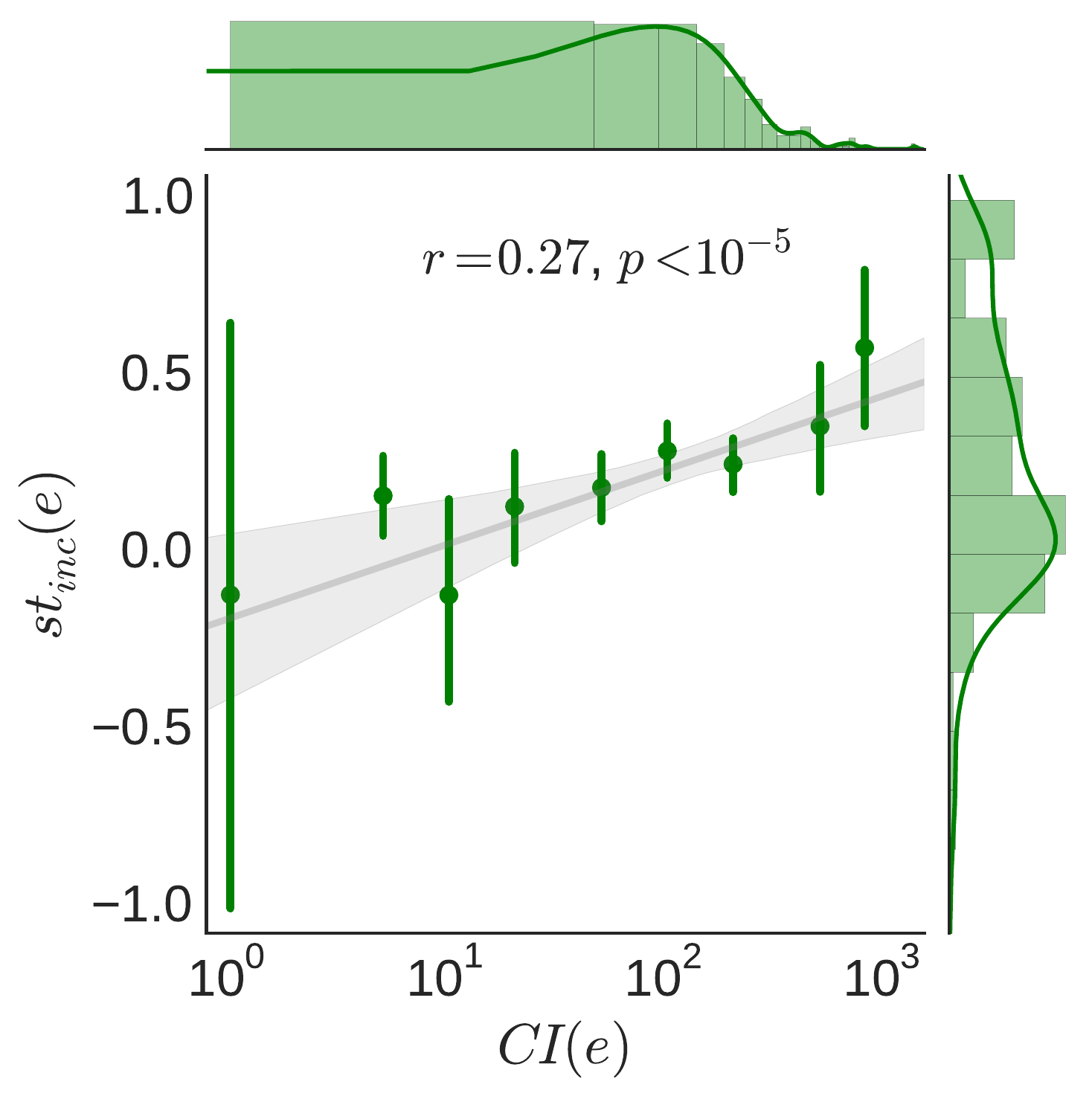}}l
   	\caption{\label{fig:edge_weight_inconsistency} \textbf{Relationship between communication intensity and link inconsistency:} (left) scatter plot; (right) linear regression visualization -- we apply logarithmic binning to account for long-tailed distribution of $CI(e)$; average value and standard deviation are shown for each bin.} 
   \end{figure*}

We indeed find significant correlation between introduced link inconsistency and communication intensity ($r = 0.27$). Results presented in Fig.\ \ref{fig:edge_weight_inconsistency} indicate following finding: the communication between two users tends to increase with status inconsistency of one or both of the users, if they are both richer on the same form of capital. If the users are status inconsistent but being rich on different forms of capital, then their communication intensity tends to decrease. As with other findings regarding social capitals, the described patterns are relevant for extreme cases (high and low edge weights), and there is a wide spectrum of edge inconsistency values taken by the medium-weight edges (Fig.\ \ref{fig:edge_weight_inconsistency}, left).  

%%%%%%%%%%%%%%%%%%%%%%%%%%%%%%%%%%%%%%%%%%%%%%%%%%%%%%%%%%%%%%%%%%%%%%%%%%%%%%%%%%%%%%%%%%%%%%%%%%%%%
\section{Temporal evolution of semantic homophily}
\label{sec:edges_temporal_SR}
%%%%%%%%%%%%%%%%%%%%%%%%%%%%%%%%%%%%%%%%%%%%%%%%%%%%%%%%%%%%%%%%%%%%%%%%%%%%%%%%%%%%%%%%%%%%%%%%%%%%%
In previous sections we performed analysis on a snapshot of Twitter network formed from the whole $6$ months dataset. In this section we investigate temporal aspects of semantic homophily by looking at different snapshots of the network for each month.
 \begin{figure}
 	\centering
 	\subfloat{\includegraphics[width=0.5\textwidth]{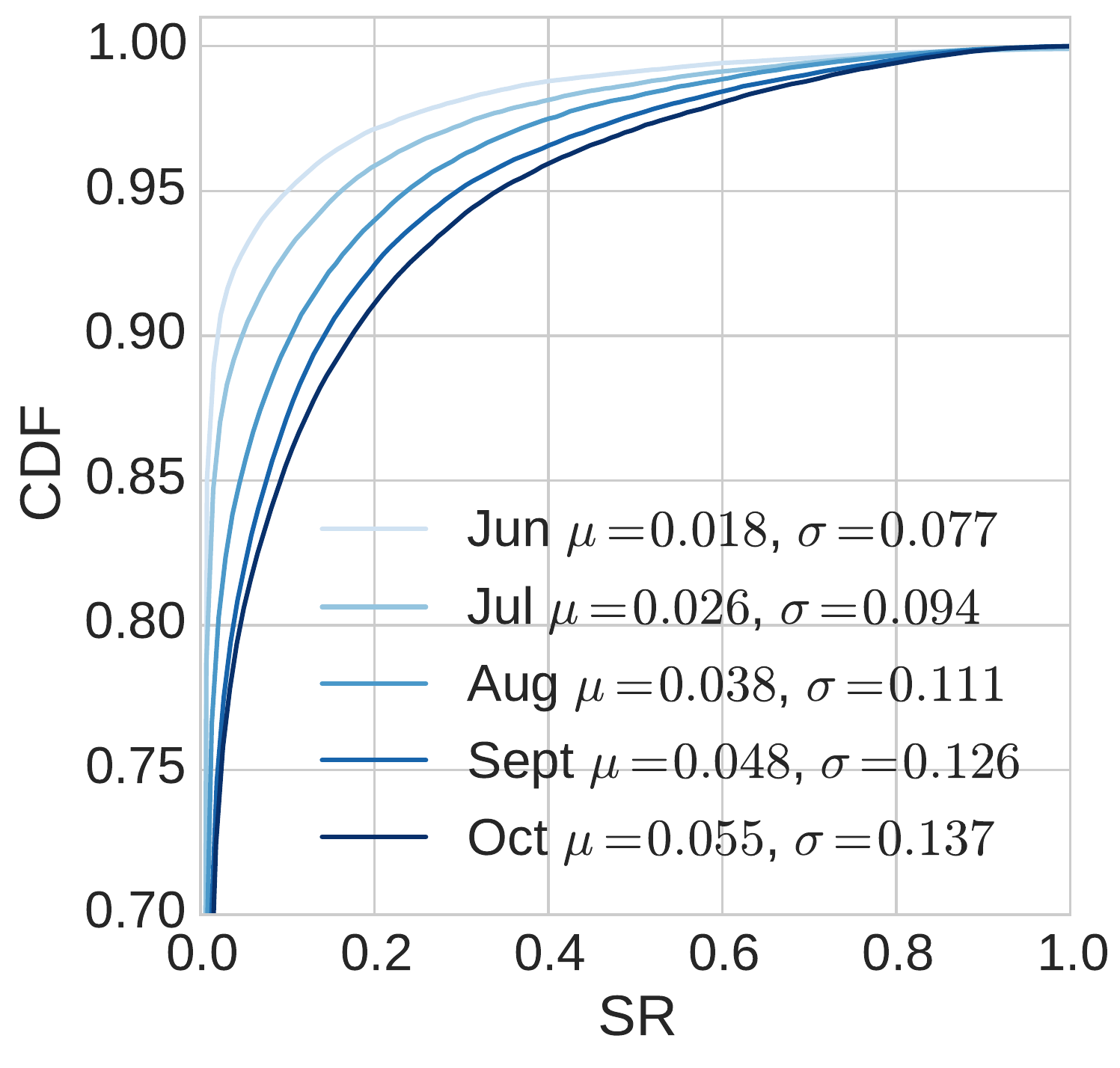}}
 	\subfloat{\includegraphics[width=0.5\textwidth]{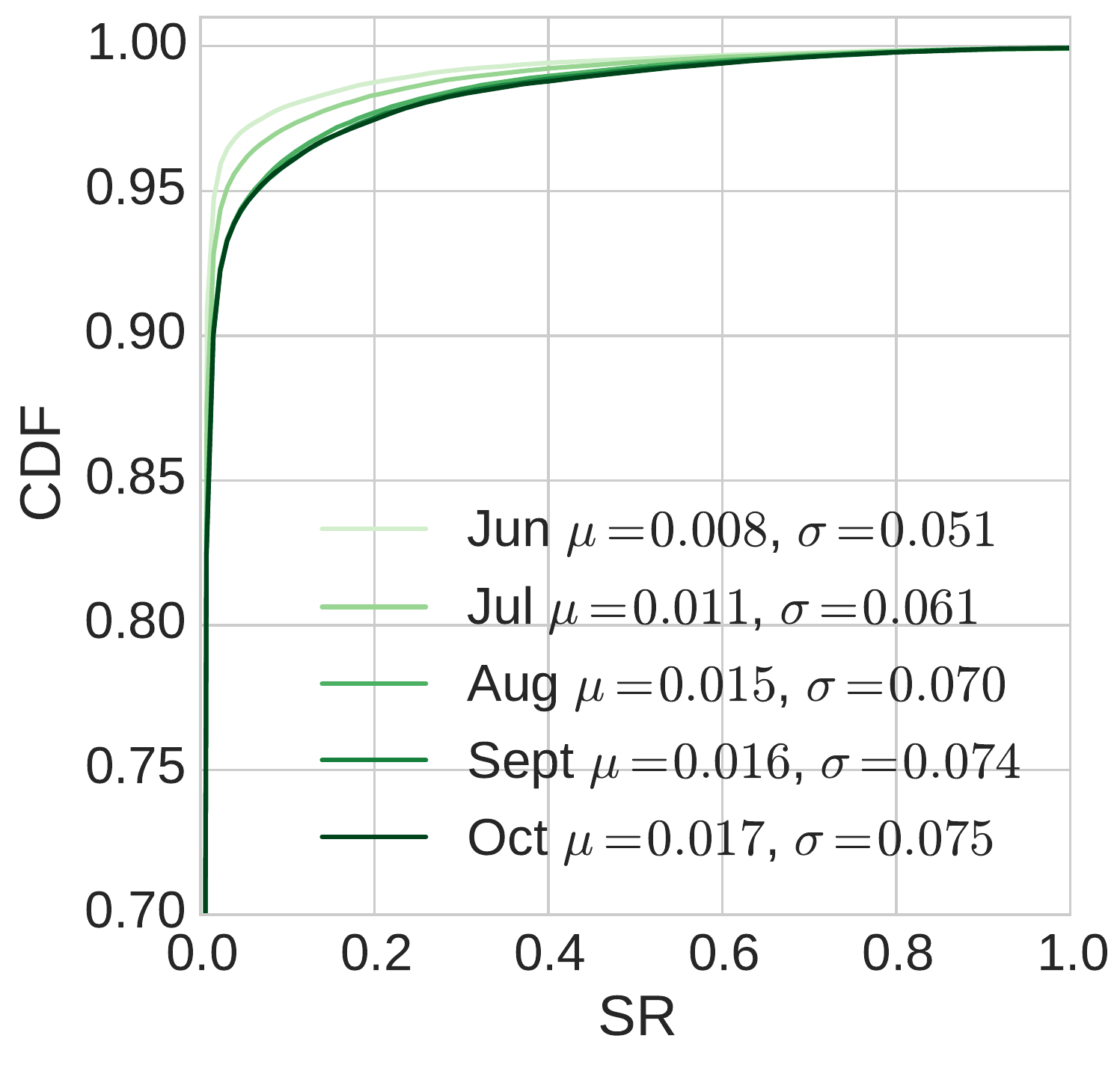}}
 	\caption{\textbf{Cumulative SR distributions} for $5$ full months in our dataset: (left) in $\mathtt{communication\ network}$ and (right) in the rest of $\mathtt{SR\ network} $. For better visualization of the differences in distributions the $y$-axis is thresholded above $0.7$. The distributions are together sharply rising up to around that point. \label{fig:sr_distributions}}
 \end{figure}
First we analyze \textit{temporal change of SR values}. In Fig.\ \ref{fig:sr_distributions}, cumulative distribution functions (CDF) of SR values for each full month in our dataset are shown for $\mathtt{communication\ network}$ and for the rest of the links in $\mathtt{SR\ network}$. Precisely, we consider all the links with mutual communication (strong ties) in $\mathtt{communication\ network}$, while for the second distribution, we take the difference between links in $\mathtt{SR\ network}$ and all communication contacts (both strong and weak). In this way we aim to distinguish between SR of user pairs affected by communication (and hence social influence) and those that are less likely to be affected (no communication of any type occurred between them in our dataset). Gradual increase in SR values takes place in both cases over time (CDF increases at higher SR values). In addition to the visualization, by applying Kolmogorov-Smirnov (K-S) test \cite{massey1951kolmogorov}  we confirm the distribution change. In particular, we compare the distributions for June and for October. For $\mathtt{communication\ network}$, K-S results in $p<e^{-24}$ and, respectively, for $\mathtt{SR\ network}$, in $p<e^{-197}$, hence in both cases strongly rejecting the hypothesis that the distributions are the same.
 
\subsection{External influences evidence}
 The increase in average SR in $\mathtt{SR\ network}$ (Fig.\ \ref{fig:sr_distributions}) among not connected pairs of users is peculiar. It indicates a possible external influence taking place during the period causing all users to talk more on a similar (external) topic. However, since the Twitter social network we investigate is not the only possible way for our users to communicate and influence each other, this does not allow us to assert whether the increase is indeed (only) due to external influence. In any case, we turn to our semantic layers to look for an evidence of common external influences in the dataset.
 
\begin{figure}
 	\centering
 	\includegraphics[width=0.82\linewidth]{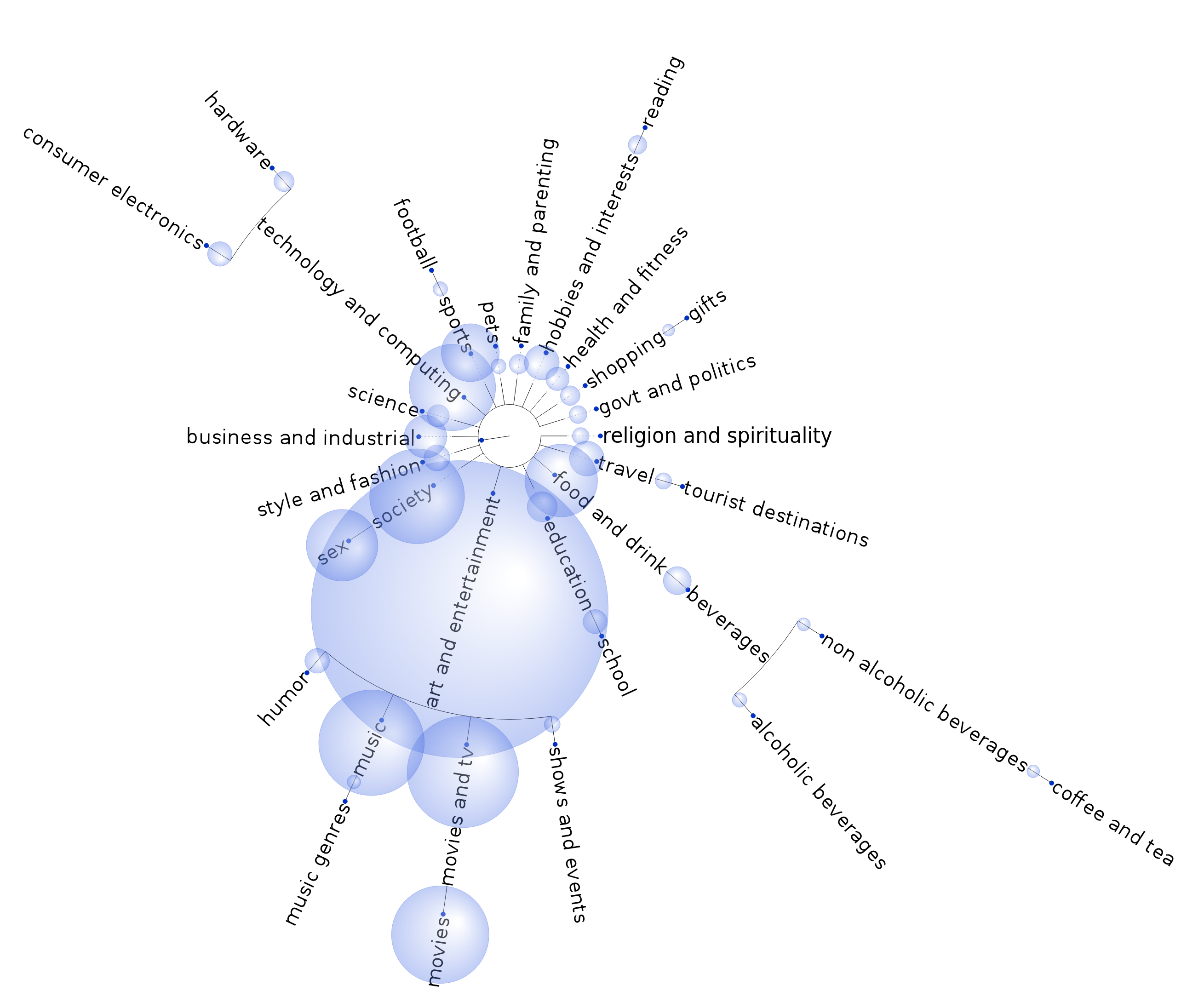}
 	\caption{\textbf{Semantic taxonomy of whole $\mathtt{communication\ network}$ visualized in a bubble-tree-map}: highest level categories are in the center. Subcategories are represented as descendants in the tree. Size of bubbles corresponds to the frequency of topics under that category in our dataset.}
 	\label{fig:comm_taxons_all}
 \end{figure}

 Using \textbf{AlchemyAPI} output, we identify overall most popular categories for topics of communication in our dataset. They are displayed in Fig.\ \ref{fig:comm_taxons_all}. \textit{Arts and entertainment}, including \textit{movies}, \textit{tv shows}, \textit{music} and \textit{humor} is the dominant category. Second set of most popular categories includes \textit{sex} (under \textit{society}), \textit{sports} and \textit{technology and computing}. 
 
Insights on common topics of communication using \textbf{Wikipedia semantic relatedness database} are consistent with those from AlchemyAPI. In Table \ref{tab:wiki-conc} we present some of top $100$ concepts (Wikipedia articles) found to describe the semantics in the dataset overall. For easier comparison, we display these concepts per (sub)categories identified using AlchemyAPI. The two seasons of TV series \textit{This Is England} that have been aired at the time corresponding to our dataset are ranked $2$nd and $3$rd. Next, we also find several musicians and bands. The concepts \textit{LOL} and \textit{Smiley Face} are in part a result of how ESA algorithm \cite{gabrilovich2009wikipedia,gabrilovich2007computing} that we used to build Wikipedia SR database works. They are also in agreement with humor being prevalent subcategory among users in our dataset. In addition to the series \textit{This is England} being aired at the time of our dataset, the death of Osama bin Laden also happened during that period, and we see an article about him describing the general conversation. ESA's output of $>300$K Wikipedia concepts describing topics in our dataset results in a fine SR metrics, as exhibited in detecting fine gradual temporal increase. At the same time, from Table \ref{t:CVs_all} we see that already the top $100$ concepts provide insights into the concrete topics of the conversation in the dataset. 	
 
 These insights, offer evidence for some external influence taking place in our dataset that could lead to global increase in SR among not connected users. Since mentioned TV series, music and events are prevalent topics in the dataset, it could mean that our users are independently watching/following and commenting on them. This in turn could lead to average increase in their SR, even if they never communicated. However, once again, we can not assert whether the increase is indeed (only) due to external influence or due to some social contacts and/or peer influence not detectable using our Twitter dataset. 
 \begin{table}
 	\centering
 	\caption{Most popular Wikipedia concepts in the dataset, per taxonomy categories: movies and TV shows, music, sports and humor}\label{t:CVs_all}
 	\begin{tabular}{l|l||l|l} \hline
 		\textbf{\textit{\specialcell{Wikipedia articles in category \\ Movies and TV shows}}}&\textbf{\textit{\specialcell{Concept \\ rank}}}&\textbf{\textit{\specialcell{Wikipedia articles in category \\ Music}}}&\textbf{\textit{\specialcell{Concept \\ rank}}}\\ 
 		\hline \hline
 		This Is England '86 (TV series) & 2 & Robert Smith (musician) & 5 \\ \hline
 		This Is England '88 (TV series) & 3 & 10cc (English rock band) & 9  \\ \hline
 		Love of Life (American soap opera) & 15 & The Cure & 10 \\ \hline
 		The Dad Who Knew Too Little & 38 & Producers & 16 \\ 
 		(Simpsons episode) &  &  (band) &  \\ \hline
 		\hline\label{tab:wiki-conc}\end{tabular}
 	
 	\vspace{3mm}
 	\begin{tabular}{l|l||l|l} \hline
 		\textbf{\textit{\specialcell{Wikipedia articles in category\\ Sports}}}&\textbf{\textit{\specialcell{Concept \\ rank}}}&\textbf{\textit{\specialcell{Wikipedia articles in category\\ Humor}}}&\textbf{\textit{\specialcell{Concept \\ rank}}}\\ 
 		\hline \hline
 		List of electronic sports titles & 22 & LOL & 1\\ \hline
 		Larry Johnson (American football) & 67 & Smiley Face & 4 \\ \hline
 		Alabama Crimson Tide football & 68 & Lolcat & 20 \\ \hline
 		Racism in association football & 82 & Pres. Obama on Death of& 36 \\
 		&  &Osama bin Laden (spoof)&  \\
 		\hline\end{tabular}
 \end{table}

\subsection{Semantic homophily, social influence and tie dissolution}
The increase in $\mathtt{communication\ network}$ can be due to homophily in its strict definition, i.e., new user pairs starting communication. Once connected they are later likely to have higher SR, as we presented in Section \ref{sec:quantifying}. This can happen due to already connected pairs becoming more related, i.e., social influence. Sociology also suggests to look for link dissolution among dissimilar individuals \cite{felmlee1990dissolution,block2014multidimensional} as one of the reasons of average network SR increase.
 
We start by investigating formation and dissolution of links through time and their SR change. The requirement for active engagement from both source and receiver allows us to define communication activation (link formation) and communication decommission (link dissolution) for reciprocal links. For each of the $69,312$ reciprocal links observed during the whole period, we define \textbf{communication activation (formation)} time to be the month when for the first time both users have mentioned each other (in our dataset period). \textbf{Communication decommission (dissolution)} time is given by the last month in our dataset that the users have both mentioned each other, after which one or both sides ceased the communication. In order to have enough data to calculate users similarity prior/after to links activation/decommission, we require the month of activation/decommission to be between July and September. With this approach, we find in total $13,492$ link activations and $10,080$ link decommissions in our dataset. As a first insight, we notice that slightly more links are activated than decommissioned. 
%This could be one cause for the observed temporal SR increase if. 
% old reciprocal links $62,757$
\begin{figure}
	\centering
	\includegraphics[width=0.7\linewidth]{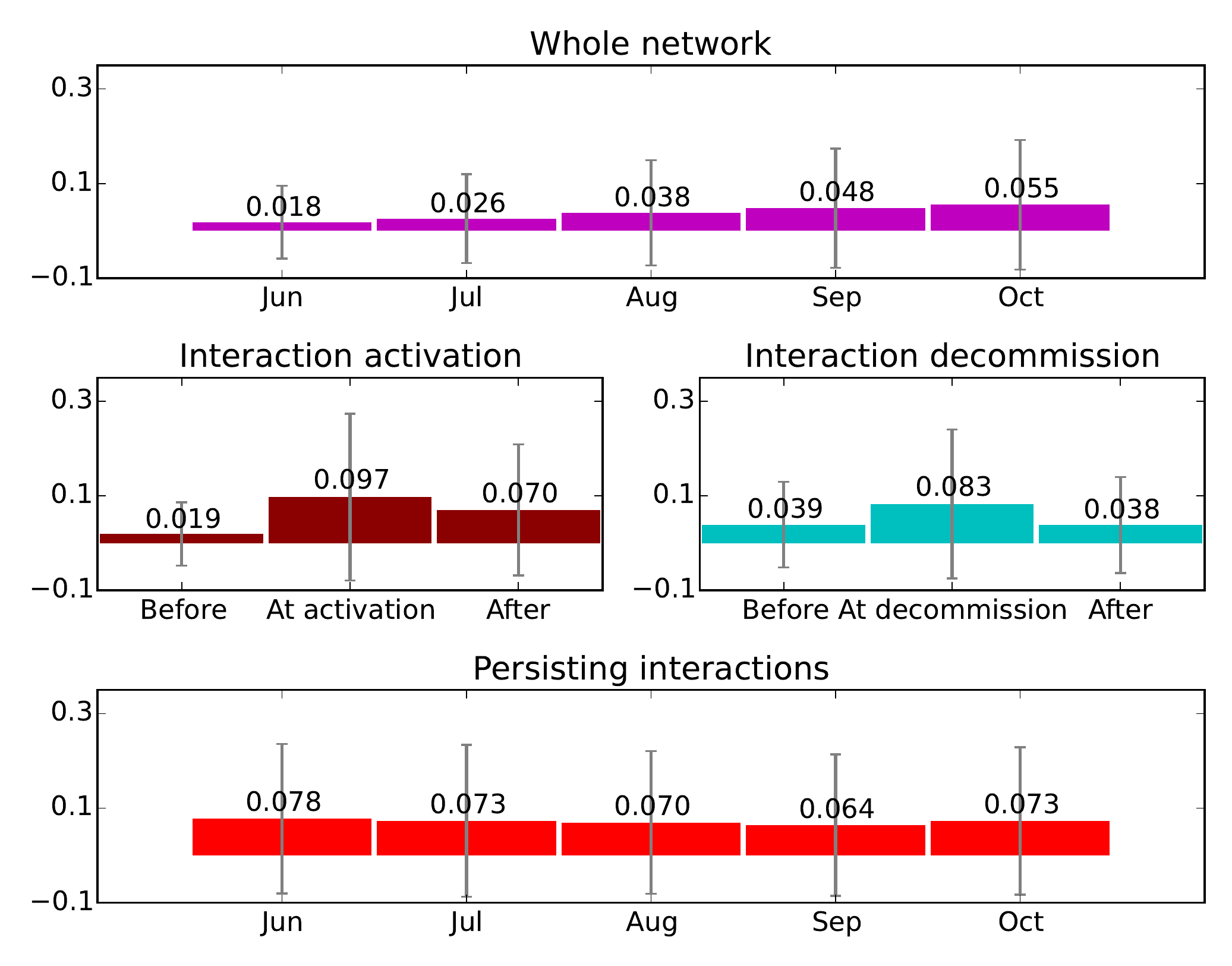}
	\caption{\textbf{Temporal SR change.} Average SR on: (top) all communication links, (mid, left) during communication activation, (mid, right) decommission, and (bottom) on persisting links; error bars show standard deviation values}
	\label{fig:edges_temporal_SR}
\end{figure}

\begin{figure}
	\centering
	\includegraphics[width=0.88\linewidth]{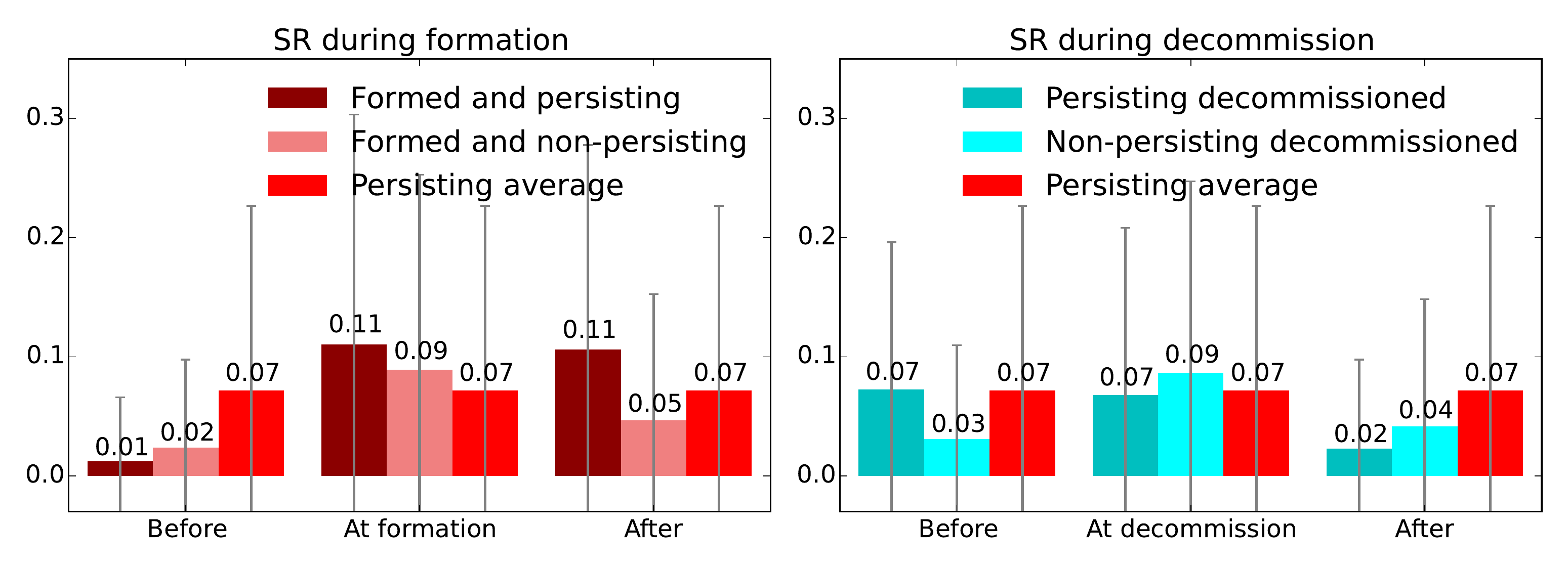}
	\caption{\textbf{SR change during formation and decommission}; (left) during link formation and (right) during link decommission. We show the differences between the links that persist after formation or not, and similarly between those that were persisting in our dataset period before decommission and those that were non-persisting. Error bars show standard deviation values}
	\label{fig:formation_deletion_SR}
\end{figure}
Temporal change of average SR on links prior to and after the \textit{activation} is shown in Fig.\ \ref{fig:edges_temporal_SR}. The SR between a user pair noticeably increases at the month of their communication activation. Similar result has been found in other networks, for instance among Wikipedia admins \cite{crandall2008feedback} and for Flickr users \cite{zeng2013social}. The drop in average SR in the period after the link activation is also reported in earlier studies \cite{zeng2013social}. To investigate the drop in our case, we look for an evidence that some interactions might not be preserved for long. This is one aspect where our approach is advantageous compared to the previous approaches, that consider a formal edge formation (adding someone as a friend or following) and do not require an active user engagement afterwards. 

Indeed, we find in total $8,166$ links that are \textit{activated and then also decommissioned} during our dataset period. 
 The SR change for such links, that are activated and then decommissioned, as well as for those that persist in our dataset after the formation is show in Fig.\ \ref{fig:formation_deletion_SR}. 
  The average SR values for formed and persisting links stay high after they are formed. It is those links that will get decommissioned soon that contribute to lowering the average SR after formation that we see in Fig.\ \ref{fig:edges_temporal_SR}. This result displays that homophily needs to be considered together with active engagement and its temporal dynamics. 

If observing only the \textit{persisting links} that were already active and persisted during the whole period in our dataset, we obtain results for their average SR change in the bottom plot in Fig.\ \ref{fig:edges_temporal_SR}. Such persisting interactions have a relatively stable average SR through time despite that the average SR in the whole network has increased from June until October. Also, SR on persisting links is higher compared to the whole network. The stability of SR for an established communication could suggest a lack of influence in our network. However, we are careful with such an interpretation, since this result might also indicate a saturation effect taking place.
If looking at  newly formed links which persist and have high SR, that indicates how at first, the users might influence each other for some time. However, their similarity is likely to stabilizes around this specific SR value ($\sim 0.07$) for persisting links in our dataset, as indicated by average SR during dissolution of previously persisting links in Fig.\ \ref{fig:formation_deletion_SR} (we discuss this result in more detail below). 

Fig.\ \ref{fig:edges_temporal_SR} also displays temporal change of SR on links that get \textit{decommissioned}. Again, in Fig.\ \ref{fig:formation_deletion_SR}, we separate persisting links (during our dataset) that get decommissioned from those that have formed during our dataset time frame (non-persisting) and get decommissioned. Indeed, we can notice how the persisting links have the above mentioned characteristic average SR of $0.07$ which does not change during the actual month of decommission, but afterwards drops significantly to $\sim 0.02$. The non-persisting links reach even higher SR during the month of decommission, but before and after their SR is lower. This can indicate a sort of short-lived active engagement/interest between such pairs, unlike more stable relationship between previously persisting links. The drop in average SR on the links that get {decommissioned} is striking: SR becomes from $2$ (on non-persisting) to $3$ (on persisting) times lower after link dissolution. Sociology suggests as one possible cause for link decommission that maintaining ties with dissimilar others might be costly \cite{felmlee1990dissolution}. However, we notice that the SR values before decommission on previously persisting links are not lower but around the same as on the links that stay persisting. 
Hence, \textit{in terms of SR there is no observable dissimilarity between users with persisting communication before they will cease communication}. We investigate other possible reasons for their link dissolution below.

\begin{figure*}
	\centering
	\subfloat{\includegraphics[width=0.88\textwidth]{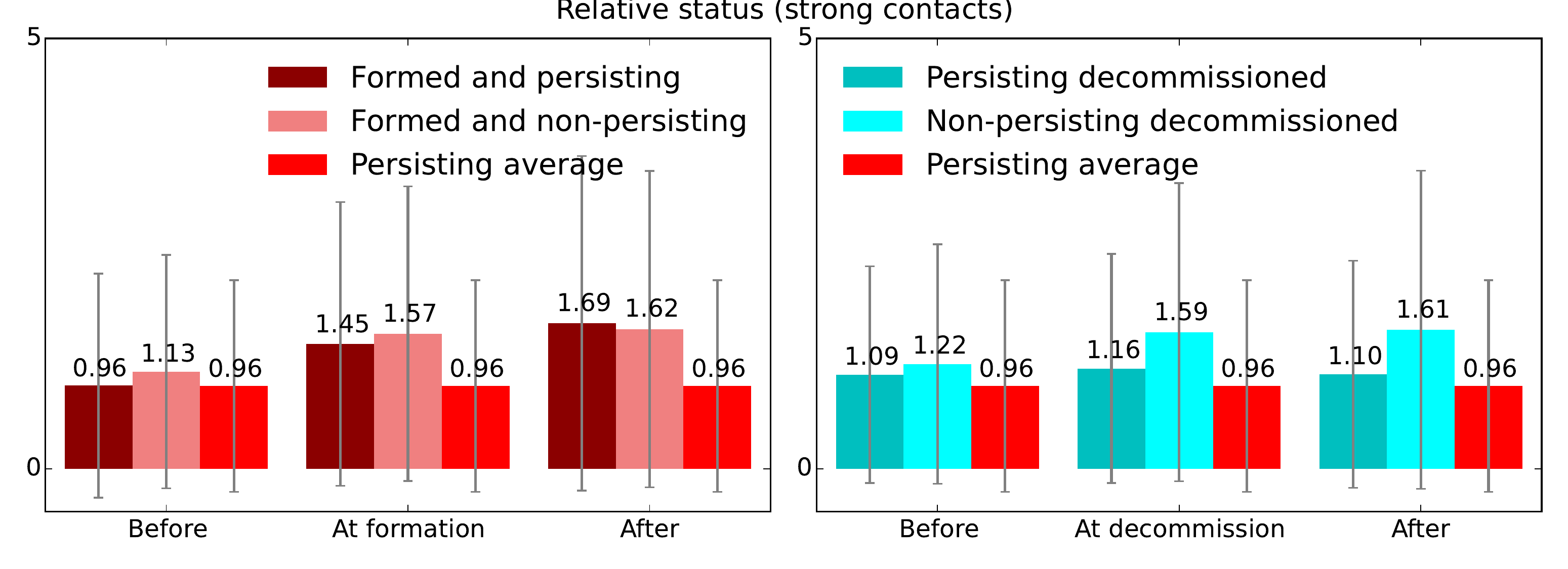}}
	\hspace{1pt}
	\subfloat{\includegraphics[width=0.88\textwidth]{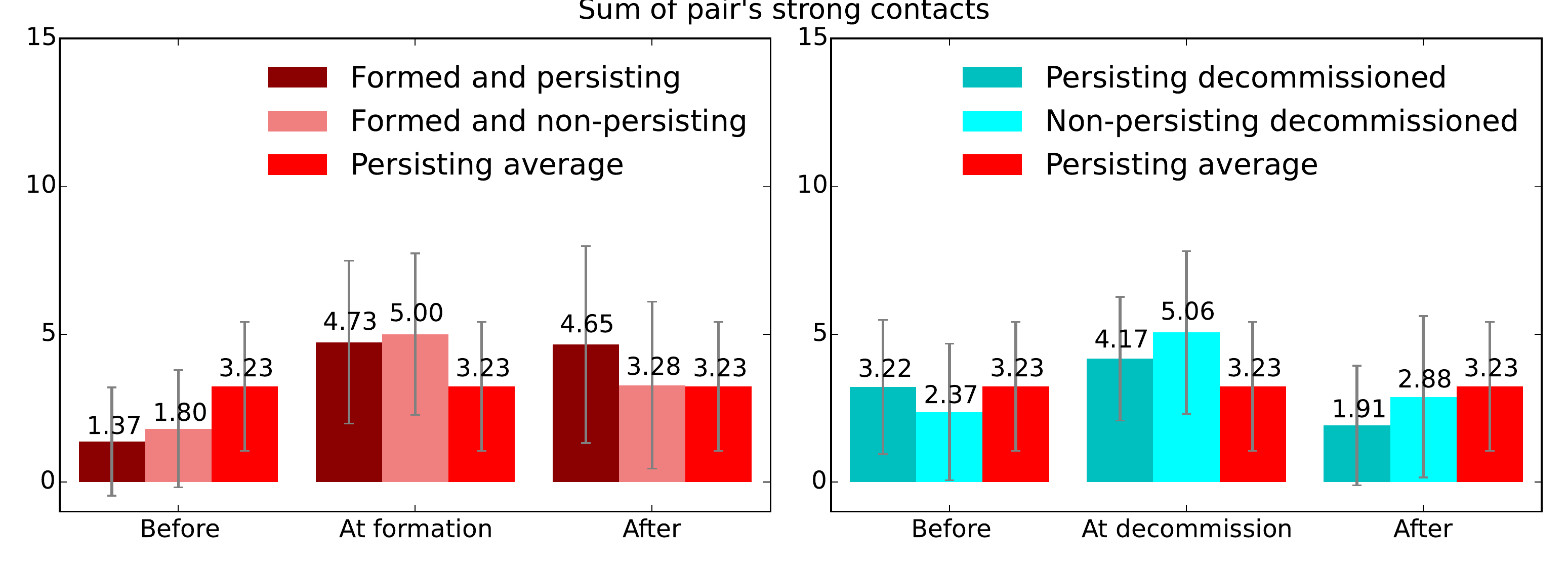}}
	\caption{\label{fig:status_diff_temporal_edges} \textbf{Temporal status differences during formation and decommission}: (top) average relative social status (number of strong contacts), (bottom) average sum of strong contacts. Error bars show standard deviation values} 
\end{figure*}
Operating on the same sets of communication links as so far, we now look at \textbf{social capital} of the communicating user pairs. As presented earlier, different forms of social capital can be assessed. Since herein we look at mutual communication, it is natural to asses social capital in terms of \textit{numbers of strong/weak contacts}. In Fig.\ \ref{fig:status_diff_temporal_edges}, we show relative status and total number of strong contacts of communicating user pairs. 

\textit{Relative social status} (discussed in Section \ref{sec:soc_sem_interplay}) is defined as the absolute \textit{difference between social capitals} of source and receiver users. Looking at relative status (top row plots in Fig.\ \ref{fig:status_diff_temporal_edges}), we first notice the difference on persisting links compared to other types of links (and also to the whole network, a result which is not displayed). Persisting links have lower relative status, i.e., users who are actively communicating tend to have similar social status rank. While homophily on the status level is not new, herein we exhibit its underlying mechanisms in $\mathtt{communication\ network}$. Namely, both types of links, those that are newly formed and those that will get decommissioned in time, have slightly, but notably higher relative social status compared to persisting links. Hence, we find evidence that \textit{link dissolution happens due to dissimilarity in social status}.
Another interesting observation is that user pairs that start with higher relative status compared to persisting also get decommissioned later (while those who start around that persisting average indeed persist communication later). 
The results are similar for relative status in terms of weak contacts so we do not present them. To reiterate, our analysis so far gives two insights about links before they get decommissioned: i) lack of semantic differences on previously persisting links (their SR is not lower at the time when link dissolution happens compared to those who consistently persist communication) and ii) higher status differences (also compared to persisting links). Hence, there is indication in Twitter network that \textit{persisting communication links dissolve in the presence of status level heterophily rather than value level heterophily}.

Findings from sociology also suggest that relationships last shorter time and are more likely to decay for pairs of individuals with lower overall \textit{social status} \cite{burt2000decay}. To assess this hypothesis in our $\mathtt{communication\ network}$, we observe social status in terms of total number of contacts for user pairs who cease communication. Results in Fig.\ \ref{fig:status_diff_temporal_edges} (plots in bottom row) do not support such hypothesis for strong contact: pairs who cease communication have around the same sum of strong contacts on average as the pairs who persist communication. Moreover, in the case of weak contacts, there is an opposite evidence: pairs prior to communication cease tend to have more weak contact compared to average of persisting links (other results for weak contacts are similar to strong so we do not show them).
In addition, the increase in the sum of pair's contacts at the month of decommission suggests that those new contacts might affect their existing link. After the decommission the sum of contacts drops, but still stays higher than would be expected after the decommission (existing link is counted as one strong contact for both users, so after the decommission, their sum of contacts would be expected to drop by $2$). Such evidence suggests that in some percent of the cases \textit{one or both of the users have established new communication links at the time of abandoning the current one between them}. This result is supported by a level of stability on the number of persisting communication links per user. Namely, most of the $5,229$ users who participate in constantly persisting links in our dataset have between one to two persisting contacts ($\mu=1.2$ and $\sigma=0.49$).

In summary, presented types of interactions show the importance of considering both homophily and influence as dynamic interdependent tendencies \cite{yavacs2014impact} in temporal networks, instead of looking at static snapshots. Our analysis on interaction decommission reveals similar results as in \cite{noel2011unfriending} where it is showed how not accounting for homophily effect on tie dissolution ('unfriending') may importantly affect social influence estimation. Precisely, we suggest that on a same communication link (interaction) at different points of time with reference to its activation/decommission time, one or the other of the tendencies might be playing a stronger role. Our dataset time-frame does not allow for that, but as a future work, we aim to look at the period in which edge formations and deletions might be happening, and whether there are some natural cycles in the human communication networks.
%%%%%%%%%%%%%%%%%%%%%%%%%%%%%%%%%%%%%%%%%%%%%%%%%%%%%%%%%%%%%%%%%%%%%%%%%%%%%%%%%%%%%%%%%%%%%%%%%%%%%
\section{Community structure and semantic foci}
\label{sec:comm_str}
%%%%%%%%%%%%%%%%%%%%%%%%%%%%%%%%%%%%%%%%%%%%%%%%%%%%%%%%%%%%%%%%%%%%%%%%%%%%%%%%%%%%%%%%%%%%%%%%%%%%%
We start by investigating what are the semantics traits that shape community structures in $\mathtt{communication\ network}$ of Twitter users. 

When dealing with representations of real-world networks one can distinguish between {structural} and {functional} communities \cite{yang2014detecting,yang2015defining}. The connectivity pattern among members in the network defines \textbf{structural communities}, whereas a common function or a role of user groups defines \textbf{functional communities}. Simply speaking, structural communities can be defined as groups of users that are more tightly connected within the group compared to the rest of the network. This definition can entail \textbf{modular} or communities with distinct users, but also, more representative of the real-world, we can think of \textbf{overlapping community structure}, where certain nodes belong to more communities.  

If we recall Feld's theory about \textit{foci of homophily} \cite{feld1981focused} that drive clustered (community) structure of social networks, then foci can be seen as one such common function or role around which communities are formed. In our case, we allow different semantic traits of user communication to define semantic foci. Our initial question can be now rephrased as \textit{whether structural communities (both modular and overlapping) can be explained in terms of their functional roles by semantic foci}. 
\begin{table}
	\caption{Largest communities in the $\mathtt{communication\ network}$ and their semantic foci}\label{t:communities}
	\centering
	\begin{tabular}{l|llllllllllll} \hline
		\textit{Num of}&\textit{2222}&\textit{686}&\textit{636}&\textit{435}&\textit{381}&\textit{343}&\textit{343}\\ 
		\textit{users}&\textit{}&\textit{}&\textit{}&\textit{}\\ \hline
		\hline
		\textit{Main}&Nigeria&Indonesia&South&Philippines,&Jamaica&U.K.&NY, LA,\\ 
		\textit{geo-entities} &  &  & Africa& Malaysia&  &  & Miami\\ 
		\hline
		\textit{\% positive users} & $0.38$ & $0.87$ & $0.67$ & $0.72$ & $0.5$ & $0.59$ & $0.71$ \\ \hline
	\end{tabular}
\end{table}	

\subsection{Modular communication communities}
\begin{figure}
	\centering
	\subfloat{\includegraphics[width=0.45\textwidth]{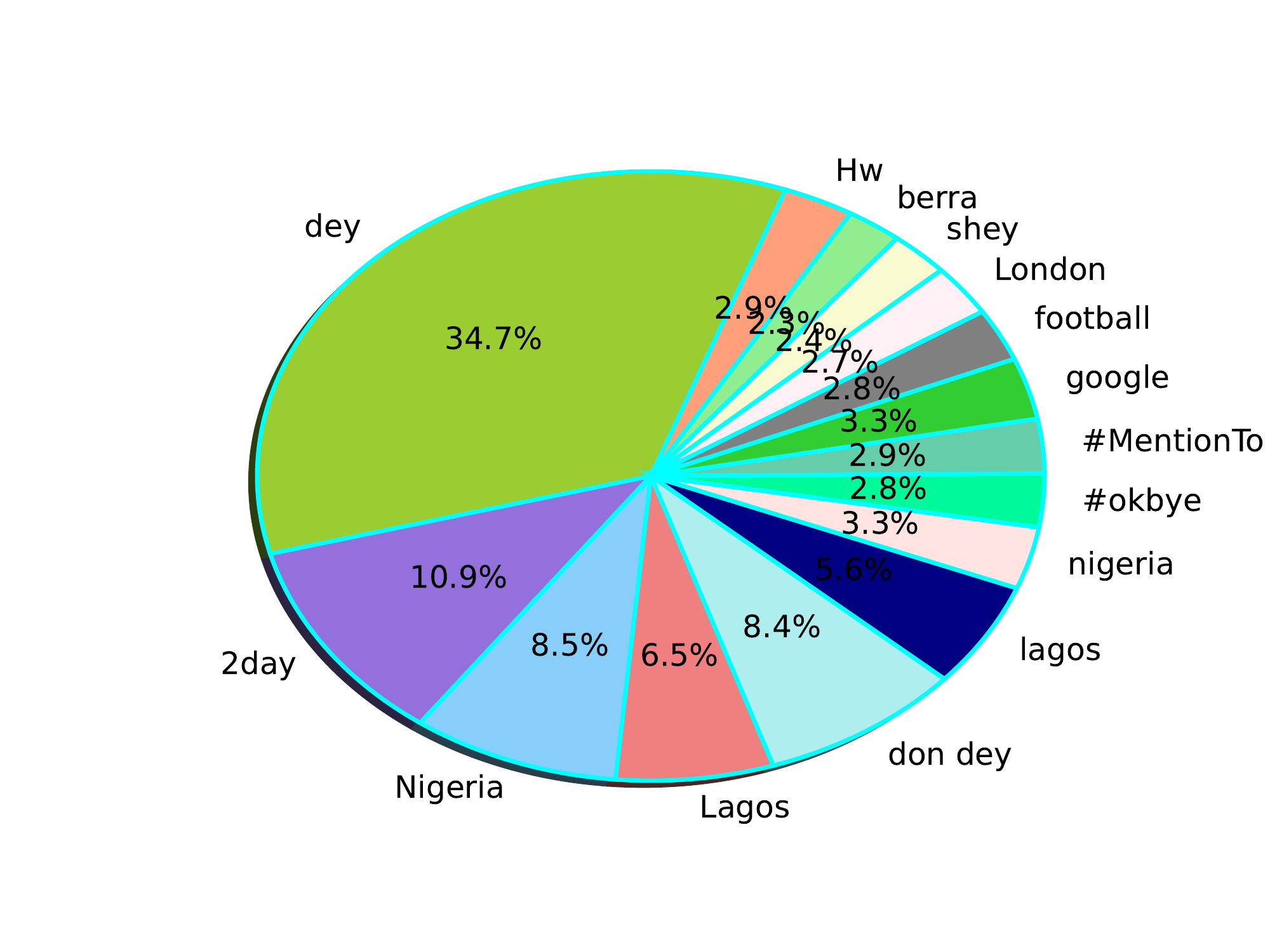}\label{fig:nigeria_entity}}
	\hfill
	\subfloat{\includegraphics[width=0.45\textwidth]{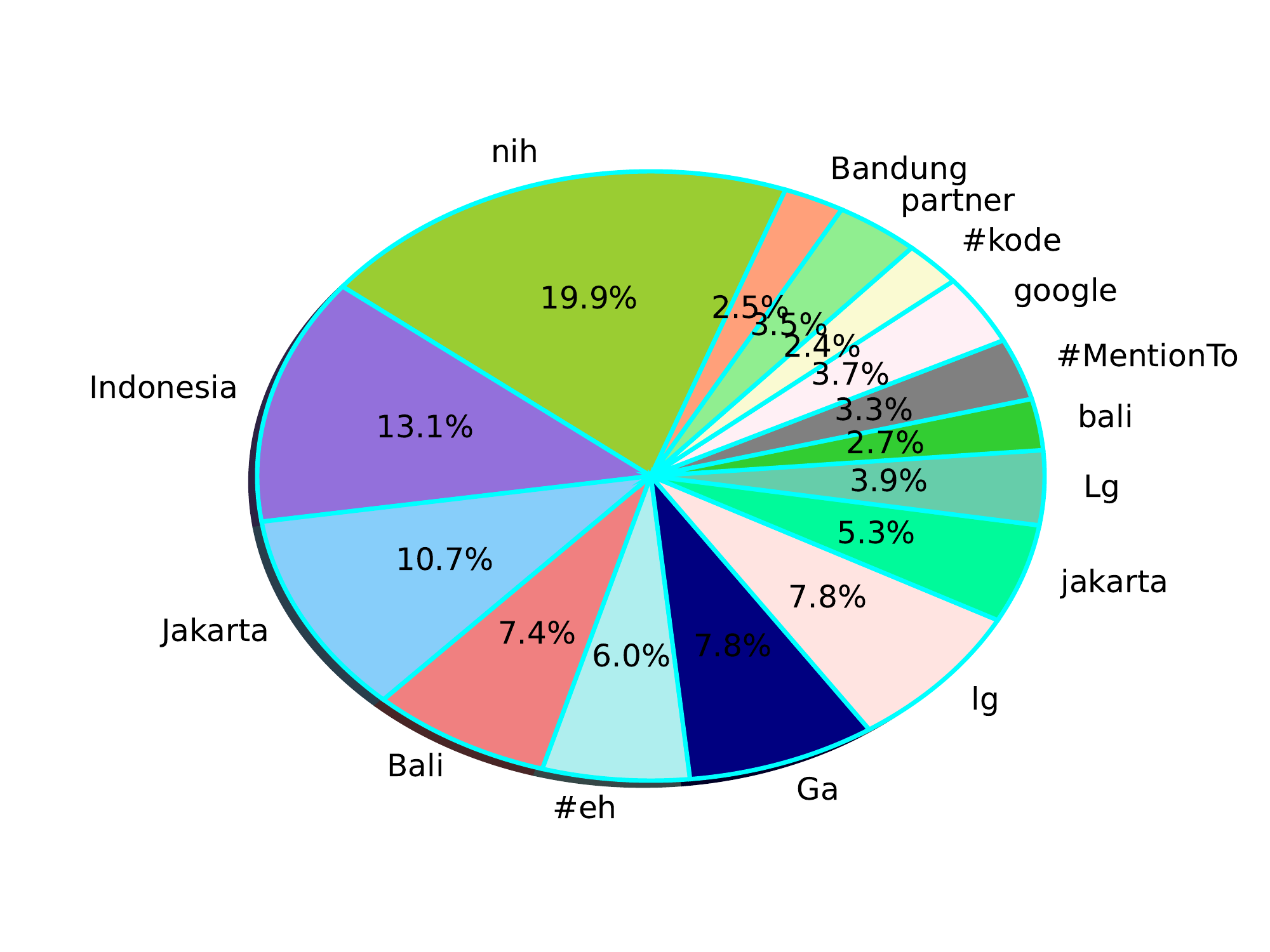}\label{fig:indonesia_entity}}
	\caption{\label{fig:comm_ent} \textbf{Most relevant entities found in the tweets of the two largest communities}: (left) Nigerian and (right) Indonesian} 
\end{figure}
A state of the art algorithm when it comes to detecting \textbf{modular community structure} is based on modularity metrics \cite{newman2006modularity}. We run its fast implementation \cite{blondel2008fast} on our $\mathtt{commnication\ network}$ and detect $2632$ communities. Statistics about the largest detected modular communities is shown in Table \ref{t:communities}. By applying semantic analysis on groups of users belonging to detected communities, we find most relevant semantic traits of the communication in each community. Precisely, we find relevant concepts, entities, categories, taxonomy tree and average sentiment for each community. Then we also apply TF-IDF analysis on the semantic traits with respect to those for the whole $\mathtt{communication\ network}$ to asses whether found semantic traits are specific to a community. After careful analysis, we conclude that only the entities of conversation can be used to explain the modular communities. As an example, in Fig.\ \ref{fig:comm_ent}, we present top entities found in tweets of the two largest communities. Thanks to those entities, we are able to conclude that they represent respectively a community of users speaking about Nigeria and about Indonesia. Importantly, in addition to a few dialect specific words (such as in this case \textit{dey} in Nigerian and \textit{nih} in Indonesian community), among most relevant entities we find geographical entities (in addition to \textit{Nigeria}, we find entity \textit{Lagos} in Nigerian and in addition to \textit{Indonesia}, we detect \textit{Jakarta} and \textit{Bali} for Indonesian community). With such analysis and additional manual inspection of the tweets, we conclude that the largest modular communities are formed around \textbf{geographic entities} as foci of communication (see Table \ref{t:communities} for the other top size communities). To reiterate, we conclude that \textit{geographic entities are homophilous foci that best explain modular communities in our $\mathtt{communication\ network}$}. Similar result are found in different types of communication networks; good predictors of cohesive communication groups in \cite{Leskovec2008,DeChoudhury2011} are geographic foci and several studies \cite{blondel2010regions,aiello2012friendship} report language foci. As a remark, the communities in our Twitter network may be formed due to the ethnicity of users or their geolocation, while in any case, their tweet contents contain relevant geo-location entities.

Another important finding regarding modular communities is that there is a wide diversity in their average \textbf{sentiment}. In Table \ref{t:communities}, we show the percent of 'positive' users in the whole community. We can see it ranges from $0.38$, for a quite 'negative' Nigerian, to $0.87$ for the most 'positive' Indonesian community. The large difference in the sentiment between these two particular communities can be also inferred from their relevant concepts: prevalent swear word-concepts in Nigeria (having negative sentiment), and, on the other hand, \textit{gratitude} and \textit{luck} being dominant in positive Indonesia. Displaying particularity of the Indonesian community, an earlier study found that Indonesian users have higher than average tweets per user ratio, which is related to higher reciprocity, and in turn a higher-reciprocity communities display a happier language \cite{poblete2011all}. 

\begin{figure}
	\centering
	\includegraphics[width=0.9\linewidth]{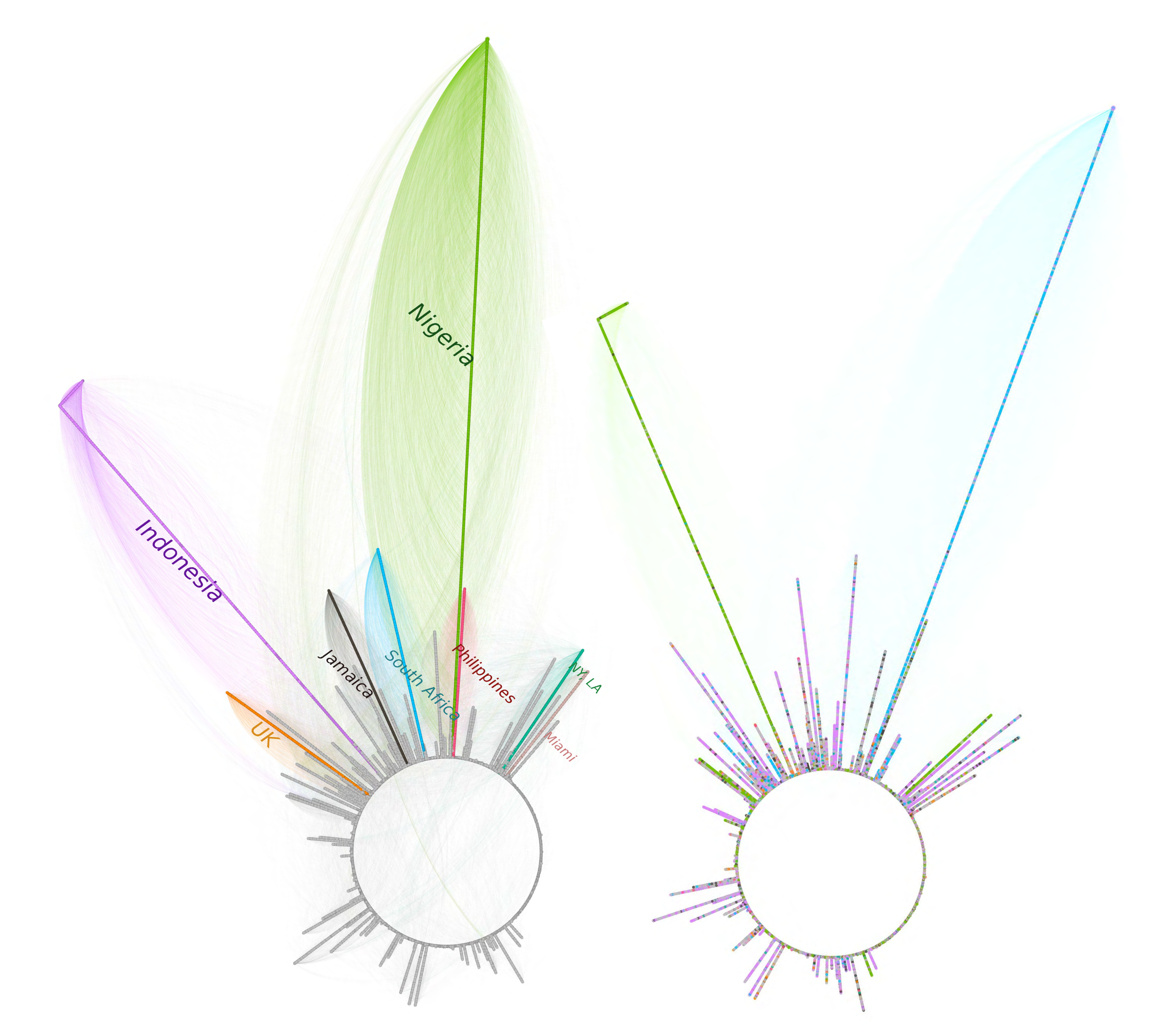}
	\caption{\textbf{Modular $\mathtt{communication\ network}$ communities;} (left)  radial axis visualization in Gephi \protect\cite{ICWSM09154} of $\mathtt{communication\ network}$ communities with displayed identified user geo-location entities in each community; (right) SR communities produced by Infomap visualized with different colors on the $\mathtt{communication\ network}$ community representation; we can see to what extent the largest modular communities from the communication layer overlap with those produced from the semantic layer}
	\label{fig:comm_radial}
\end{figure}
If modular structural communities were not formed around foci as suggested by Feld's theory, but if instead they were simply a result of semantically related users connecting more often, then we would expect to see similar communities when running community detection on the $\mathtt{SR\ network}$. We test such \textit{hypothesis} by detecting communities on a several $\mathtt{SR\_x}$ networks. In order to evaluate how well the sets of communities from communication ($P$) and semantic layer ($L$) match, we apply the procedure used in \cite{yang2014detecting,yang2012community,yang2015defining} to find the matching score:

$$S=\max_{P_{j}\in P, L_{i}\in L}F_{1}(L_{i}, P_{j}),$$ where $F_{1}()$ uses $F_{1}$ as a score for similarity between the two sets. Resulting $S \in [0,1]$, where $1$ indicates perfect matching.

Best matching score we find when running InfoMap algorithm \cite{rosvall2008maps} on the $\mathtt{SR\_0.2}$ network. The threshold $x=0.2$ matches with and is explained by the analytical analysis of $\mathtt{SR\ network}$ that we discussed earlier (see Section \ref{sec:SR_comm}). InfoMap is not modularity-based community detection and the rationale why it performs better on the \textit{semantic layer} is because $\mathtt{SR\_x}$ networks are so dense. Modularity metric, which is optimized by modularity-based algorithms, evaluates existence of dense connections among nodes within communities but \textbf{sparse} connections with nodes in different communities. Hence it can not work well on dense networks, such as $\mathtt{SR\_0.2}$. Best matching scores for biggest communities (with more than $50$ users) are presented in Table \ref{t:f1_sim}. We also visualize modular communication communities and their respective SR community counterparts in Fig.\ \ref{fig:comm_radial}. The matching scores reveal that SR communities can only to a moderate extent explain the communication community structure. Such conclusion, in turn, supports Feld's theory about foci of homophily, in particular when he states that \textit{similarities need not lead to focused (clustered) interaction, and focused interaction can exist apart from similarity of individual characteristics} \cite{feld1981focused}.
\begin{table}
	\centering
	\caption{Community similarity between communication and semantic layer}\label{t:f1_sim}
	\begin{tabular}{c|l|l} \hline
		\textbf{\textit{P communities}}&\textbf{\textit{L communities}}&\textbf{$S$}\\ \hline
		$P_{0}$ - Philippines & $L_{326}$&  0.41\\ 
		$P_{8}$ - Nigeria &$L_{2}$&  0.45\\ 
		$P_{10}$ - Indonesia &$L_{159}$&  0.18\\ 
		$P_{11}$ - Nigeria &$L_{2}$& 0.18\\ 
		$P_{102}$ - UK &$L_{211}$& 0.13 \\
		\hline\end{tabular}
\end{table}

\subsection{Overlapping communication communities}
Next we analyze overlapping structural communities in $\mathtt{communication\ network}$. We select the algorithm BigClam \cite{yang2013overlapping} because it detects overlapping communities as the groups of nodes with denser links presence, in agreement with sociological theories, such as the Feld's \cite{yang2014overlapping}. BigClam automatically detects $198$ communities in our network, largest in size consisting of $586$ users. \textbf{Community membership} of a user (defined as the number of communities in which it belongs), ranges from the minimum $1$, for a majority of users, to the maximum $14$, for a small number of users, and it exponentially decreases. 
\begin{figure}
	\centering
	\subfloat{\includegraphics[width=0.5\textwidth]{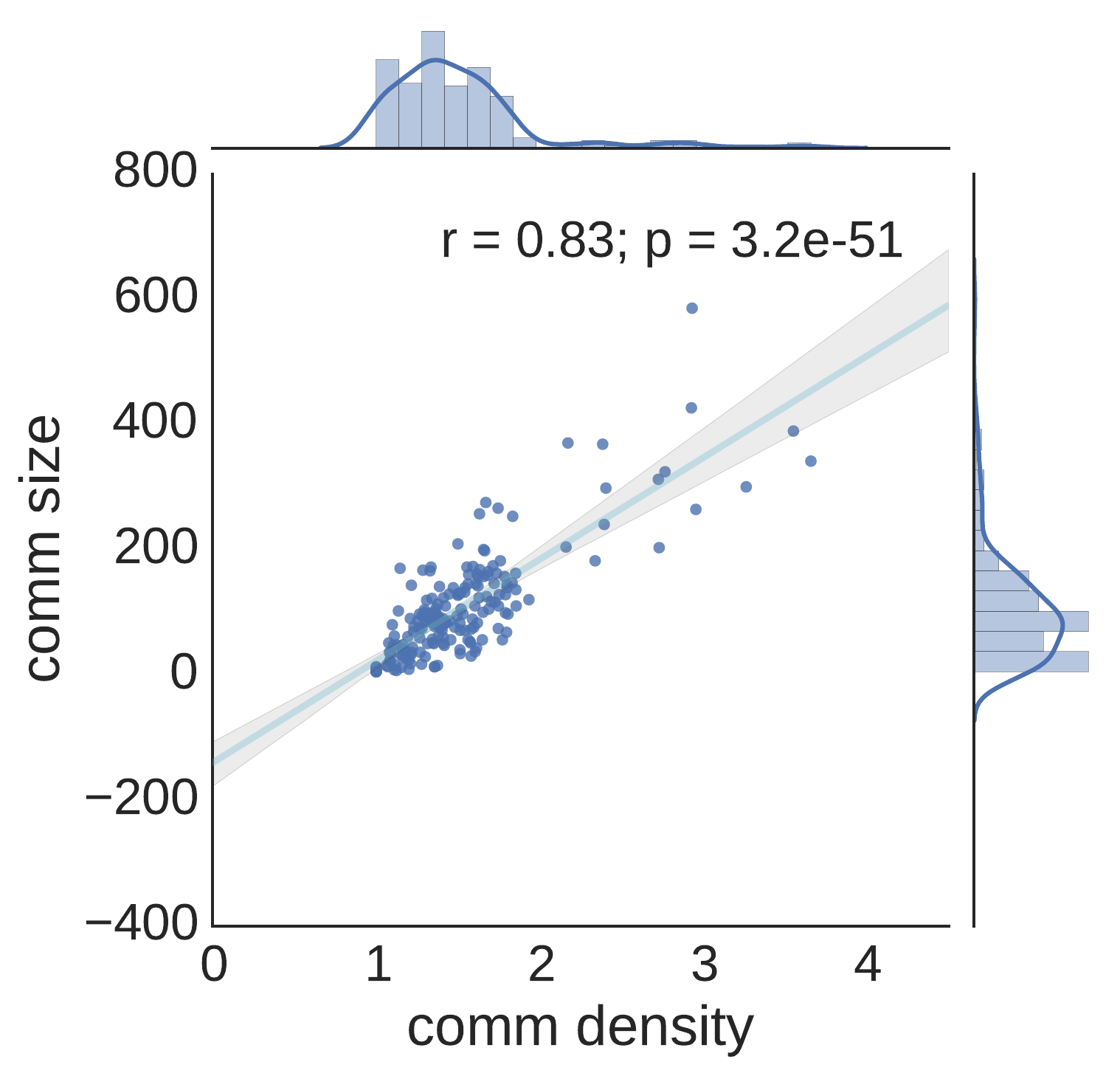}}
	\subfloat{\includegraphics[width=0.43\textwidth]{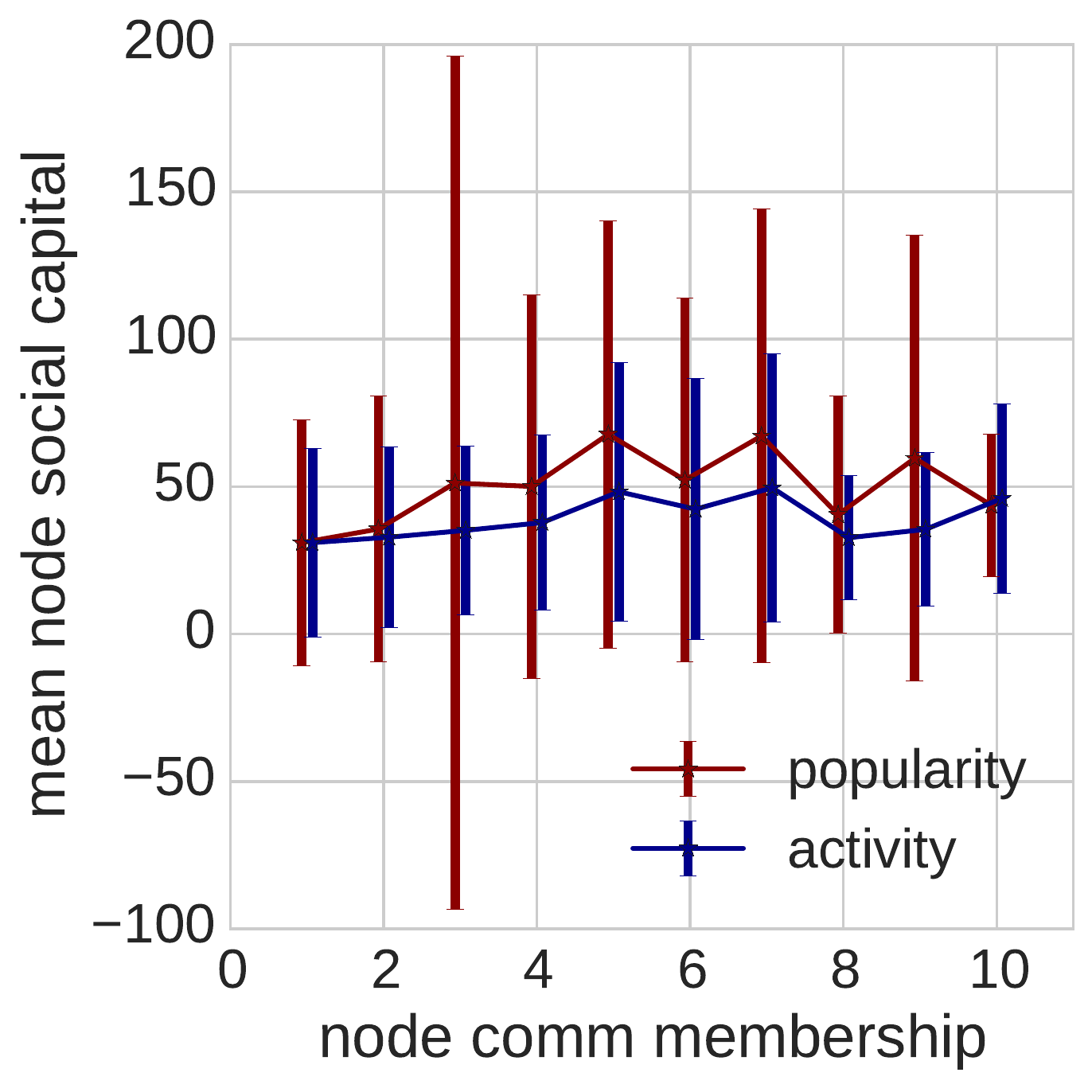}}
	\caption{\textbf{Overlapping communities}: (left) relationship between community size and density; (right) forms of social capital as a function of node community membership. We average values for $10$ and more community memberships, due to data scarcity. Error bars show one standard deviation.\label{fig:comm_membership_soc_capital}}
\end{figure}
Similar semantic analysis as with modular communities reveals that geographic foci are again the strongest predictor of communities. The subtle difference, however, is seen in modular communities being broken apart in several overlapping communities. For instance, the largest Nigerian modular community now has 7 overlapping counterparts. Many of the nodes from one modular community will belong to several such counterparts. By careful analysis, we reveal other foci, behind the overarching geographic, that drive such overlapping communities within the modular (these foci can again be geographic or not). For example, withing the Nigerian group, we find subgroups discussing different geo-entities, in addition to common Nigeria: some talk about Ghana, some about Zambia and others about London. That not only geographic foci drive these overlapping sub-communities, we can see from the case for Malaysia where one subcommunity of $260$ users has the predominant entity \textit{selamat hari raya}, or Muslim greeting for Happy Eid. We also find communities around specialized topics, such as one of $144$ users talking predominantly about NASCAR (auto racing). Hence, \textit{our semantic analysis of overlapping community structure reveals that geographic and language foci are the largest foci, in terms of number of users connected.  Within these foci as enablers, we can find other more focused and overlapping foci, with smaller number of users discussing more specific topics}.

Additionally, we look into which communities are featuring most overlaps with others. To this purpose, we introduce \textbf{community density} as the average number of community memberships for the nodes in the community. As presented in Fig.\ \ref{fig:comm_membership_soc_capital} (left), there is a strong positive correlation between the size of the community and its introduced density. Such result exhibits that the largest communities are those that feature most overlaps with other (sub)communities. Thinking of foci, such result can be interpreted also in the following way. The largest foci are as well enablers for participating users to develop more additional foci of homophily. A related result in an analysis on Twitter is reported by Halberstam et al.\ \cite{halberstam2014homophily} who found that users \textit{affiliated with majority political groups, relative to the minority group, have more connections, and are more densely connected}.
%%%%%%%%%%%%%%%%%%%%%%%%%%%%%%%%%%%%%%%%%%%%%%%%%%%%%%%%%%%%%%%%%%%%%%%%%%%%%%%%%%%%%%%%%%%%%%%%%%%%%
\subsection{Pluralistic homophily}
\begin{figure}
	\centering
	\subfloat{\includegraphics[width=0.33\textwidth]{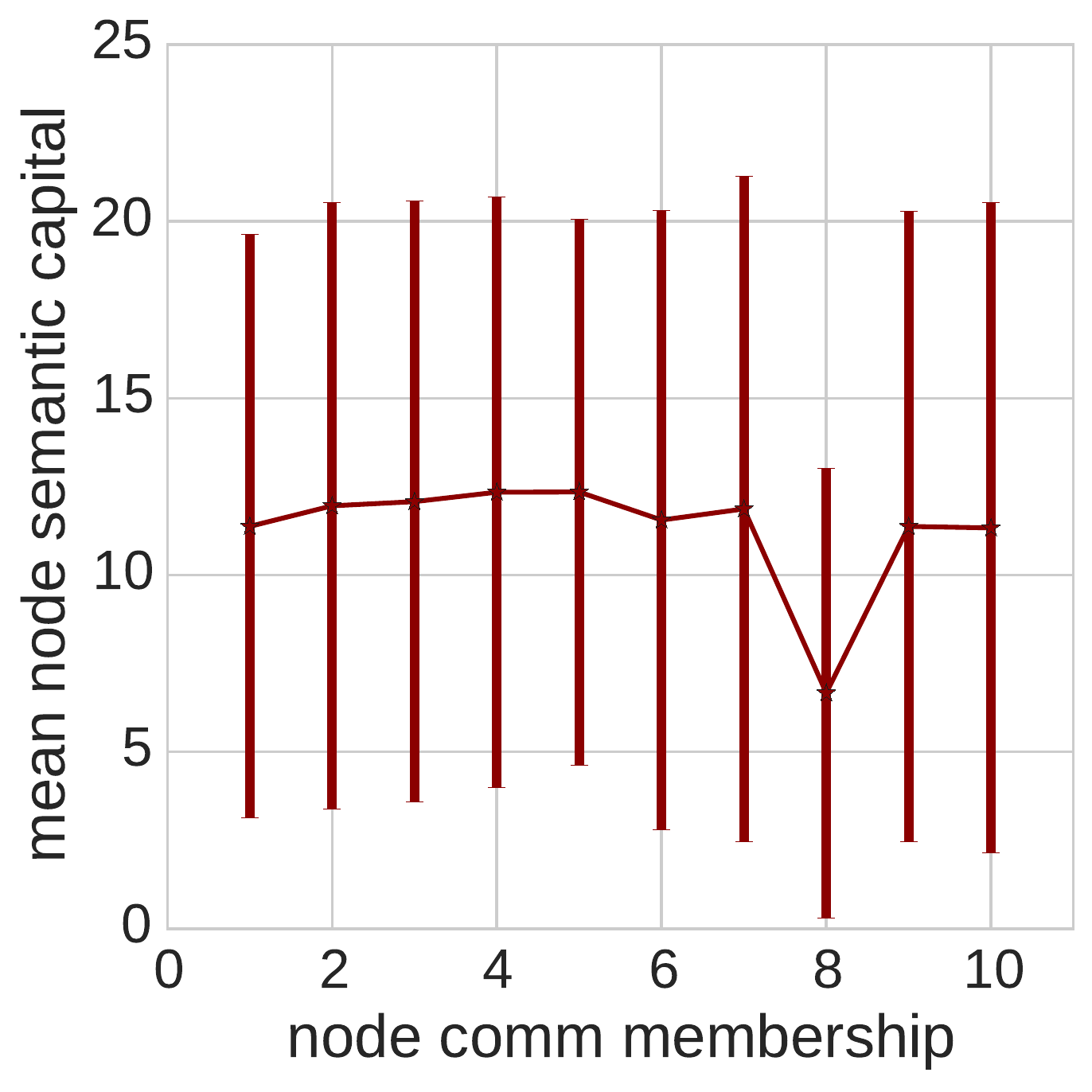}}
	\subfloat{\includegraphics[width=0.33\textwidth]{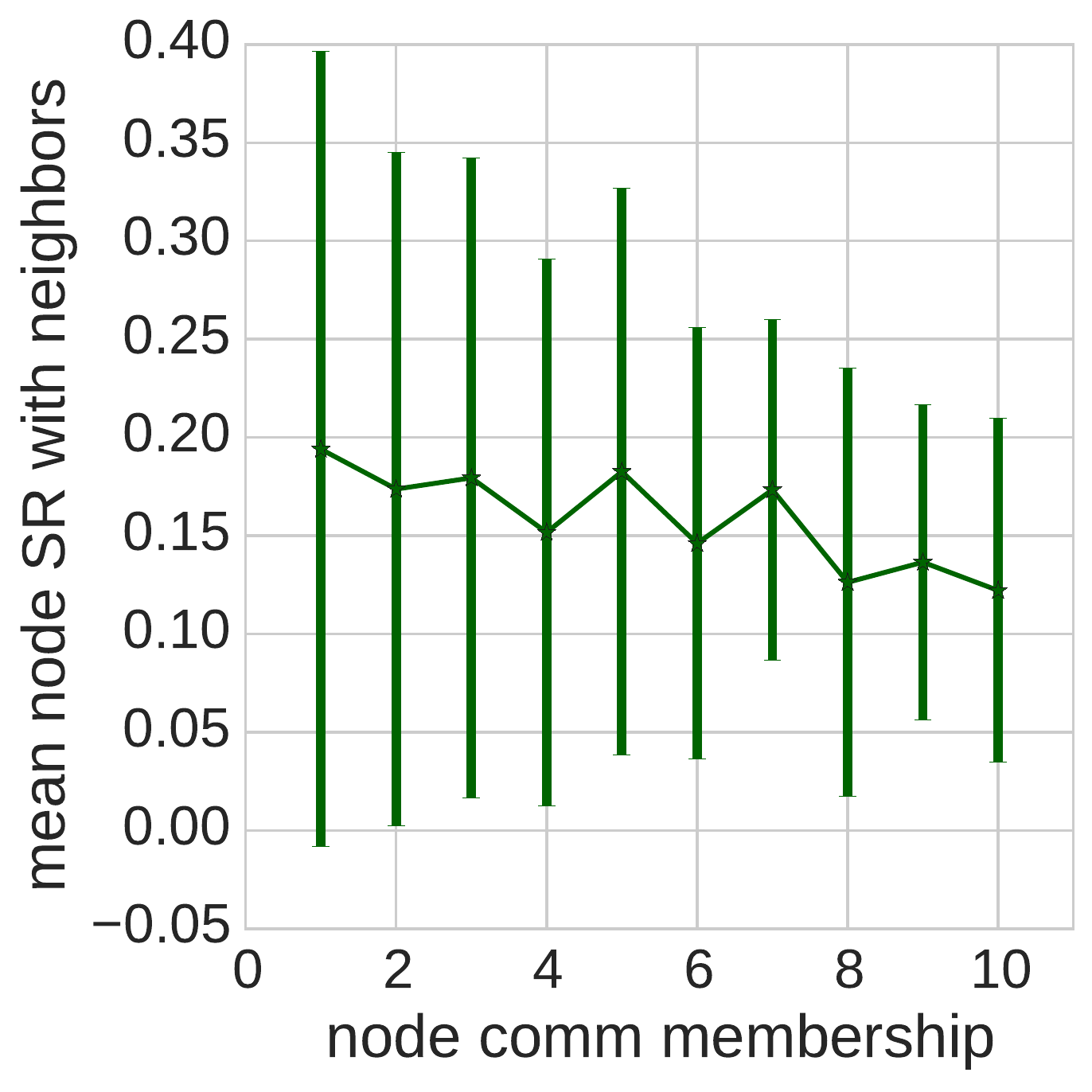}}
	\subfloat{\includegraphics[width=0.33\textwidth]{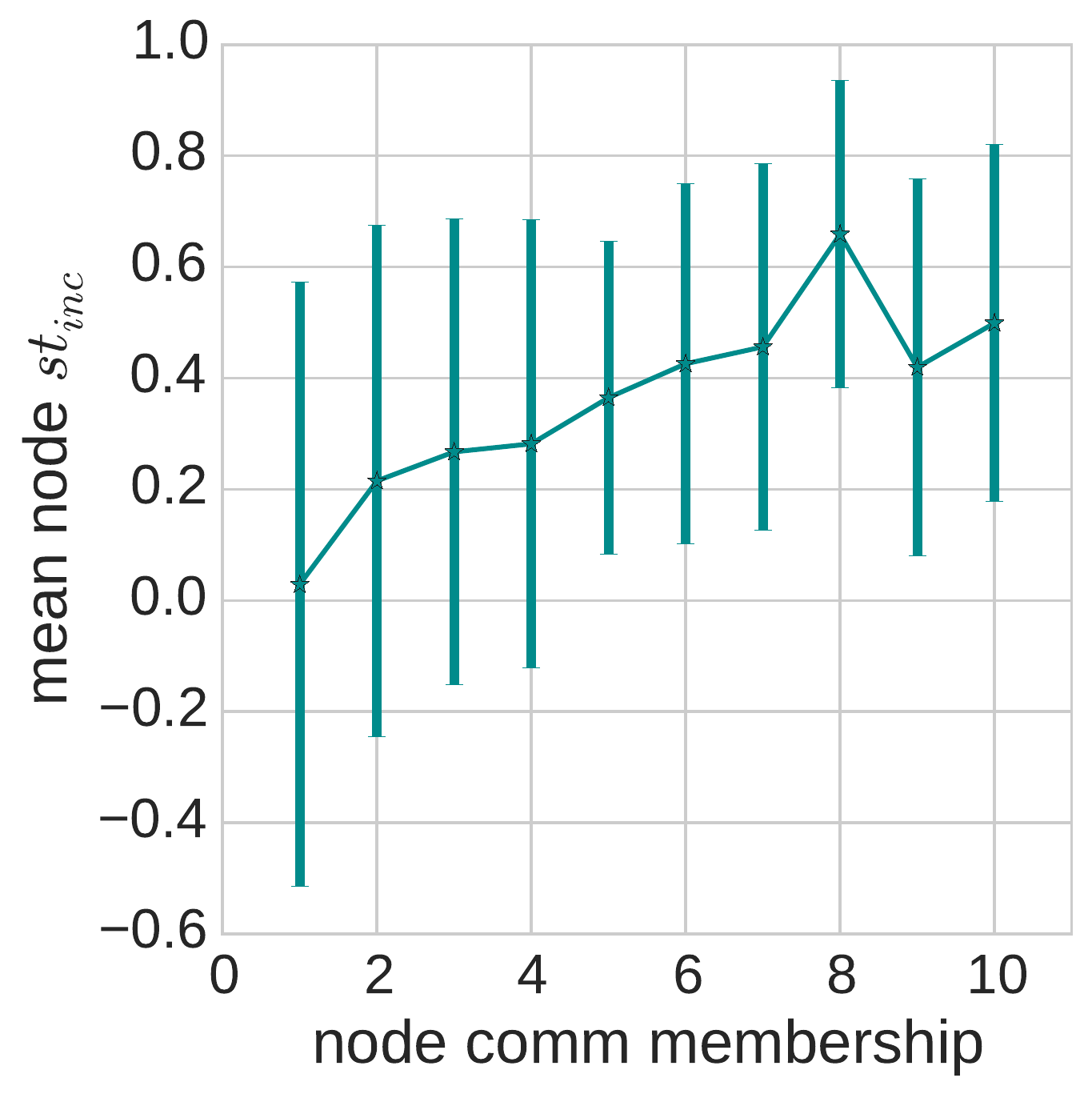}}
	\caption{\label{fig:PLHom_7s} \textbf{Pluralistic homophily and semantic capital}: (left) semantic capital, (middle) mean neighborhood SR and (right) status inconsistency in function of community memberships. Error bars show one standard deviation.}
\end{figure}

%%%%%%%%%%%%%%%%%%%%%%%%%%%%%%%%%%%%%%%%%%%%%%%%%%%%%%%%%%%%%%%%%%%%%%%%%%%%%%%%%%%%%%%%%%%%%%%%%%%%%
Pluralistic homophily results from several different foci. The users that share more communities (they are found in overlapping parts) have more homophilous foci in common, i.e., they feature aspects of pluralistic homophily. Such users are more densely connected \cite{yang2013overlapping} forming a network core \cite{yang2014overlapping}. Hence we ask whether the nodes in these parts tend to have higher \textit{social capital}. However, we find no correlation between community membership and social capital, except that the nodes in more communities tend to be slightly more popular than active (see Fig.\ \ref{fig:comm_membership_soc_capital}, right). Interestingly, the same holds for \textit{semantic capital}: nodes with higher community membership are no more likely to be semantically rich than those belonging to less communities (see Fig.\ \ref{fig:PLHom_7s}, left).
This is a particularly surprising result, as such nodes, with higher community membership, are tied with their friends from different communities around different foci, according to the theory of focused interaction \cite{feld1981focused}. We would expect them to be semantically richer, since they talk on several additional topics, as their community membership grows. However, as they are not semantically richer, then our next assumptions is that such nodes must be less similar to their neighbors on average. Indeed, this is true, as presented in Fig.\ \ref{fig:PLHom_7s} (middle). The correlation between similarity to an average neighbor and community membership is highly negative and significant $-0.83, p=0.003$. A concept of \textbf{opinion leaders} \cite{rogers1970homophily} defines them as \textit{the members of the group sought by others for opinion and advice and they are said to posses features and conformity to the norms that make them super-representative or similar to their average follower}. Hence, the users with increased community membership in our network potentially represent opinion leader withing their different communities.
So far we find no correlation between social or semantic capital and community membership. However, we ask what about \textit{status inconsistency}. Since status inconsistency can be negative, we correlate its median value against community membership (although, similar result holds for mean value, as well). As presented in Fig.\ \ref{fig:PLHom_7s} (right), status inconsistency grows with community membership (r = $0.87$ and p = $0.001$). With this we reveal that the users in more communities do not have higher social or semantic status, but they can be characterized by increased status inconsistency. As mentioned in introduction, status inconsistency is suggested to be an attribute of individuals who are \textbf{drivers of social change} \cite{lenski1954status}. Therefore, we conclude that individuals featuring pluralistic homophily in communication networks are likely to be the opinion leaders and drivers of social change within their communities.

\section{Discussion and conclusion}
\label{sec:discussion}
Despite the vast and growing literature and research on what interweaves people in social networks, the interplay of homophily and influence as the main factors for social correlation with the network is still not fully explored and understood. Our first set of findings quantify to what extent semantic homophily and social influence affect the communication, its propensity and intensity in online social networks, though we are not trying to distinguish between these two factors. Concretely, we analyze interplay of semantic relatedness and communication intensity and show that while their correlation is low, their relationship is strongly captured by subtle communication network properties. 

Next we show that several types of homophily are present in $\mathtt{communication\ network}$, such as value (topics, sentiment) and status (social and semantic capital) homophily. Introduced social and semantic status metrics allow us to exhibit their growth with strength of the links (both, in terms of reciprocal communication and with increase in intensity). 
Assessment on how the two types of capital are affecting each other in communication network reveals diversity of relationships depending on which exact form of the two types of capitals is considered. While popularity and semantic capital are positively correlated, sentiment, inversely, is negatively correlated with social capital. In any case, we exhibit large diversity among users on the existing combinations of capitals they posses. Additional investigation on sociological concept of relative status reveals strong preference for communication with users of similar status. However, for relative social status particularly, we notice pattern of less popular users initiating more communication towards higher popularity users. We also find evidence for sociological proposition that status inconsistency of one or both of the parties increases communication effectiveness. Moreover, our data suggest a new hypothesis: this proposition holds only when both users are higher on the same status type, otherwise, communication intensity decreases compared to average.

Using temporal communication network we show that the tendencies of homophily and influence are dynamic and change their role and magnitude in time. In addition to confirming previous finding in other types of networks that similarity of users sharply grows before their link formation, we also explain in part the following decrease in similarity -- as a result of link decommission. A novel insight we make is that relative difference in social status is a stronger predictor for link decommission compared to differences on a value homophily level. 

We analyze modular and overlapping community structure of the communication layer and find evidence for Feld's theory about focused organization of social ties. Comparison of best matching between community structure in communication and in semantic layer shows that cohesive communities cannot be explained only by semantic relatedness of users, instead there need to be a foci of homophily present around which communities are formed. Further analyses reveal that geographic foci are the largest predictor for both modular and overlapping communities. However, in the case of overlapping community structure, we find that such large foci also give space for smaller but stronger foci around which sub-communities within are formed. Precisely, larger foci tend to create denser communities (i.e., those with more overlapping parts within). Explanation from sociology is a tendency of people who are connected around one foci to find or create new foci to strengthen the interaction.

Finally, we also exhibit that pluralistic homophily does not correlate with social or semantic capital; instead the users who are connected with others around several different foci tend to have lower average similarity to those neighbors, while at the same time being increasingly status inconsistent. 
\begin{table} [htp]
	\caption{\textbf{Summary of our contributions}: for each theory or question from sociology that defined the analysis we describe found evidence and/or some novel hypotheses or open questions that arise from the analysis.}\label{t:contr_summ}
	\hspace{-30pt}\begin{tabular}{|c|llll|lllll} \hline
		
		\specialcell{\textbf{Sociology;}\\ \textbf{theory and}\\ \textbf{questions}}&
		\multicolumn{4}{|c|}{\textbf{Experimental evidence}} &
		\multicolumn{4}{|c|}{\textbf{Novel hypotheses/open questions}}\\ 
		
		\hline \hline \multicolumn{9}{|c|}{\textbf{\textit{Quantification}}}\\ \hline \hline
		
		\specialcell{\textbf{Semantic}\\\textbf{homophily}}&
		\multicolumn{4}{|c|}{\specialcell{Comm. propensity ($\hat{{cp}}$) and intensity\\($CI$) increase with SR.}} &

		\multicolumn{4}{|c|}{\specialcell{CI increases with \textbf{status inconsistency},\\when both users are high on the same\\status dimension; otherwise CI decreases.}}\\ \hline
	
		\specialcell{\textbf{Status level}\\\textbf{homophily}}&
		\multicolumn{4}{|c|}{\specialcell{(Un)directed degree assortativity;\\increases with tie strength and CI. $\implies$\\ 
				%\textbf{With higher CI} \\
		\textbf{On higher CI, Twitter is more a}\\ \textbf{social than information network.}}} &
		\multicolumn{4}{|c|}{\specialcell{\textbf{in-in} and \textbf{out-out} deg. assortativity\\coefficients obey a different pattern\\to others with increase in CI.\\Tendency of users with a \textbf{particular}\\ \textbf{popularity difference} to interact.}}\\	 \hline
		
		\specialcell{\textbf{Value level}\\\textbf{homophily}}&
		\multicolumn{4}{|c|}{\specialcell{Attribute assortativity on semantic\\aspects of comm., such as topics,\\sentiment, semantic diversity.\\}} &
		\multicolumn{4}{|c|}{\specialcell{Semantic diversity and negative\\sentiment increase with comm. activity.}}\\ \hline

		\hline \hline\multicolumn{9}{|c|}{\textit{\textbf{Temporal evolution}}}\\ \hline \hline

		\specialcell{\textbf{Semantic}\\ \textbf{homophily}\\\textbf{evolution}}&
		\multicolumn{4}{|c|}{\specialcell{Average increase through time in SR\\among communicating users.\\The increase is driven by semantic homophily\\and social influence.}} &
		\multicolumn{4}{|c|}{\specialcell{Average increase through time in SR\\among users who \textbf{never communicated}.\\The increase is driven by external influence.}}\\ \hline
		
		\specialcell{\textbf{Heterophilous}\\ \textbf{links}\\ \textbf{dissolution}}&
		\multicolumn{4}{|c|}{\specialcell{Dissolution more due to social status and\\less due to semantic value heterophily.\\ Persisting pairs having more weak\\ contacts are increasingly likely\\to stop communicating.}} &
		\multicolumn{4}{|c|}{\specialcell{At the time of a link dissolution,\\one or both of the participating users\\are likely to have found a new contact\\that will replace the one being\\disconnected.}}\\ \hline
		
		\hline \hline	\multicolumn{9}{|c|}{\textbf{\textit{Community foci}}}\\ \hline \hline
		
		\specialcell{\textbf{Theory of}\\ \textbf{focused}\\\textbf{interaction}}&
		\multicolumn{4}{|c|}{\specialcell{Semantic similarity in terms of SR\\only moderately explains structural\\communities. Modular communities\\ explained by \textbf{geolocation entities} as\\comm. foci.}} &
		\multicolumn{4}{|c|}{\specialcell{}}\\ \hline
		
		\specialcell{\textbf{Pluralistic}\\ \textbf{homophily}\\\textbf{}}&
		\multicolumn{4}{|c|}{\specialcell{Overlapping communities\\formed around other foci enabled\\by overarching geolocation foci.}} &
		\multicolumn{4}{|c|}{\specialcell{High correlation between size of a\\ community and its density of overlap.\\Pluralistic homophily is not explained\\by social or semantic capital.\\On the other hand, individuals\\exhibiting pluralistic homophily\\are increasingly \textbf{status inconsistent.}}}\\ \hline
		
		\end{tabular}
\end{table}

\subsection{Limitations}
A limitation of our work posed by the restricted dataset is that we are not considering the entire Twitter channel for information flow, as there are also considerable amount of information flowing along the retweet network, which is not taken into consideration in this work. Besides this, the mention mechanism in Twitter can be sometimes biased towards specific target audiences for specific information \cite{tang2015locating}.

Another limitation is that our results are solely about computer-mediated communication and we do not tackle the impact of Internet (online medium) on social interaction.

Further investigation is needed on the influence of the threshold for semantic relatedness on the semantic homophily, as we show in this work that the semantic layer became disassortative after threshold equal to $0.6$. Additional and improved sentiment analysis is needed to understand how the social reinforcement influences communication between users and if there exists happiness paradox while people communicate in social network.

\section{Methods}\label{sec:methods}
In this section we describe how we build Wikipedia-based semantic database using an English pages dump (52GB in size, uncompressed). The first step is to take the article texts as the algorithm builds on the large amount of knowledge they provide. We then apply an open-source script \textit{wikiextractor} \cite{wikiextractor} to pre-process and clean the texts.
The ESA algorithm is based on the TF-IDF (term frequency - inverse document frequency) \cite{baeza1999modern} scores of words in different articles in the Wikipedia corpus. As a result a word $w_1$ is mapped to the \textit{concept vector} $CV(w_1)=\{(C_1^1,V_1^1),\ (C_2^1,V_2^1),\ (C_3^1,V_3^1),\ ...,\ (C_{M1}^1,V_{M1}^1)\}$. $C_j^1$ represent Wikipedia concepts and $V_j^1$ are TF-IDF scores for the word $w_1$ in those articles and are calculated as follows:
\begin{equation}
\centering
V_j^1= TF \cdot IDF = (1 + \log(f_{1,j})) \cdot \log(\frac{N}{n_t}) ,
\end{equation}
where $TF$ is the log-normalized raw frequency ($f_{1,j}$) of the word $w_1$ in article $j$, and $IDF$ is the inverse document frequency, $N$ is the number of articles, and $n_t$ is the number of articles in which the word $w_1$ is present.

The algorithm was implemented in Python with application of the scikit-learn machine learning library \cite{scikit-learn} and the resulting database was stored in a MongoDB collection. Since some of the concept vectors might have tens of thousands of terms; prior to storing, we apply the pruning process \cite{gabrilovich2009wikipedia} that for each word keeps only important $CV$ elements. The algorithm implementation needs tuning several parameters, and in this process we also consult some of the existing implementations of the ESA algorithm. Our implementation of ESA is open-source and published on Github \cite{sanja7s}.
%%%%%%%%%%%%%%%%%%%%%%%%%%%%%%%%%%%%%%%%%%%%%%%%%%%%%%%%%%%%%%%%%%%%%%%%%%%%%%%%%%%%%%%%%%%%%%%%%%%%%
\subsubsection{Word Semantic Relatedness}
\label{sec:words}
%%%%%%%%%%%%%%%%%%%%%%%%%%%%%%%%%%%%%%%%%%%%%%%%%%%%%%%%%%%%%%%%%%%%%%%%%%%%%%%%%%%%%%%%%%%%%%%%%%%%%
The semantic relatedness (SR) between words is not measured directly, but it is rather determined through a set of concepts highly related to them \cite{gabrilovich2009wikipedia,hieu2013extracting}. Let us assume that the SR between words $w_1$ and $w_2$ is requested. The word SR calculation follows the two steps below.
\begin{itemize}
	\item
	\textbf{Determining the corresponding CVs derived from Wikipedia for the words $w_1$ and $w_2$}. The CVs are based on concepts (or articles) of Wikipedia which are related to the words. Let us assume that $w_1$ is mapped to \textit{concept (tf-idf) vector}: $CV(w_1)=\{(C_1^1,V_1^1),\ (C_2^1,V_2^1),\ (C_3^1,V_3^1),\ ...,\ (C_{M1}^1,V_{M1}^1)\}$ and $w_2$ is mapped to \textit{concept (tf-idf) vector}: $CV(w_2)=\{(C_1^2,V_1^2),\ (C_2^2,V_2^2),\ (C_3^2,V_3^2),\ ...,\ (C_{M2}^2,V_{M2}^2)\}$. These are the sets of Wikipedia concepts, $C_j^1$ and $C_j^2$, which are related to the word $w_1$ and $w_2$ and their TF-IDF scores, $V_j^1$ and $V_j^2$, respectively. In the following, we will assume that $N$ is the number of common concepts in $CV(w_1)$ and $CV(w_2)$.
	\item \textbf{Calculating the SR between words using cosine similarity between obtained CVs}.
	For measuring the degree of semantic relatedness, cosine similarity between the $CV$s for two words $w_1$ and $w_2$ is calculated. This measure gives the cosine of the angle between the two vectors $CV(w_1)$ and $CV(w_2)$. The cosine measure can be re-formulated for our purpose as follows:
	\begin{equation}\label{eq:WSRel}
	SR(w_1, w_2) = \cos(CV(w_1), CV(w_2)) =\frac{\sum_{i=1}^{N} V_i^1 \cdot V_i^2}{\sqrt{\sum_{k=1}^{M1} \left(V_k^1\right)^{2}} \cdot \sqrt{\sum_{l=1}^{M2} \left(V_l^2\right)^{2}}},
	\end{equation}
	where $i$ iterates over the common concepts.
\end{itemize}
The $SR(w_1, w_2)$ values range from 0 (i.e., no semantic relatedness) to 1 (i.e., perfect semantic relatedness) as the TF-IDF weights can not be negative.
%%%%%%%%%%%%%%%%%%%%%%%%%%%%%%%%%%%%%%%%%%%%%%%%%%%%%%%%%%%%%%%%%%%%%%%%%%%%%%%%%%%%%%%%%%%%%%%%%%%%%
\subsubsection{Document Semantic Relatedness}
\label{sec:SR_docs}
The semantic relatedness (SR) between documents is measured through the SR of the words found in the documents. Let us assume that the SR between documents $d_1$ and $d_2$ is requested. The document SR calculation follows the three steps below.
\begin{itemize}
	\item
	\textbf{Analyzing documents using the term frequency (TF) approach which finds the frequency of words in the document}. The result of this step is a list of important words with their corresponding TF scores. Let us assume that:
	
	$d_1$ is analyzed to \textit{term (tf) vector}:  $T(d_1)=\{(t_1^1,v_1^1),\ (t_2^1,v_2^1),\ (t_3^1,v_3^1),$ $ \ ...,\ (t_m^1,v_m^1)\}$, 
	
	$d_2$ to \textit{term (tf) vector}:   $T(d_2)=\{(t_1^2,v_1^2),\ (t_2^2,v_2^2),\ (t_3^2,v_3^2), $ $ \ ...,(t_n^2,v_n^2)\}$, and $m<n$.
	\item
	\textbf{Determining the corresponding $CV$s derived from Wikipedia for the documents $d_1$ and $d_2$.} For each term in the lists $T(d_1)$ and $T(d_2)$ we derive their individual $CV$s (as described for words in Section \ref{sec:words}). For instance, the $t_1^1$ term is mapped to \textit{concept (tf-idf) vector}: $CV(t_1^1)=\{(C_1^1,(v_1^1 \times V_1^1)),\ (C_2^1,(v_1^1 \times V_2^1)),\ (C_3^1,(v_1^1 \times V_3^1))$ , $...,\ (C_M^1,(v_1^1 \times V_M^1))\}$. The other terms in $T(d_1)$ can be represented in a similar way.  When summarizing the $CV$s for one document, the $CV$ for each term is multiplied with its \textit{TF} score in the document (found in the previous step). If the terms in $T(d_1)$ have the same concepts in their $CV$s, we sum the weighted TF-IDF scores of those concepts. After this process we obtain $CV(d_1)$, the list of Wikipedia concepts and TF-IDF scores which are related to all the terms in $T(d_1)$. Similarly, for $d_2$ the list of relevant Wikipedia concepts and TF-IDF scores is found in $CV(d_2)$.
	\item
	\textbf{Calculating the SR between documents using cosine similarity between obtained CVs}.
	Finally, we obtain the $SR(d_1, d_2)$ between documents by calculating the cosine similarity of $CV(d_1)$ and $CV(d_2)$ (see Eq.~\ref{eq:WSRel}).
\end{itemize}
%%%%%%%%%%%%%%%%%%%%%%%%%%%%%%%%%%%%%%%%%%%%%%%%%%%%%%%%%%%%%%%%%%%%%%%%%%%%%%%%%%%%%%%%%%%%%%%%%%%%%
\subsubsection{SR database evaluation}
\label{sec:eval_SR}
%%%%%%%%%%%%%%%%%%%%%%%%%%%%%%%%%%%%%%%%%%%%%%%%%%%%%%%%%%%%%%%%%%%%%%%%%%%%%%%%%%%%%%%%%%%%%%%%%%%%%
The English version of Wikipedia used includes over 2.5 million articles. Since many of the articles are highly specialized, and due to the described pruning process, we find only around $15\%$ of those articles ($387,992$) relevant for our tweets corpus. In a similar manner as in the original paper \cite{gabrilovich2009wikipedia}, we evaluate the quality of the SR database that we built against available datasets with human judgment for word pairs relatedness. We use several such datasets available online, as one of the most comprehensive current resources \cite{faruqui-2014:SystemDemo}. The results of the evaluation are presented in Table \ref{t:SR_eval}. We do not provide herein a comparison with the existing implementations, since not all of them  provide their evaluation on the same datasets with human judgments, and since a previous study comparing them has shown that some of these results are incompatible \cite{cramer2008well}. However, our evaluation scores are comparable to the original implementation \cite{gabrilovich2009wikipedia} and to the ESA implementations available online.
\begin{table}
	\centering
	\caption{SR knowledge database evaluation}\label{t:SR_eval}
	\begin{tabular}{c|l|l} \hline
		\textit{Human judgments dataset}&\textit{Spearman's rank}&\textit{Pearson's correlation}\\ 
		\hline
		WordSim-353 &0.51&0.45\\
		Miller and Charles&0.79&0.82\\ 
		Word pair similarity, MTurk&0.53&0.45\\ 
		Rubenstein and Goodenough&0.81&0.74\\ 
		MEN dataset of word pair sim.&0.73&0.44\\ 
		\textbf{Average}&\textbf{0.67}&\textbf{0.58}\\
		\hline\end{tabular}
\end{table}
%%%%%%%%%%%%%%%%%%%%%%%%%%%%%%%%%%%%%%%%%%%%%%%%%%%%

\begin{comment}
	content...

% Appendix
\appendix
\section*{APPENDIX}
\setcounter{section}{1}
In this appendix, we measure
the channel switching time of Micaz [CROSSBOW] sensor devices.
In our experiments, one mote alternatingly switches between Channels
11 and 12. Every time after the node switches to a channel, it sends
out a packet immediately and then changes to a new channel as soon
as the transmission is finished. We measure the
number of packets the test mote can send in 10 seconds, denoted as
$N_{1}$. In contrast, we also measure the same value of the test
mote without switching channels, denoted as $N_{2}$. We calculate
the channel-switching time $s$ as
\begin{eqnarray}%
s=\frac{10}{N_{1}}-\frac{10}{N_{2}}. \nonumber
\end{eqnarray}%
By repeating the experiments 100 times, we get the average
channel-switching time of Micaz motes: 24.3$\mu$s.

\appendixhead{ZHOU}
\end{comment}

% Acknowledgments
\begin{acks}
S.\v{S}. research was partially financed by CIVIS EU FP7 project (FP7- SMARTCITIES-2013). I.M. work was partially financed by the Faculty of Computer Science and Engineering at the University ''Ss. Cyril and Methodius''. S.\v{S}. and I.M. also gratefully acknowledge the CyberTrust research project for their support. B.G. thanks the Moore and Sloan Foundations for support as part of the Moore-Sloan Data Science Environment at New York University. P.H. thanks General Research Fund 26211515 from the Research Grants Council of Hong Kong.
S.\v{S}. thanks A. Ukkonen for the help with the SR database implementation. S.\v{S}. also acknowledges collaboration with P. T. Trung during his MSc thesis project when we performed a similar type of SR analysis on Twitter data. The authors also thank A. Gionis for the helpful discussion and for reviewing the manuscript.
\end{acks}

% Bibliography
\bibliographystyle{ACM-Reference-Format-Journals}
\bibliography{the}
                           
% Electronic Appendix
%\elecappendix

\begin{comment}
\medskip

%\section{This is an example of Appendix section head}

Channel-switching time is measured as the time length it takes for
motes to successfully switch from one channel to another. This
parameter impacts the maximum network throughput, because motes
cannot receive or send any packet during this period of time, and it
also affects the efficiency of toggle snooping in MMSN, where motes
need to sense through channels rapidly.

By repeating experiments 100 times, we get the average
channel-switching time of Micaz motes: 24.3 $\mu$s. We then conduct
the same experiments with different Micaz motes, as well as
experiments with the transmitter switching from Channel 11 to other
channels. In both scenarios, the channel-switching time does not have
obvious changes. (In our experiments, all values are in the range of
23.6 $\mu$s to 24.9 $\mu$s.)

%\section{Appendix section head}

The primary consumer of energy in WSNs is idle listening. The key to
reduce idle listening is executing low duty-cycle on nodes. Two
primary approaches are considered in controlling duty-cycles in the
MAC layer.
\end{comment}

\end{document}